%% file: supercdms_snowmass_2021.tex
\documentclass[10pt,onecolumn]{article}
\usepackage{graphicx}
\usepackage{amsmath,amssymb,wasysym}
\usepackage{fancyhdr}
\usepackage[compact]{titlesec}
\usepackage{floatflt}
\usepackage{wrapfig}
\usepackage{xcolor}
\usepackage{ulem}
\usepackage{hyperref}
\usepackage{enumitem}
\usepackage[font={footnotesize,sl}]{caption}
\usepackage{textcomp}
\usepackage{multicol}

\input{ltptf_macros}

\newcommand\snowmass{
\begin{center}
  \rule[-0.2in]{\hsize}{0.01in}\\
  \rule{\hsize}{0.01in}\\
  \vskip 0.1in
  Submitted to the Proceedings of the US Community Study\\
  on the Future of Particle Physics (Snowmass 2021)\\
  \rule{\hsize}{0.01in}\\
  \rule[+0.2in]{\hsize}{0.01in}\\[-2em]
\end{center}
}

\usepackage[firstpage=true]{background}
\backgroundsetup{contents={\parbox{6.5in}{\snowmass}}, scale=1,placement=top,opacity=1,color=black,position={3.25in,1.2in}}

\usepackage{fancyhdr}
\fancypagestyle{plain}{%
  \fancyhf{}%
  \fancyhead[C]{}
  \fancyfoot[C]{}
}

\fancypagestyle{empty}{%
  \fancyhf{}%
  \fancyhead[C]{{\it \titletext{} }}
  \fancyfoot[C]{\thepage}
}
\newcommand\titletext{A Strategy for Low-Mass Dark Matter Searches \\
with Cryogenic Detectors in the SuperCDMS SNOLAB Facility}
\title{\titletext{}}

\date{}
\usepackage{authblk}

\input{authors}

\begin{document}

\maketitle

%\cfoot{}

% Executive summary is sufficient
%\begin{abstract}
%\noindent abstract
%\end{abstract}

\vspace{-24pt}
\begin{center}
\begin{tabular}{l}
Version 1: March 15, 2022 \\
Version 2: October 19, 2022 \\
Version 3: April 1, 2023 \\
\end{tabular}
\end{center}

\clearpage

\setcounter{page}{1}
\pagenumbering{roman}

%\lhead{SuperCDMS Snowmass 2021}
%\rhead{\thedate}
%\cfoot{\thepage}

\section*{Executive Summary}    
The SuperCDMS Collaboration is currently building SuperCDMS SNOLAB, an experiment designed to search for nucleon-coupled dark matter in the 0.5--5~\GeV mass range.  Looking to the future, the Collaboration has developed a set of experience-based upgrade scenarios, as well as novel directions, to extend the search for dark matter using the SuperCDMS technology in the SNOLAB facility.  The experienced-based scenarios are forecasted to probe many square decades of unexplored dark matter parameter space below 5~\GeVnosp, covering over 6 decades in mass: 1--100~eV for dark photons and axion-like particles, 1--100~\MeV for dark-photon-coupled light dark matter, and 0.05--5~\GeV for nucleon-coupled dark matter.  They will reach the nucleon-coupled neutrino fog in the 0.5--5~\GeV mass range, and they will test a variety of benchmark models and sharp targets for electron-coupled dark matter.  The novel directions involve greater departures from current SuperCDMS technology but promise even greater reach in the long run, and their development must begin now for them to be available in a timely fashion.

The experienced-based upgrade scenarios rely mainly on dramatic improvements in detector performance based on demonstrated scaling laws and reasonable extrapolations of current performance.  Importantly, \textbf{these improvements in detector performance obviate significant reductions in background levels beyond current expectations for the SuperCDMS SNOLAB experiment.}  Given that the dominant limiting backgrounds for SuperCDMS SNOLAB are cosmogenically created radioisotopes in the detectors, likely amenable only to isotopic purification and an underground detector life-cycle from before crystal growth to detector testing, the potential cost and time savings are enormous and the necessary improvements much easier to prototype.

This executive summary outlines the detector upgrades, the new dark matter search modes they enable, the resulting expected science reach, and the novel directions under consideration.  In the full document that follows, Section~\ref{sec:overview} reviews the SuperCDMS SNOLAB experiment under construction, including updated sensitivity expectations, and it provides a high-level overview of future opportunities, both experienced-based upgrades and novel directions.  Section~\ref{sec:experience_forecasts} provides detailed sensitivity forecasts for the experience-based upgrades.  Section~\ref{sec:novel_forecasts} presents specific activities being undertaken now to explore specific novel directions.

\subsection*{Detector Upgrades}

The detector upgrades contemplated are summarized as follows:

\begin{description}

\item[Detector size:] Smaller detectors can quickly provide improvements in energy resolution and threshold based on current technology while providing sufficient target mass to explore significant new parameter space.  We consider detectors roughly 25$\times$ and 25$^2\times$ smaller in size than the kg-scale SuperCDMS SNOLAB detectors.

\item[Phonon Energy Resolution:] Physically motivated, experimentally validated scaling laws promise up to three orders of magnitude improvement in phonon energy resolution.  Such vast gains open up new modes of searching for dark matter, described below, and also help to mitigate ionization collection non-idealities.

\item[Ionization Energy Resolution:] An order of magnitude improvement in ionization resolution is possible via reductions in detector size and modifications to readout electronics.  Such an advance would make traditional nuclear-recoil discrimination via ionization yield effective for masses as low as 0.5--1~\GeV (vs.\ 5--10~\GeV currently).

\item[Improvement of Ionization Collection Non-Idealities:] Improvements in ionization leakage, leakage pileup, and impact ionization and charge trapping in HV detectors will reduce effective recoil energy thresholds.

\end{description}

\subsection*{New Dark Matter Search Modes}

These upgrades enable new modes of searching for dark matter:

\begin{description}

\item[Spectral Shape Discrimination:] Below about 0.5~\GeV dark matter mass, the spectrum of energy depositions by nucleon-coupled dark matter becomes steeper than conventional particle backgrounds, including coherent elastic neutrino-nucleus scattering (\cevnsnosp) of solar neutrinos. 

\item[Event-by-Event Nuclear Recoil Discrimination with Phonons Only:] Traditional SuperCDMS iZIP detectors measure ionization yield via a direct measurement of the ionization signal.  The new ``phonon iZIP'' architecture may make it possible to encode the ionization yield in the much higher signal-to-noise phonon signal, providing another means to extend nuclear recoil discrimination to the tens of eV energy deposition scale and to reach the neutrino fog in the 0.5--5~\GeV mass range.  This architecture may even enable such discrimination down to the eV bandgap in Si and Ge, below which recoils of all types no longer create ionization.

\item[Nuclear Recoil Spectral Discrimination with HV Detectors:] Future SuperCDMS HV detectors will achieve fine enough phonon energy resolution to identify production of single electron-hole pairs.  This advance will cause nuclear and electron recoils to reside in different places in the spectrum of collected phonon energy, yielding nuclear-recoil discrimination down to recoil energies at which nuclear recoils create a single electron-hole pair.

\end{description}

\subsection*{Expected Science Reach}

These upgrade scenarios and new dark matter search modes will enable the following science goals:
\begin{description}
\item[SG-1: Nucleon couplings of sub-\GeV (0.05--0.5~\GeVnosp) dark matter ] 
\hfill \\ 
New, small detectors operated without ionization bias are able to probe the mass range 0.05--0.5~\GeV for nucleon-coupled dark matter, down to the neutrino fog over most of this regime, by use of spectral shape discrimination of dark matter from particle backgrounds, including \cevns of solar neutrinos.

\item[SG-2: Nucleon couplings of \GeV (0.5--5~\GeVnosp) dark matter down to the neutrino fog] 
\hfill \\
Three different detector types --- full-size HV, medium-size iZIP, and medium-size piZIP detectors --- all have the potential to search for nucleon-coupled dark matter in the 0.5--5~\GeV mass range down to the neutrino fog, and, with large iZIP or piZIP detector counts, well into it.

\item[SG-3: Electron couplings of kinetically mixed \eV (1--100~\eVnosp) dark photon dark matter] 
\hfill 
Small and medium-size detectors operated without bias voltage, and full-size HV detectors, can probe for dark photons a decade in mass and up to 3 decades in kinetic mixing parameter beyond current constraints in the 1--100~\eV mass range.

\item[SG-4: Electron couplings of \eV (1--100~\eVnosp) axion and axion-like particle dark matter] 
\hfill \\
The same SG-3 detectors can probe for axion-like-particle dark matter 1--1.5 decades in mass and up to 3 decades deeper in coupling strength $g_{ae}$ than current particle dark matter search constraints in the 1--100~\eV mass range, exceeding even the sensitivity of astrophysical constraints and more robustly testing the stellar cooling hint that can be interpreted as evidence for such particles.

\item[SG-5: Dark-photon-mediated couplings of \MeV (1--100~\MeVnosp) light dark matter] 
\hfill \\ 
The same SG-3 detectors can extend the reach for light dark matter in the 1--100~\MeV mass range, coupled via a heavy or light dark-photon mediator, by two-thirds of a decade in mass and 1--2 decades in cross section.  This reach would test a wide range of sharp targets and benchmark models --- ELDER, SIMP, Elastic Scalar, Asymmetric Fermion, Freeze-in, and Freeze-out models.

\end{description}

\subsection*{Novel Directions}

Beyond the above experience-based upgrade scenarios, novel directions in detector development will also be initiated.  High-sound-speed materials may provide enhanced nucleon couplings to sub-\GeV dark matter.  Polar materials can extend the reach for dark-photon-coupled fermionic light dark matter to the few \keV thermal limit.  Small-gap materials may extend ionization-yield-based electron recoil rejection to lower energies and thus masses. Large-gap materials may improve ionization collection non-idealities in HV detectors.  The potential for diurnal modulation in a variety of detector responses --- ion displacement energy, displaced ion ionization yield, and collective excitation coupling --- could extend the search for dark matter deep into the neutrino fog and provide a smoking gun signal in the case of a detection.

There is strong potential for collaboration in the continued development of these technologies and the mitigation of new low-energy particle and environmental backgrounds. 

\clearpage

\begin{figure}
\begin{center}
\begin{tabular}{cc}
\multicolumn{2}{c}{SG-1, SG-2: Nucleon-Coupled DM, 0.05--5~\GeV} \\
%\multicolumn{2}{c}{\red{Tarek to add Ge piZIP detB 10cm3 x20 curve}} \\
%& \textcolor{magenta}{TS: 10/18 Done}\\
\includegraphics*[width=0.48\textwidth,viewport=0 0 576 410]{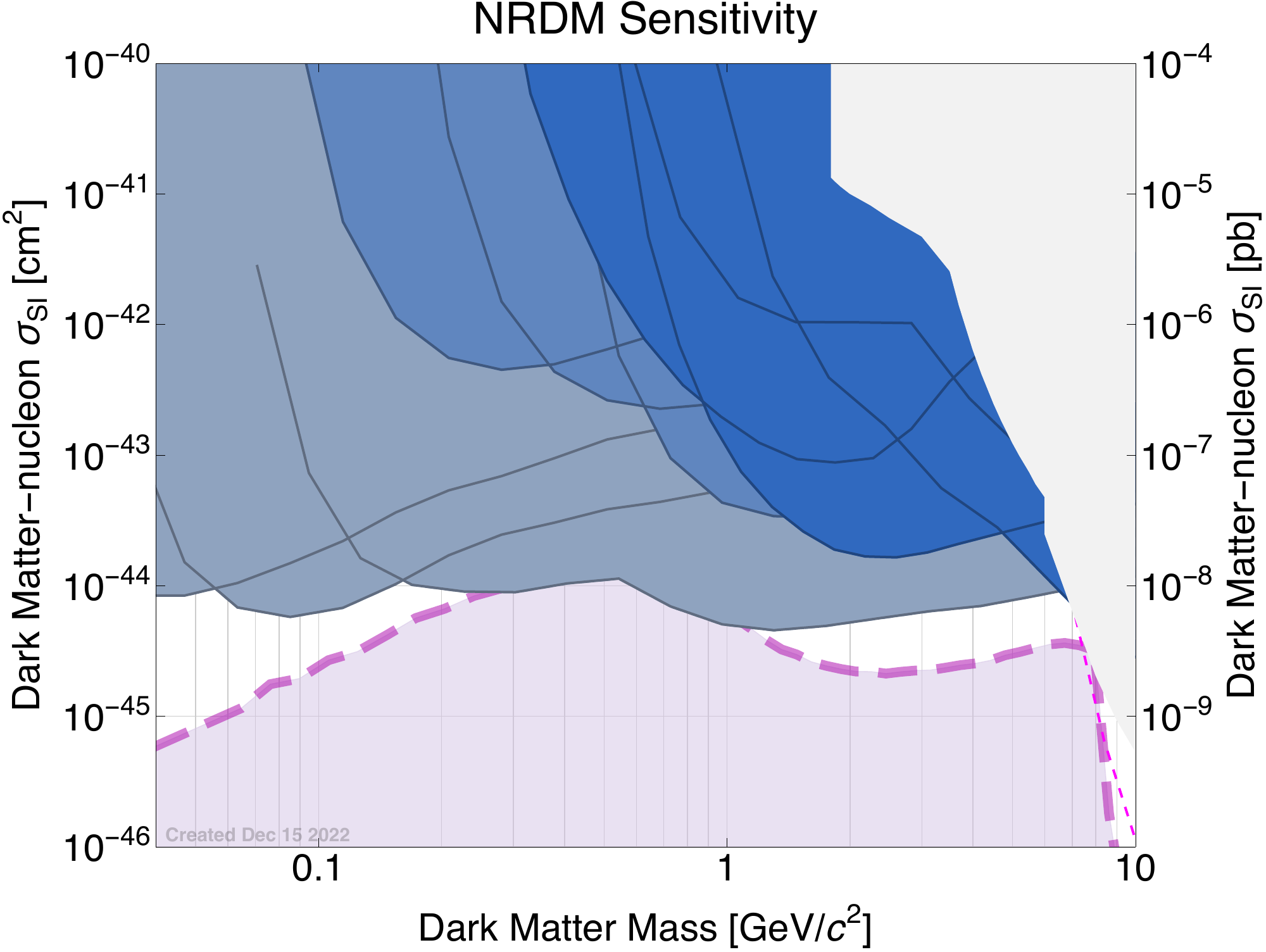} &
\includegraphics*[width=0.48\textwidth,viewport=0 0 576 410]{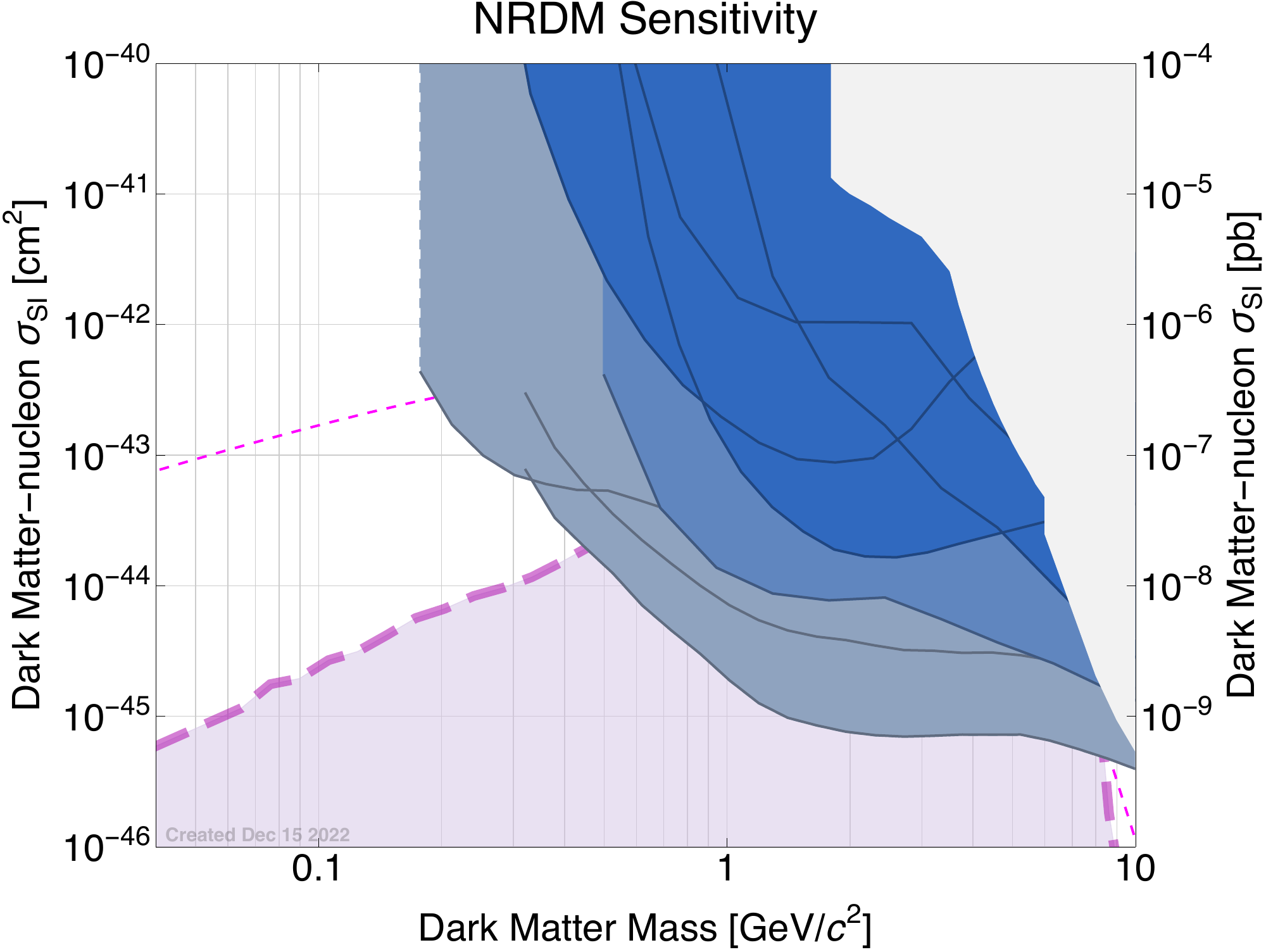}\\
\vspace{-6pt} & \\

SG-3: Dark Photon DM, 1--100~\eV & SG-4: Axion-Like-Particle DM, 1--100~\eV \\
\includegraphics*[width=0.48\textwidth,viewport=0 0 576 410]{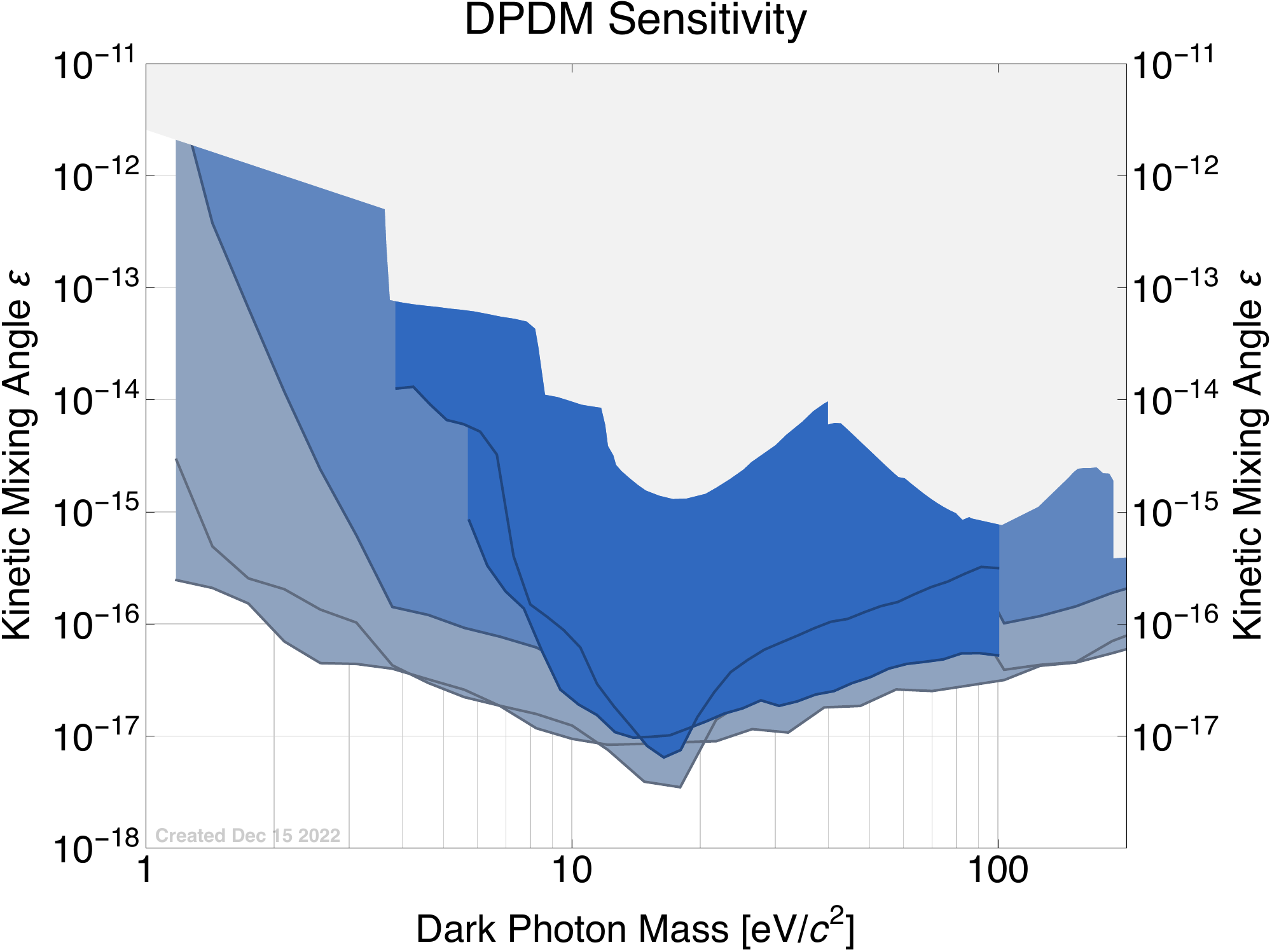} &
\includegraphics*[width=0.48\textwidth,viewport=0 0 576 410]{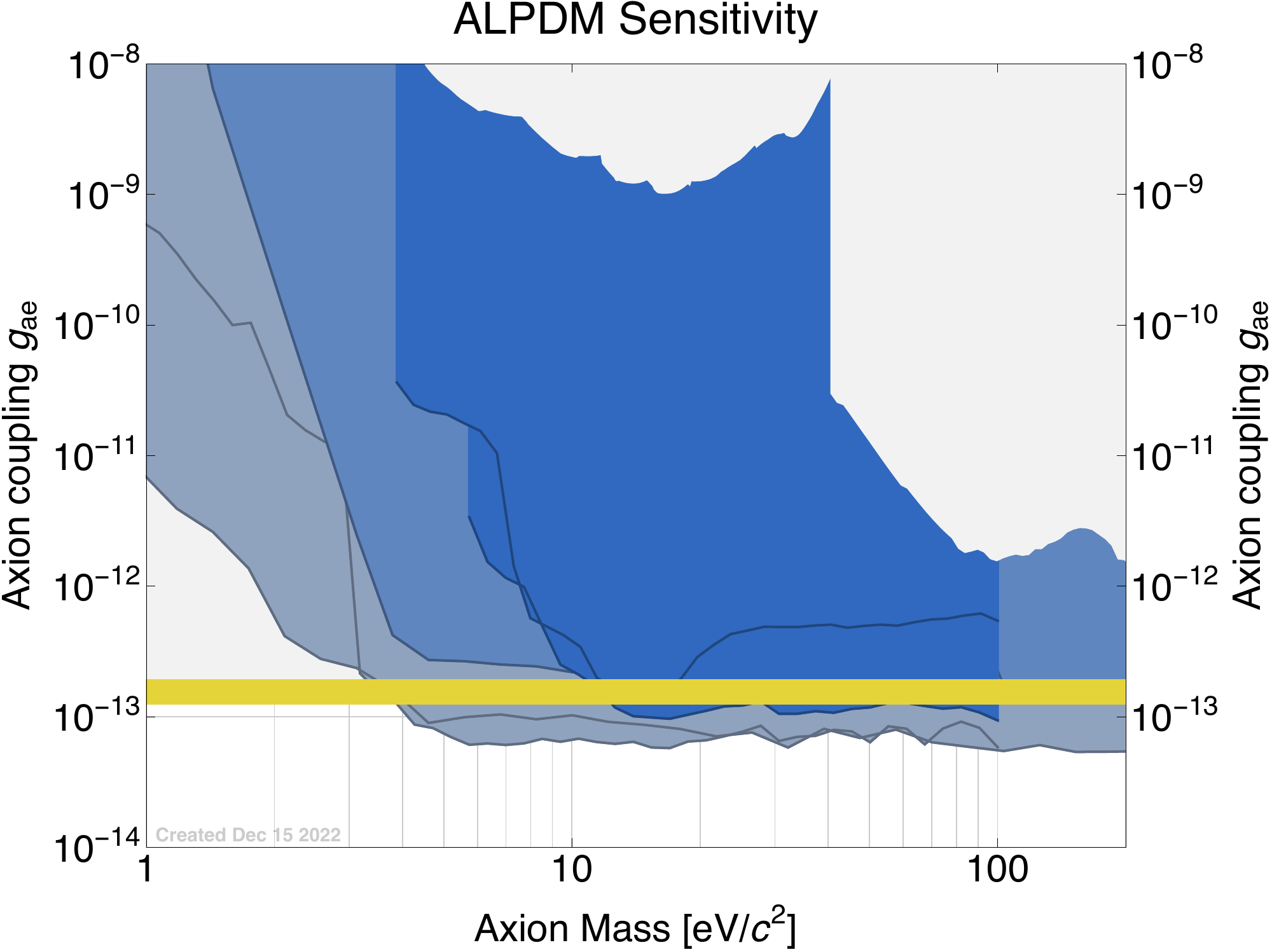}\\
\vspace{-6pt} & \\

\multicolumn{2}{c}{SG-5: Dark-Photon-Coupled Light Dark Matter, 1--100~\MeV} \\
Heavy mediator ($F(q) = 1$)
& Light mediator ($F(q) = 1/q^2$) \\
\includegraphics*[width=0.48\textwidth,viewport=0 0 576 410]{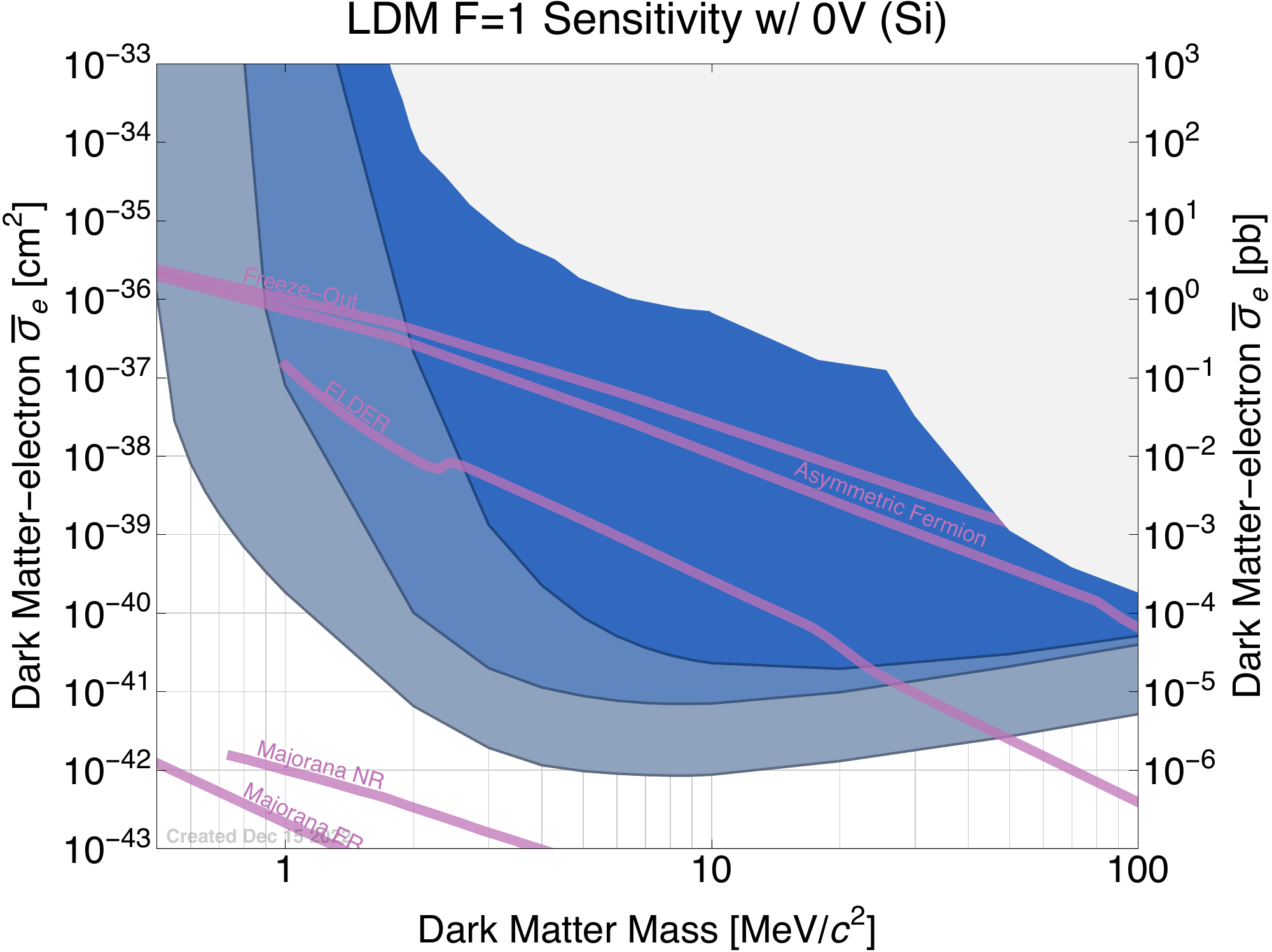} &
\includegraphics*[width=0.48\textwidth,viewport=0 0 576 410]{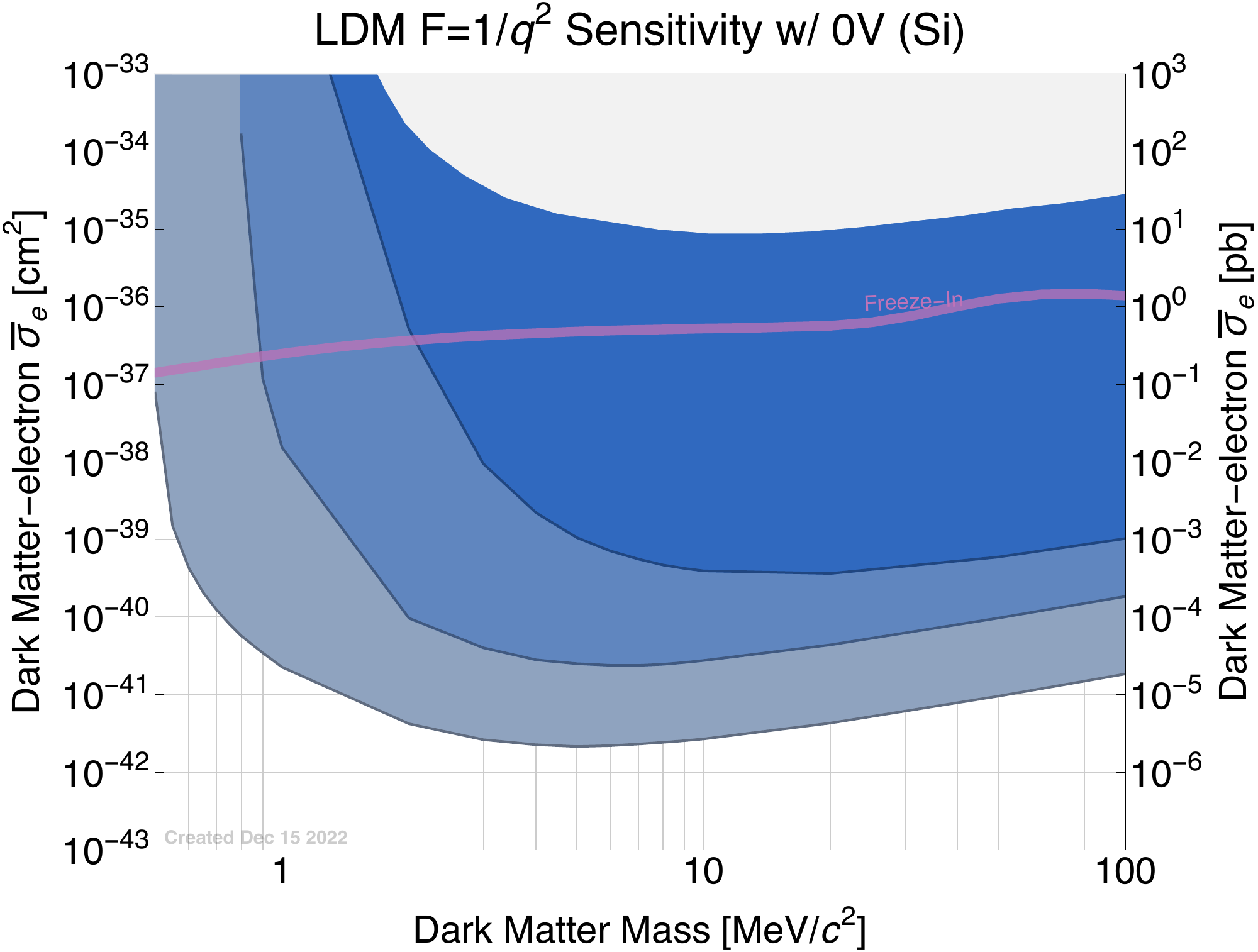} \\
\end{tabular}
\end{center}
\vspace{-6pt}
\caption[Summary of future science reach of the SuperCDMS technology via upgrades in the SNOLAB facility]{\textbf{Summary of future science reach of the SuperCDMS technology via upgrades in the SNOLAB facility.}  For each of the science goals (SG), we display a subset of sensitivity forecasts summarizing the potential of the SuperCDMS technology, ranging from scenarios could be implemented with in-hand detector performance and background improvements (light blue) to scenarios that exploit the full reach possible with the current cryostat and facility but that rely only on the same in-hand background improvements~($\mathit{T = 15}$~mK; grey-blue) .  Section~\ref{sec:experience_forecasts} details these forecasts, by science goal.   Where multiple detector types yield a particular shaded region, lines are shown for each detector type.  Also shown are current constraints~(light grey shading) and the expected reach of the SuperCDMS SNOLAB experiment under construction~(dark blue).  For axion-like-particle dark matter, the yellow band signifies the parameter space consistent with the hint from stellar cooling~\cite{stellarcooling2016}.  For light dark matter, the sharp targets presented in~\cite{cvdm2017} are shown, assuming $\mathit{M_{A'} = 3 \mdm}$ where $A'$ is the dark photon and $\chi$ is the dark matter.}
\label{sec:exec_summary_plots}
\end{figure}

\clearpage

\newpage
\setcounter{secnumdepth}{5}
\setcounter{tocdepth}{4}
\tableofcontents 
\listoftables
\listoffigures

\newpage

\setcounter{page}{1}
\pagenumbering{arabic}

\section{Overview}
\label{sec:overview}

This section provides a summary of the SuperCDMS SNOLAB experiment under construction, including updated sensitivity expectations, followed by an overview of future opportunities. 

\subsection{SuperCDMS SNOLAB}
\label{sec:supercdms_snolab}

The planned SuperCDMS SNOLAB experiment will be located approximately 2~km underground within SNOLAB in Sudbury, Ontario, Canada. The SNOLAB rock overburden provides shielding against cosmic-ray secondaries equivalent to 6010 meters of water. The experiment will be located within the ``ladder lab'' drift at SNOLAB~\cite{Duncan:2010zz}.

For the sake of providing a self-contained document, we reprise, at a qualitative level, much of the summary of the experimental design and backgrounds discussed in~\cite{sensitivity2016}, updating where necessary.

\subsubsection{The SuperCDMS Detectors}
\label{sec:Detectors}

SuperCDMS SNOLAB will include a mixture of detectors composed of silicon (Si) and germanium (Ge).   These detectors consist of cylindrical crystals, 100
mm in diameter and 33.3~mm thick. Each Ge~(Si) crystal has a mass of 1.39~(0.61)~kg. Two detector designs, denoted iZIP and HV, have common physical dimensions and are fabricated from the same materials using the same techniques.  iZIP and HV detectors are differentiated by details of the superconducting sensors (patterned lithographically on the top and bottom surfaces), the operating bias voltages, and whether ionization is sensed directly or via its drift phonon production.  iZIP detectors~\cite{Agnese:2014PRL} discriminate nuclear recoils (NRs) due to nucleon-coupled dark matter, neutrons, and neutrinos from electron recoils (ERs) due to other backgrounds on the basis of ionization yield, while HV detectors~\cite{Agnese:2015nto} use the phonon signal generated by drifting charges to achieve a lower recoil threshold at the cost of ionization-yield-based discrimination.  
%The HV and iZIP detector technologies complement one another by providing, respectively, access to lower-energy recoils without recoil-type discrimination and, for recoil energies greater than 2 k\evnrnosp, the ability to discriminate the primary recoil type. 

\begin{figure}[b!]
\vspace{12pt}
\begin{center}
\includegraphics[width=.45\textwidth]{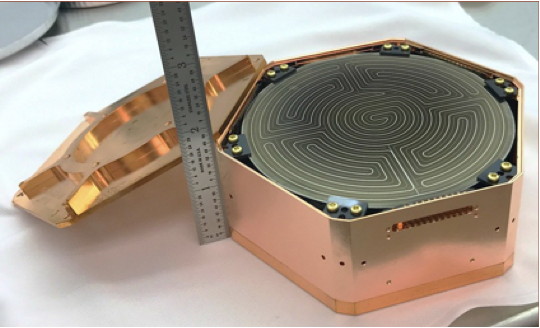}
\includegraphics[width=.45\textwidth]{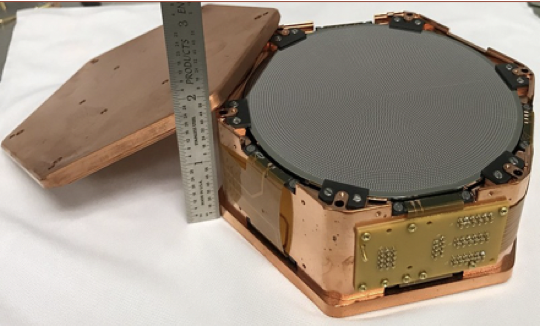} \\
\includegraphics[width=.45\textwidth, trim={600 200 600 350},clip]{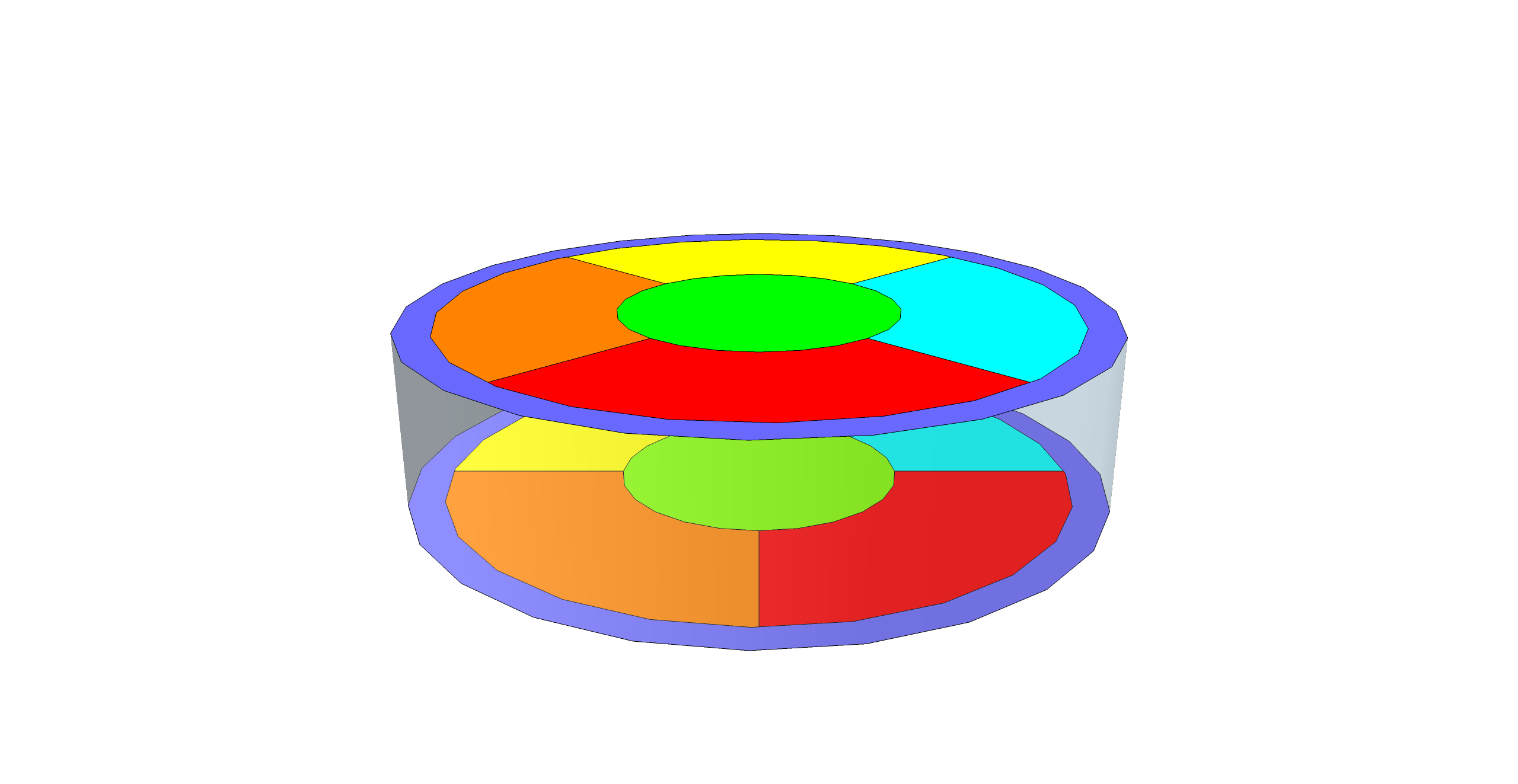}
\includegraphics[width=.45\textwidth, trim={600 200 600 350},clip]{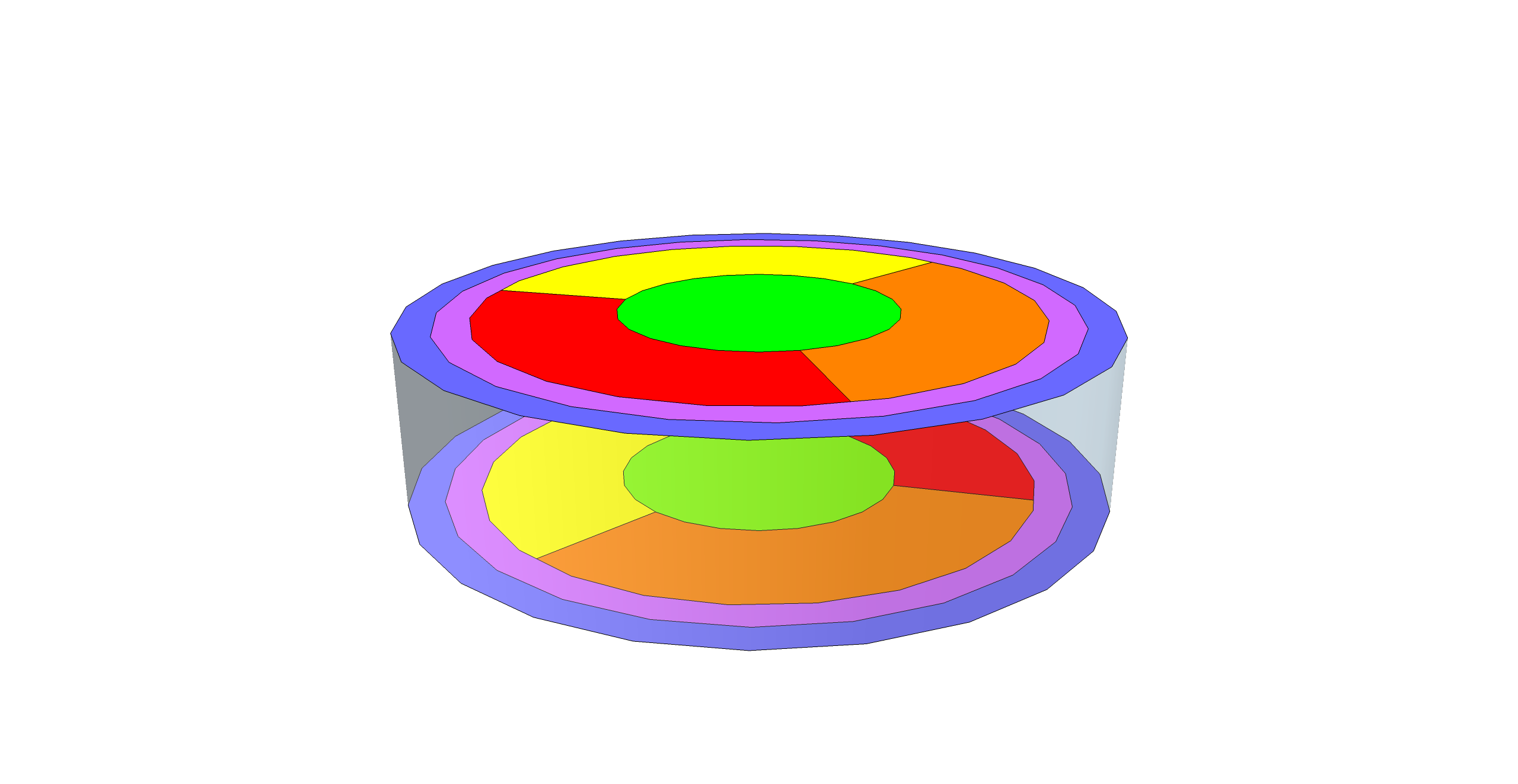}
\end{center}
\caption[SuperCDMS SNOLAB detectors.]{\textbf{SuperCDMS SNOLAB detectors.} (Left)~iZIP.  (Right)~HV.  (Top)~Prototype detectors in their housings.  (Bottom)~Channel layout. 
The interleaved Z-sensitive Ionization Phonon (iZIP) detector has six phonon channels on each side, arranged as an inner core surrounded by four wedge-shaped channels and one outer ring. Each channel contains hundreds of lithographically defined superconducting sensors. An ``outer'' ionization channel shares the same area and is interleaved with the outermost phonon ring, and an ``inner'' ionization channel is interleaved with the remaining phonon channels. The wedge channels on the bottom surface are rotated by 45$^\circ$ with respect to those on the top.
The HV detector has six phonon channels on each side: an inner core surrounded by three wedge-shaped channels and two outer rings designed to reject events near the edge. The wedge channels on the bottom surface are rotated by 60$^\circ$ with respect to those on the top.}
\label{fig:DetectorSchematic}
\vspace{-12pt}
\end{figure}

On the iZIP detectors (Figure~\ref{fig:DetectorSchematic}), the top and bottom surfaces are each instrumented with six phonon sensors interleaved with inner and an outer ionization collection electrodes.  Voltage biases of $\pm 3$~V (Ge; $\pm$4~V for Si) are applied to the top and bottom ionization electrodes while the phonon sensors are maintained near ground voltage, resulting in a vertical electric field in the bulk and a strong, transverse field near the surface.  Collection of ionization provides the iZIP detectors with the ability to distinguish beta- or gamma-induced ERs from neutron-, neutrino-, or dark matter-induced NRs through the ionization yield\footnote{Formally, the number of electron-hole pairs generated per unit recoil energy, but frequently quoted as \nehtxtnosp\epsehtxt per unit recoil energy, where \epsehtxt is the average ER energy needed to create an electron-hole pair, 3.0~eV in Ge (3.8~eV in Si).}. This yield-based discrimination rejects all ER backgrounds in the bulk of the detector for recoil energies above which NRs generate detectable ionization.  The strong, transverse field provides excellent rejection of surface backgrounds~\cite{Agnese:2013APL} because the ionization signal is asymmetric between the two sides.\footnote{Both electrons and holes are collected on the biased ionization electrodes for bulk events, while one charge sign is lost to the grounded phonon sensors for surface events.}  This ability to reject the vast majority of backgrounds on an event-by-event basis means the iZIP detectors can be operated in a nearly background-free mode.

HV detectors have six phonon sensors on each face with no ionization sensors, as shown in Figure~\ref{fig:DetectorSchematic}.  The HV detectors are intended to be operated at a bias of up to $\sim$100~V.  This bias makes it possible to take advantage of the Neganov-Trofimov-Luke (NTL) effect~\cite{neganov1978, Luke:1988JAP} to increase the phonon signal by the amount of work performed by the electric field on the charge carriers produced by an interaction as they move across the detector.  This transduction of the ionization signal into phonons reduces the recoil-energy threshold thanks to the large phonon signal produced (\nehtxtnosp\evbtxtnosp) and the fact that the phonon energy threshold is independent of the bias applied (for low-enough ionization leakage, \S\ref{sec:ionization_leakage_upgrades}).  Without the ability to identify the type of recoil from an interaction, the data from the HV detectors will be dominated by ER backgrounds.  The NTL effect, however, decreases the rate of ER backgrounds relative to NR dark matter signal: the larger ionization yield for ERs causes any bin in recoil energy to be spread out over a larger range of ``total phonon energy'' (recoil + NTL phonons) relative to NRs of the same recoil energy~\cite{Pyle:2012LTD14}.  The phonon-only sensor layout is optimized to provide phonon-based position information, which is critical for rejecting some surface backgrounds that can suffer degraded NTL amplification~\cite{Kurinsky:2016ichep}.

%\begin{table}[htp]
% TDR resolutions for charge -- don't know any better at this point
% phonon resolutions based on expected Tc and f_Al = 0.35 for HV, 0.04 for iZIP
% Ge HV: Tc = 60 mK
% Si HV: Tc = 40 mK
% Ge iZIP: Tc = 60 mK
% Si iZIP: Tc = 40 mK
% Model scaling from 220308PyleGolwalaDetEnergyScalings.xlsx
%                    TDR Fig 13       PD2      PD2x0.5    PD2x0.25
%       Si iZIP 40 mK      8 eV     19 eV*       10 eV        5 eV
%       Si HV 40 mK        5 eV     26 eV        13 eV*       6 eV
%       Ge iZIP 40 mK     15 eV     39 eV        21 eV       10 eV
%       Ge HV 40 mK        8 eV     43 eV        20 eV       11 eV
%       Ge iZIP 60 mK     50 eV    133 eV        65 eV       33 eV*
%       Ge HV 60 mK       25 eV    134 eV        67 eV       34 eV*
% We pick Si iZIP PD2 because Si iZIP, HV 40 mK actually are the same thing as PD2 and
%     these will still meet KPP requirements (50 for iZIP)
% We pick Si HV PD2x0.5 because MCP says that Si HV incorporates QET improvements that give a factor
%     of 2 improvement over PD2.
% We pick Ge iZIP, HV 60 mK PD2x0.25 because excess TFN and parasitic power seen in PD2 at 40 mK probably 
%     not an issue at 60 mK.  Also, meets KPP requirements (100 eV, 50 eV for iZIP, HV).
\begin{wraptable}[15]{r}{3.25in}
\vspace{-18pt}
\begin{center}
{\small
\begin{tabular}{  l  c  c  c  c } \hline
&\multicolumn{2}{c}{iZIP} & \multicolumn{2}{c}{HV}\\ 
&  Ge & Si    & Ge & Si \\ \hline
Number of detectors               & 10  & 2   & 8  & 4   \\
Total exposure [kg\(\cdot\)yr]    & 45  & 3.9 & 36 & 7.8 \\
Phonon resolution [eV]            & 33  & 19  & 34 & 13  \\
Ionization resolution [\eveenosp] & 160 & 180 & -- & --  \\
Voltage Bias ($V_+ - V_-$) [V]    & 6 & 8 & 100 & 100\\ \hline
\end{tabular}
}
\end{center}
\caption[Anticipated exposures and detector parameters for the SuperCDMS SNOLAB experiment]{\textbf{Anticipated exposures and detector parameters for the SuperCDMS SNOLAB experiment.} The exposures are based on 4 years of operation with an 80\% live time. The quoted phonon energy resolutions represent the rms values of the total measured quantity (i.e., combining all active sensors) while the ionization energy resolutions represent the rms values for a single inner-electrode channel.  The outer electrode would be used only as a veto, and a $\sqrt{2}$ improvement may be available from averaging inner-electrode electron and hole signals.}
\label{tab:ExpOperations}
\end{wraptable}
%\end{table}

The planned payload, detector performance, and anticipated total exposures for the SNOLAB experiment are summarized in Table~\ref{tab:ExpOperations}. The detectors will be deployed in four towers of six detectors each. The phonon resolutions listed in Table~\ref{tab:ExpOperations} are based on the detector noise analysis described in~\cite{Pyle:2012PhDThesis,Kurinsky:2016ichep}.  A prototype detector operated at a test facility demonstrated noise performance consistent with the predictions of~\cite{Pyle:2012PhDThesis}.  These resolutions incorporate current best estimates, based on witness fabrication wafers for the actual detectors to be deployed, for the superconducting transition temperature (\Tctxtnosp) of the tungsten transition-edge sensors used in the phonon sensors, combined with resolutions measured in prototype detectors and the scaling laws summarized in \S\ref{sec:phonon_resolution_upgrades}.  The ionization resolutions are derived from the readout electronics equivalent noise charge value of 53~\textit{e} in Ge (47~\textit{e} in Si) and \epsehtxt and represents the rms energy resolution of a single inner-electrode channel for electron recoils.

\subsubsection{Cryostat and Shielding}
\label{sec:SNOBOX}

\begin{figure*}[t!]
\begin{center}
\includegraphics[width=\textwidth]{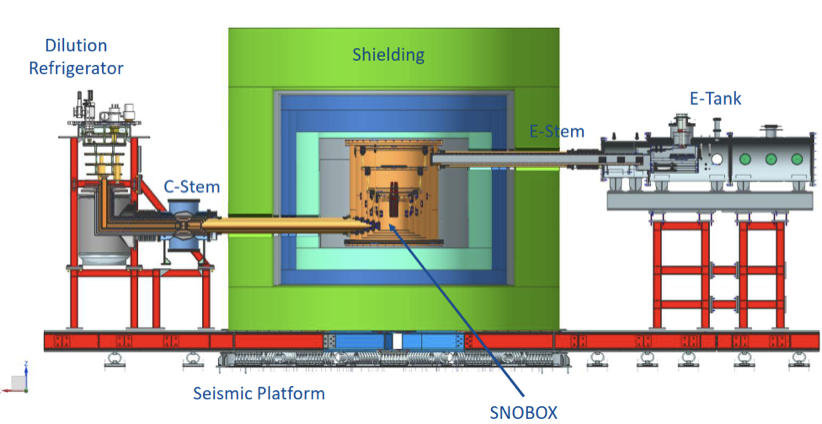}
\end{center}
\caption[Schematic of the SuperCDMS SNOLAB experiment]{\textbf{Schematic of the SuperCDMS SNOLAB experiment.} The assembly rests on top of a seismic platform to provide isolation from major seismic events. The outer water tanks provide protection from cavern neutrons. A gamma shield protects from external gamma-rays and the inner polyethylene layers serve to absorb radiogenic neutrons emitted from the cryostat and gamma shield.}
\label{fig:snolab}
\end{figure*}

The detector towers will be cooled to 15\,mK using a dilution refrigerator that utilizes cryocoolers to establish 50\,K and 4\,K thermal stages.  The cold region of the full experiment is referred to as the SNOBOX. As shown in Figure~\ref{fig:snolab}, the SNOBOX consists of six cylindrical copper cans suspended by Kevlar ropes. Each SNOBOX can is mapped onto a thermal stage of the refrigerator. The outermost can along with the stems and E-tank form the vacuum system. 

A 40-cm-thick layer of polyethylene surrounds the SNOBOX and serves to moderate and absorb neutrons produced by radiogenic contamination, both internal to the apparatus and from the cavern. This inner polyethylene layer is surrounded by a 23-cm-thick gamma shield made from commercially available low-activity lead. The lead shield layer is surrounded by a thin Rn diffusion barrier made of aluminum.  The volume inside the barrier will be purged with boil-off nitrogen gas to reduce the overall Rn levels and the backgrounds caused by prompt Rn daughters.  The outermost shield layer consists of polyethylene and water tanks that provide additional shielding from the cavern neutron flux. A mu-metal magnetic shield resides just outside the SNOBOX, inside all the other shielding layers, to attenuate Earth's field, which can degrade the performance of the phonon sensors and SQUIDs.

The shield and SNOBOX are penetrated in two locations opposite each other. The electronics stem (E-stem) provides a path for twisted-pair cables to run between the cold hardware and the electronics tank, which forms the vacuum bulkhead where signals emerge. The cryogenics stem (C-stem) connects the various layers of the SNOBOX to external cryogenic systems. 

The detectors reside in copper housings attached to the bottom of ``Towers'' that are mounted to the lid of the lowest-temperature can of the SNOBOX.  Each Tower accommodates six detectors and provides readout wiring and hosts cryogenic components of the electronics readout (SQUIDs for phonons, HEMTs for ionization, bias resistors, LEDs\footnote{Used to ``neutralize'' the detectors on cooldown and between periods of data-taking, see \S\ref{sec:ionization_leakage_upgrades}.}, and ancillary thermometry).  The Towers have four temperature stages, connected to corresponding stages of the SNOBOX between 15~mK and 4~K, both to thermally sink the wiring to the detectors and to provide appropriate temperatures for various electronics components.  Connection to the room-temperature component of the electronics is via the twisted-pair cabling in the E-stem.  For later reference, each detector has 12 phonon sensors, and, for iZIPs, 4 ionization sensors, yielding 72 phonon and up to 24 ionization signals per Tower.

The cryogenic system's thermal design can accommodate up to 31 Towers, a number set by the thermal load on each cryostat stage due to loads conducted by mechanical structures and wiring and dissipated by the electronics.  During final design, it was necessary for budgetary reasons to downscope the cryostat diameter so it will only accommodate 7 Towers, but it retains the cooling power needed to accommodate the channel count of 31 towers, a point that is important for some of the upgrade scenarios (\S\ref{sec:live_time_channel_counts}).

\begin{figure*}[t!]
\begin{center}
\includegraphics[width=\textwidth]{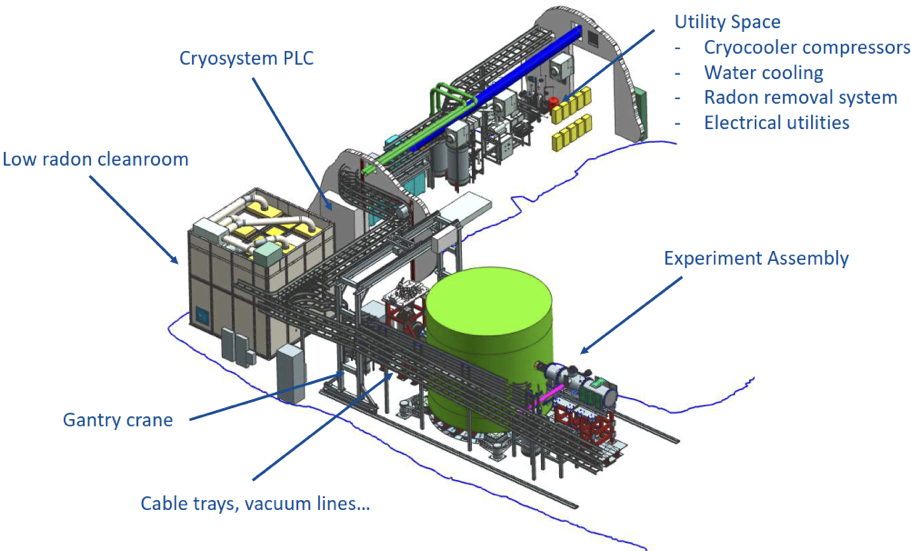}
\end{center}
\caption[Layout of the SuperCDMS SNOLAB experiment in the SNOLAB Ladder Lab]{\textbf{Layout of the SuperCDMS SNOLAB experiment in the SNOLAB Ladder Lab.}  The low-radon cleanroom is visible at the left. Support facilities are located in an adjacent drift at the top.}
\label{fig:snolab_layout}
\end{figure*}

For later estimation of background levels, it is important to note that the inner four cans of the cryostat are removable and installation of Towers will occur inside a low-radon cleanroom adjacent to the experiment (Figure~\ref{fig:snolab_layout}; the Towers will be shipped and stored in radon-purged containers and only opened inside the cleanroom.)  These four cans will be sealed during their transport to and installation in cryostat except for short periods of time when the lids must be partially removed to make thermal connections to the cryostat.  A purge will be employed at all times to prevent radon ingress.

%%
%% BACKGROUND SOURCES
%%

\subsubsection{Background Sources}
\label{sec:Backgrounds}

In this section, we review at a qualitative level the background sources anticipated for the SuperCDMS SNOLAB experiment and accounted for in our sensitivity forecasts: (1)~sources that produce energy depositions throughout the detector crystal volume and (2)~sources that produce energy depositions primarily on or very near the surfaces of the detector crystal.  These categories are further divided into ER (\textit{e.g.},\ betas or Compton scatters) and NR events. Each of these event types must be tracked separately for each type of detector because of different detector response functions, fiducial efficiencies, and analysis efficiencies, as discussed in Section~\ref{sec:forecast_procedure} and Appendix~\ref{sec:forecast_procedure_detail}.  Please refer to \cite{sensitivity2016} for quantitative information on raw background levels assumed and for details of simulation implementation.  Some radioactivity levels have been updated since that publication and will be presented in a future publication.

\paragraph{Bulk Event Background Sources}
\label{sec:volumesources}
The background sources described in this subsection can produce events that occur throughout the detector volume (``bulk'' events).

\subparagraph*{Detector Contamination}
The dominant backgrounds expected for the HV detectors are due to radioactive impurities within the detector crystals:

%Table~\ref{tab:detectorcontamination} presents the detector contamination levels assumed in the calculation of the SuperCDMS SNOLAB sensitivities shown in Fig.~\ref{fig:OI-Limits}.
%\begin{table}[htp]
%\begin{tabular}{ c c c c c c} \hline
% &         & Production Rate & \multicolumn{2}{c}{Concentration}\\
% &  & (atoms/kg/day)  & \multicolumn{2}{c}{(decays/kg/day)} \\
%Material & Isotope &   & HV & iZIP \\ \hline
%Ge & $^{3}$H & 80  & 0.7 & 1.5 \\
%Si & $^{3}$H & 125 & 1 & 2 \\
%Si & $^{32}$Si & --  & 80 & 80\\ \hline
%\end{tabular}
%\caption{Assumptions used to determine the $^{3}$H and $^{32}$Si detector contamination levels for the SuperCDMS SNOLAB sensitivities shown in Fig.~\protect{\ref{fig:OI-Limits}}.  The assumed sea-level cosmic-ray exposure for the HV(iZIP) detectors is 60(125) days, followed by a 365 day underground ``cooldown'' period before acquisition of science data. $^{32}$Si is intrinsic to the production process and is expected to be the same for iZIP and HV detectors. }
%\label{tab:detectorcontamination}
%\end{table}

\begin{description}

\item[Cosmogenically produced $^{3}$H:]
For both the Ge and Si detector crystals, exposure to high-energy cosmic-ray secondaries (\textit{i.e.}, neutrons, protons, and muons) results in the production of tritium ($^{3}$H) as a spallation product from interaction of the cosmic-ray secondaries with the nuclei in the detector crystals~\cite{Avignone:1992jg}. The long half-life of tritium ($t_{1/2} = 12.3$\,years) results in an accumulation of this radioactive impurity, whose \mbox{$\beta$-decay} product has an endpoint energy of 18.6\,keV. 
%The tritium background is modeled using a generic $\beta$-decay energy spectrum~\cite{Neary:1971}, the production rates and activation times shown in Table~\ref{tab:detectorcontamination} for tritium in Ge~\cite{Armengaud:2016aoz} and in Si~\cite{tritium:paper}, and the detector crystal masses given in Table~\ref{tab:ExpOperations}.

\item[Naturally occurring $^{32}$Si:]
This radioactive isotope is produced as a spallation product from cosmic-ray secondaries on argon in the atmosphere~\cite{Lal:1960}. The $^{32}$Si atoms make their way into the terrestrial environment through aqueous transport (i.e., rain and surface water). Consequently, the exact source and location of the silicon used in the production and fabrication of silicon detectors may impact the concentration level of $^{32}$Si observed in future detectors. The long $\sim$153\,year half-life of $^{32}$Si~\cite{Ouellet:2011nds} means its concentration, measured in decays/kg/day, is essentially fixed once the Si detector crystal is grown.  We assume the central value of the $^{32}$Si concentration measured by the DAMIC collaboration in their CCD detectors, \(80^{+110}_{-65}\)\,decays/kg/day at a 95\% confidence level~\cite{Aguilar-Arevalo:2015lvd}.  This rate is conservative, as the literature~\cite{plaga:1991} suggests low-radioactivity silicon is in principle available for rare-event searches, and later measurements yielded a rate of 11.5$\pm2.4$/kg/day~\cite{damic_32si_2021}.  
%The $^{32}$Si background is modeled using a generic $\beta$-decay energy spectrum~\cite{Neary:1971}.

\item[Ge activation lines:]
Exposure of the Ge detector substrates to high-energy cosmic-ray secondaries results in the production of several radioisotopes that decay by electron capture. We include the eight isotopes observed in the CoGeNT experiment~\cite{Aalseth:2010vx} that are sufficiently long-lived to contribute background in the SuperCDMS SNOLAB Ge detectors: $^{68}$Ge and $^{68}$Ga daughter, $^{65}$Zn, $^{73}$As, $^{57}$Co, $^{55}$Fe, $^{54}$Mn and $^{49}$V. Each decay can proceed via electron capture from the K, L or M shell, giving rise to a total of 24 spectral peaks.
%(\textit{cf.}~Fig.~\ref{fig:HVBackgroundSpectraRaw}). 
%We scale the K-shell peak rates by the ratio of the sea-level exposure for the SuperCDMS detectors and the reference CoGeNT detector, and the rates of the L- and M-shell lines are scaled according to their relative branching fractions.

\end{description}

\subparagraph*{Material Activation}
Exposure to high-energy cosmic-ray secondaries results in the production of long-lived radioisotopes in the construction materials surrounding the detectors. In particular, the cosmogenic activation of copper presents a background source for the SuperCDMS SNOLAB experiment. Copper is used both for the detector tower mechanical assembly and the nested cylindrical cryostat canisters. %Table~\ref{tab:copperexposure} presents the assumptions used to assess the emission rates due to cosmic-ray activation of these components.
%\begin{table}[htp]
%\begin{tabular}{c c c c} \hline
%        & \multicolumn{1}{c}{Production Rate} & \multicolumn{2}{c}{Contamination Rate ($\mu$Bq/kg)}\\
%Isotope & (atoms/kg/day)  &  Housings/Towers & Cryostat\\ \hline
%$^{46}$Sc & 4.6  & 0.88 & 0.62\\
%$^{48}$V  & 9.5  & 0.76 & 0.25\\
%$^{54}$Mn & 19   & 7.9 & 12\\
%$^{56}$Co & 20   & 3.5 & 2.3\\
%$^{57}$Co & 155  & 62 & 89\\
%$^{58}$Co & 143  & 23 & 13\\
%$^{59}$Fe & 39   & 2.9 &  0.9\\
%$^{60}$Co & 181  & 47 & 90 \\ \hline
%\end{tabular}
%\caption{Assumptions used to determine the cosmogenic exposure and activation of copper for the SuperCDMS SNOLAB sensitivities shown in Fig.~\protect{\ref{fig:OI-Limits}}.  The sea-level activation rates are taken from \cite{Cebrian:2010ApP}, except for $^{48}$V, which is taken from~\cite{Laubenstein2009}. A sea-level exposure of 90~days is assumed for copper in the detector housings and towers, followed by a 90~day underground ``cooldown'' period before acquisition of science data. The sea-level exposure and underground-cooldown periods for the copper cryostat cans are both assumed to be 180\,days. At the time of this publication only $^{57}$Co, $^{58}$Co, $^{60}$Co, and $^{54}$Mn have been simulated for the tower and housing copper; the decay rate for the other listed isotopes is at least 5$\times$ lower. For the cryostat cans, only $^{60}$Co is presently included; emissions from the other isotopes are lower in energy and thus less penetrating and can be neglected.}
%\label{tab:copperexposure}
%\end{table}

\subparagraph*{Material Contamination}
\label{sec:materialcontamination}
Radioactive impurities are introduced in all materials at some level during the manufacturing process.  The $^{238}$U and $^{232}$Th isotopes are unstable but long-lived and are present in most materials at low concentrations. Both of these isotopes have a chain of decay daughters that are assumed to be in secular equilibrium except where noted. Additionally, isotopes such as $^{40}$K and $^{60}$Co are naturally present in many materials because of their long half-lives, but they do not have accompanying series of daughter radioisotopes. 

\subparagraph*{Non-Line-of-Sight Surfaces}
\label{sec:nonlossurfaces}
Materials accumulate concentrations of radioactive isotopes on surfaces exposed to air containing dust and radon. Airborne dust typically contains relatively high concentrations of $^{238}$U, $^{232}$Th, and $^{40}$K. Daughters from the decay in air of $^{222}$Rn may implant shallowly into a material surface, resulting in a buildup of the long-lived $^{210}$Pb that later decays through a short chain and produces a roughly constant emission rate of X-rays, betas and alphas.

We consider separately surfaces with and without a clear line of sight to the detector; surfaces with line of sight are generally of much greater concern and are discussed in Section~\ref{sec:surfacesources}. For surfaces without line of sight, we are concerned primarily with gamma- and X-ray emission, and to a lesser extent neutron emission, as those are the only radiation types capable of reaching the detectors. $^{210}$Pb produces some soft X-rays that may reach the detectors if emitted from surfaces very nearby, and its daughter $^{210}$Bi has a moderately high energy $\beta$-decay that may in turn produce bremsstrahlung that may be more penetrating.  Finally, the alpha produced by the subsequent $^{210}$Po decay may produce neutrons via an $(\alpha,n)$ reaction on $^{13}$C, and so $^{210}$Pb accumulation on hydrocarbon surfaces such as polyethylene is a potential concern.  Estimates of radon-daughter contamination levels are made based on typical post-etching exposure times to air with typical above-ground or underground radon levels.  While~\cite{sensitivity2016} did not incorporate estimates for dust, forecasts presented here do.
%For this analysis we assumed a $^{210}$Pb activity of 850~nBq/cm$^2$ for non-line-of-sight surfaces inside the cryostat, roughly corresponding to 100 days exposure to air with radon concentrations of 10 Bq/m$^3$.  For the outer cryostat and shielding surfaces we assume an activity of 11,000~nBq/cm$^2$ corresponding to 100 day exposure to 130~Bq/m$^3$ air. Dust has not been included at this time, but preliminary estimates indicate that it should contribute less than or comparably to background from non-line-of-sight $^{210}$Pb surface contamination.

\subparagraph*{Cavern Environment}
The cavern environment background sources include naturally occurring radioactivity in the underground environment leading to gamma-rays or neutrons that potentially pass through the SuperCDMS shield and interact in the detectors.
The experiment cavern is surrounded by Norite rock that has been coated with a layer
of shotcrete. The cavern floor is concrete. The wall and floor layers have variable thicknesses
but are on the order of a few inches thick~\cite{Duncan:2010zz}.

\begin{description}

\item[Gamma-rays]
The gamma-ray background is modeled as a $^{40}$K decay along with decay chains in secular equilibrium for $^{238}$U and $^{232}$Th. 
Those gamma-ray emission spectra were simulated and evaluated for their leakage through the shielding, in particular the E- and C-stem penetrations (Figure~\ref{fig:snolab}).  It was determined that this source was subdominant to other sources and, because of the challenge of simulating a background sourced from such a large area, a full simulation of this background component is not included in our forecasts.
%The gamma-ray flux was estimated using results from assays of rocks collected in the SNOLAB ladder labs and Monte Carlo simulations. This source, however, is not included in the present model because of limited simulation statistics and because preliminary results indicated that it is subdominant to other sources. The gamma shield design is being optimized with a design goal of allowing $<1$\,counts/kg/keV/year of a single--scatter background rate in the Ge HV detectors (see for comparison Table~\ref{tab:backgroundbudget}).

\item[Neutrons]
The neutron background from the cavern environment is modeled as two components: neutrons from $^{238}$U spontaneous fission and neutrons produced through $(\alpha,n)$ reactions in the rock due to U- and Th-chain alpha-emitting isotopes. 
%The shape of the neutron spectra are calculated using a modified version of {\sc Sources 4c}, which calculates neutron spectra for spontaneous fission and the material specific $(\alpha,n)$ process in Norite and shotcrete~\cite{sources4,sources4c:IDM}. The neutrons are propagated through the materials using Monte Carlo simulations. The overall normalization of the spectrum is taken from Ref.~\cite{Smith:2012fq}, which specifies a flux of 4000 fast neutrons/(day$\cdot \mathrm{m}^2$).

\item[Radon]
Radon decays in the mine air produce moderately high-energy gamma-rays via the $^{214}$Pb and $^{214}$Bi daughters. Decays occurring outside the shield contribute to the total gamma-ray flux already considered for the cavern as a whole and are not considered separately.  If air in the region between the lead gamma shield and the SNOBOX were allowed to mix freely with the mine air, radon decays in this region would produce a significant background.  However, this contribution is assumed to be made negligible %, $< 0.1$\,counts/kg/keV/year rate of single--scatters in the HV detectors (\textit{cf.}Table~\ref{tab:backgroundbudget}), 
by the hermetic radon purge surrounding the gamma shield. 

\end{description}

\subparagraph*{\textit{In situ} Cosmic-Ray-Induced Backgrounds}
The overburden provided by SNOLAB significantly reduces, but does not eliminate, cosmic rays.  Muons may pass directly through a detector or create secondary particles through interactions with the surrounding materials; high-energy neutrons produced via spallation are our primary concern.  We simulated muons with the angular and energy distribution appropriate for SNOLAB depth. 
%parameterized by Mei and Hime~\cite{Mei:2006PRD} from a $\sim$10~meter diameter plane. 
%Unlike SuperCDMS Soudan, the detector at SNOLAB will not have a muon veto. 
Given the depth, SuperCDMS SNOLAB will not have a muon veto.
%The background rate estimated from this source of neutrons is listed as Cosmogenic Neutrons in Table~\ref{tab:backgroundbudget}.

\subparagraph*{Coherent Neutrino Interactions}
Although not expected to be a significant background for the initial SuperCDMS SNOLAB experimental program, the interaction of solar neutrinos through coherent elastic neutrino-nucleus scattering (\cevns~\cite{Freedman:1977xn}) currently presents a limiting background source to future low-mass dark matter search experiments~\cite{Billard:2013qya,Ruppin:2014prd}.  We include all the components of the solar neutrino spectrum, including the $pp$, \besevennosp, CNO, $pep$, \beightnosp, and $hep$ neutrinos.

%The decay of $^{8}$B at the end of the pp-III solar fusion reaction chain produces the primary solar neutrino background for future expansions of the SuperCDMS SNOLAB experiment. The background is estimated using a theoretical value for the solar neutrino fluxes~\cite{Bahcall:2004pz}, the theoretical $^{8}$B solar neutrino energy spectrum~\cite{Strigari:2009bq}, and the coherent elastic neutrino-nucleus scattering cross-section~\cite{Freedman:1977xn}.

\paragraph{Surface Event Background Sources}
\label{sec:surfacesources}
The second broad category of background sources produces energy depositions on or very near the surfaces of the detector substrates. These backgrounds are explicitly related to the exposure of the detectors and their housing materials (primarily Cu) to $^{222}$Rn and its progeny during fabrication, testing and installation. When radon decays in air, for example, its daughters can plate out onto a surface and the subsequent Po alpha decays can cause the long-lived $^{210}$Pb daughter to become implanted into the surface. If the implantation occurs far enough along in the fabrication process, it is no longer practical to remove and is thus a source of background that will be present for the duration of the experiment. 

The $^{210}$Pb decay chain produces a variety of radiation types that are generally not very penetrating. Consequently, aside from the few exceptions pointed out in Section~\ref{sec:nonlossurfaces} above, $^{210}$Pb surface contamination is a background concern only if there is a clear line of sight between the location of the contamination and a detector surface. There are three principle radioisotopes in the decay chain: $^{210}$Pb, $^{210}$Bi and $^{210}$Po.  $^{210}$Pb yields a combination of low-energy betas and X-rays, resulting in a near-surface ER background in all detector types, whose spectrum resembles a $\beta$-decay spectrum with a 63~keV endpoint but superimposed with several Auger electrons and spectral lines (most notably at 46.5~keV from internal conversion). $^{210}$Bi undergoes $\beta$-decay with a 1.16~MeV endpoint, also resulting in a near-surface ER background but with a harder spectrum and no lines. Finally, $^{210}$Po decays by emitting an alpha so energetic that it is generally outside the dark matter signal region. Unlike in the preceding two decays, however, the $^{206}$Pb daughter nucleus recoils with sufficient energy to potentially create a NR in the dark matter signal region.  If the decay occurs on a detector surface such that the $^{206}$Pb recoil is directed into the surface, the full 103~keV recoil energy is deposited in the detector.  If the decay occurs on a nearby surface (\textit{e.g.}, detector housing), the energy of the $^{206}$Pb nucleus may be degraded because of the implantation depth of the $^{210}$Po parent, yielding a continuum of NR energies up to 103~keV.

$^{14}$C or $^{39}$Ar are also likely present due to detector fabrication processes and citric acid passivation of nearby copper surfaces.  Both beta decay (with 163~keV and 565~keV endpoints) to stable daughter nuclei and are included on the basis of rough estimates (not assay results). 

The detector response and detector type are particularly important considerations when evaluating the impact of these surface backgrounds on the experimental sensitivity. $^{206}$Pb recoils incident on a detector face will predominantly be tagged by a large energy deposition in the adjacent detector from the associated alpha, and thus such events will not contribute to the background of a dark matter search. Similarly, $^{210}$Pb decays often result in simultaneous energy depositions in adjacent detectors, allowing them to be rejected as dark matter candidates.  As discussed in~\cite{Agnese:2013APL}, interleaving the iZIP phonon and ionization sensors enables discrimination of surface events at the detector faces. Sensor modularity enables fiducialization of the signal to reject surface events incident at the sidewalls~\cite{Agnese:2014PRL}. This ``radial'' fiducialization is expected to be effective for HV detectors as well as iZIPs. The SuperCDMS SNOLAB detector response is discussed in more detail below in Section~\ref{sec:forecast_procedure} and in Appendix~\ref{sec:forecast_procedure_detail}.

%For the evaluation of the SuperCDMS SNOLAB sensitivities shown in Fig.~\ref{fig:OI-Limits}, we assume a total $^{210}$Pb surface activity of 50\,nBq/cm$^2$ for line-of-sight surfaces (the detector surfaces and inner surfaces of the copper housings). This is the same level of activity that was observed on the surfaces of the Ge iZIP detectors in the SuperCDMS Soudan experiment, inferred from the rate \textit{versus} time of 5.3\,MeV $^{210}$Po alphas incident on the detector faces.

\subsubsection{Expected Sensitivity of SuperCDMS SNOLAB}

\paragraph{Forecasting Procedure}
\label{sec:forecast_procedure}

Our forecasting procedure involves a spectrum-level simulation: given a candidate DM model and known backgrounds, we generate many realizations of event spectra on the basis of the assumed live time, detector counts and masses, and full Monte Carlo simulations of particle backgrounds, we determine allowed regions for each realization using a profile likelihood-ratio (PLR) method\footnote{This is a departure from \cite{sensitivity2016}, which used the optimum interval (OI) method~\cite{yellin_optimum_interval}, which does not attempt to statistically subtract backgrounds.  See \S\ref{sec:updated_nrdm} for further discussion.\label{footnote:oiplr}}, and then we take the median over the ensemble to obtain our expected sensitivity.  The procedure obviously depends on background scenario via the background model and the detector scenario via the detector response model.  

This procedure does \textit{not} simulate individual event timestreams and pass them through an event reconstruction pipeline and then a set of signal-selection/background-rejection cuts, nor does it simulate reconstructed event parameters and pass them through such a set of cuts.   Instead, the procedure uses simplified models to account for analysis cuts we apply to reject backgrounds, yielding spectra of signal and backgrounds that have been modified by the acceptance (efficiency or misidentification) as a function of energy, which then form the input to the above simulation.  We undertake this process with spectra in phonon energy, rather than recoil energy, because it is easier to forward-model the (position-dependent) addition of NTL phonon energy to recoil energy than to backward-model reconstruction of recoil energy from phonon energy.

We describe this treatment in more detail in Appendix~\ref{sec:forecast_procedure_detail}.  

\paragraph{Updated Forecasts for Nucleon-Coupled Dark Matter}
\label{sec:updated_nrdm}

\begin{figure}[t!]
\begin{center}
\textbf{SuperCDMS SNOLAB \GeV Nucleon-Coupled DM}\\
\vspace{3pt}
\begin{tabular}{cp{1.7in}c}
2016--2022 OI comparison & & OI-PLR comparison  \\
\end{tabular}\\
\includegraphics*[width=0.48\textwidth,viewport=0 0 576 410]{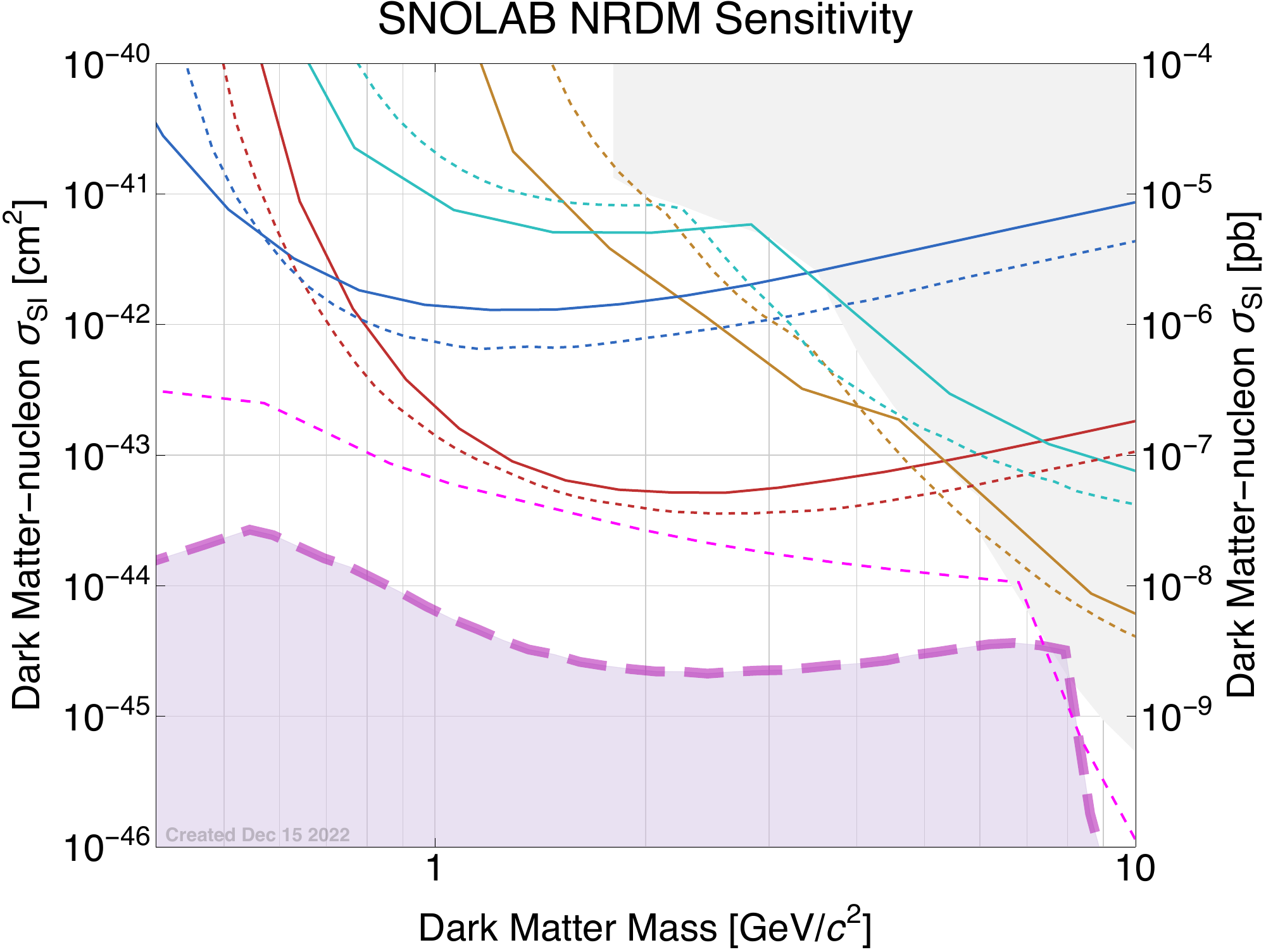}
\hfill
\includegraphics*[width=0.48\textwidth,viewport=0 0 576 410]{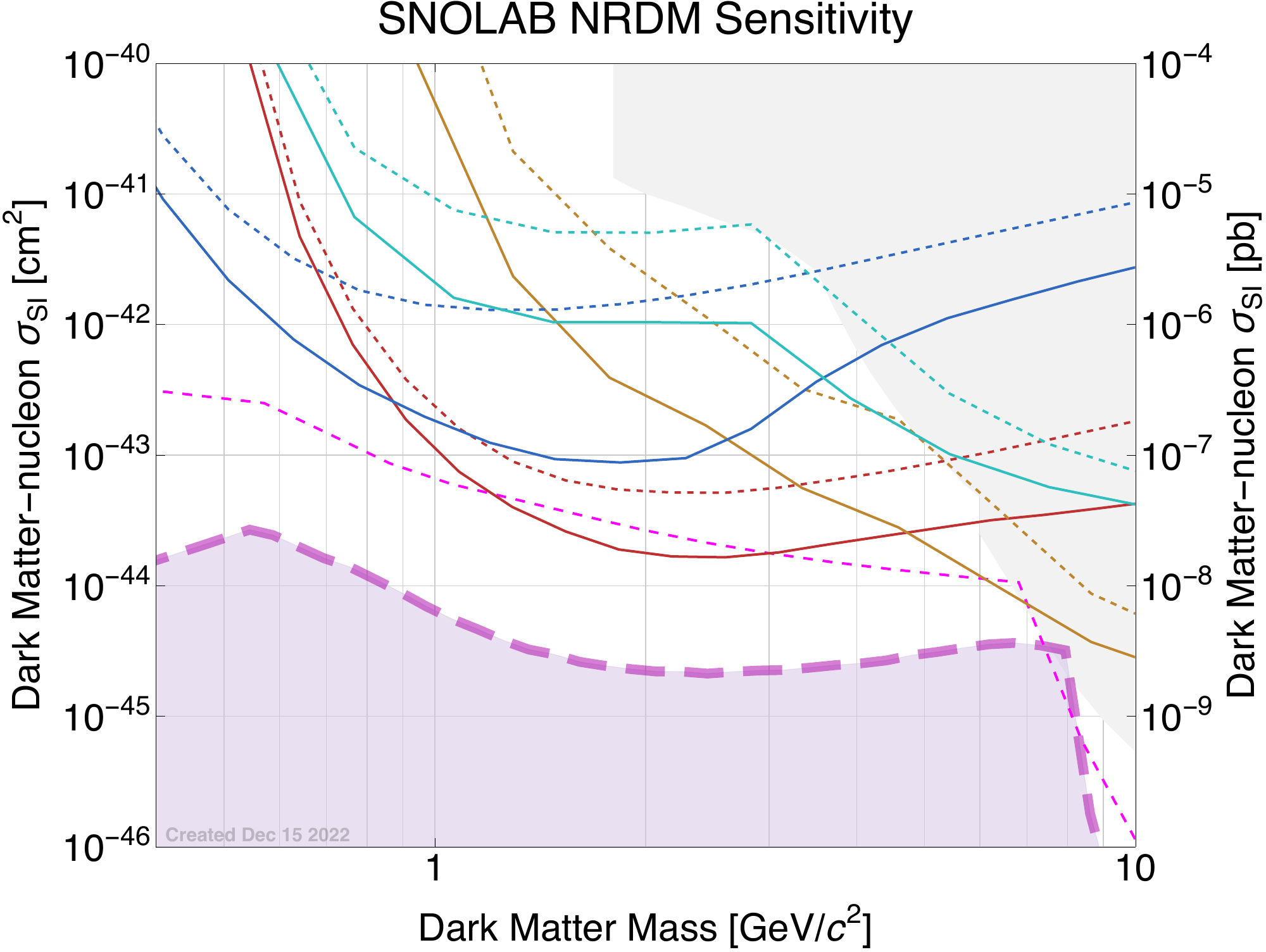}\\
\vspace{3pt}
\noindent
\hspace{0.6in}Ge iZIP \hspace{3in}Si iZIP\hfill \\
\includegraphics*[width=0.46\textwidth,viewport=0 0 420 322]{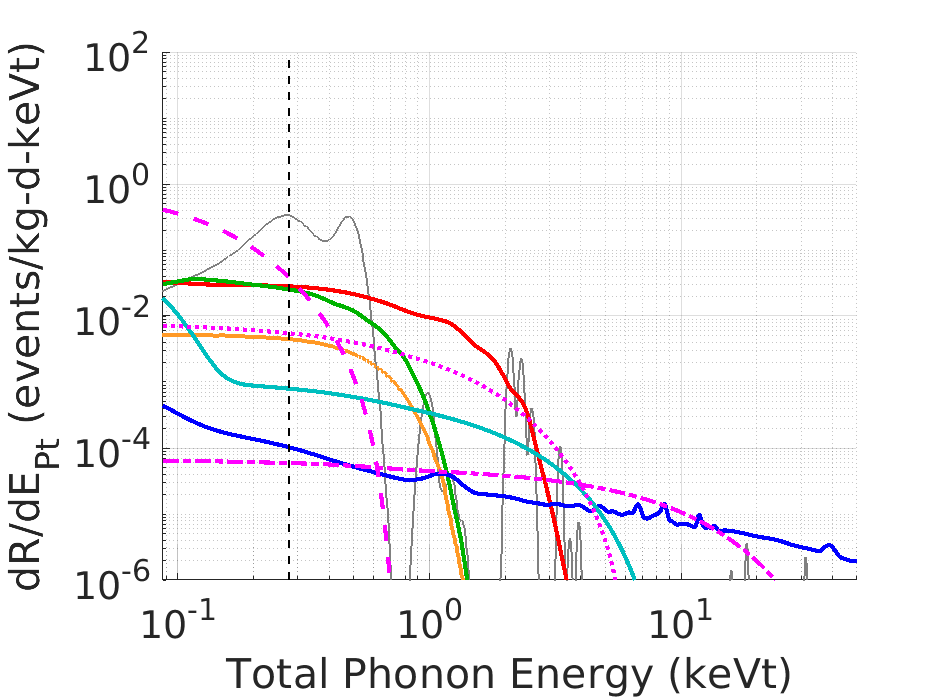}
\hfill
\includegraphics*[width=0.46\textwidth,viewport=0 0 420 322]{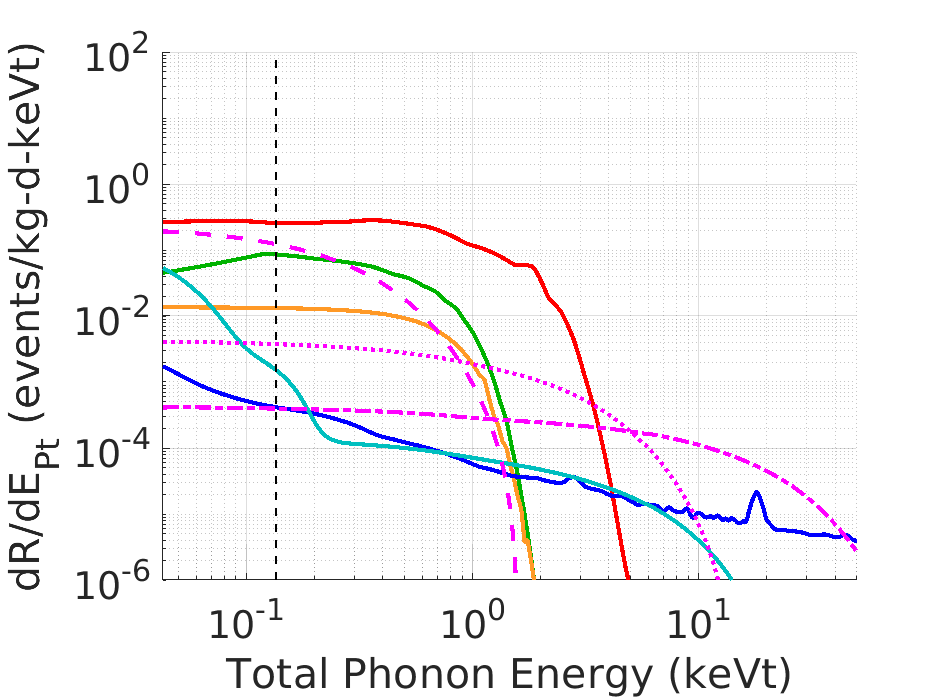}
\\
\noindent
\hspace{0.5in}Ge HV \hspace{3in}Si HV\hfill \\
\includegraphics*[width=0.46\textwidth,viewport=0 0 420 322]{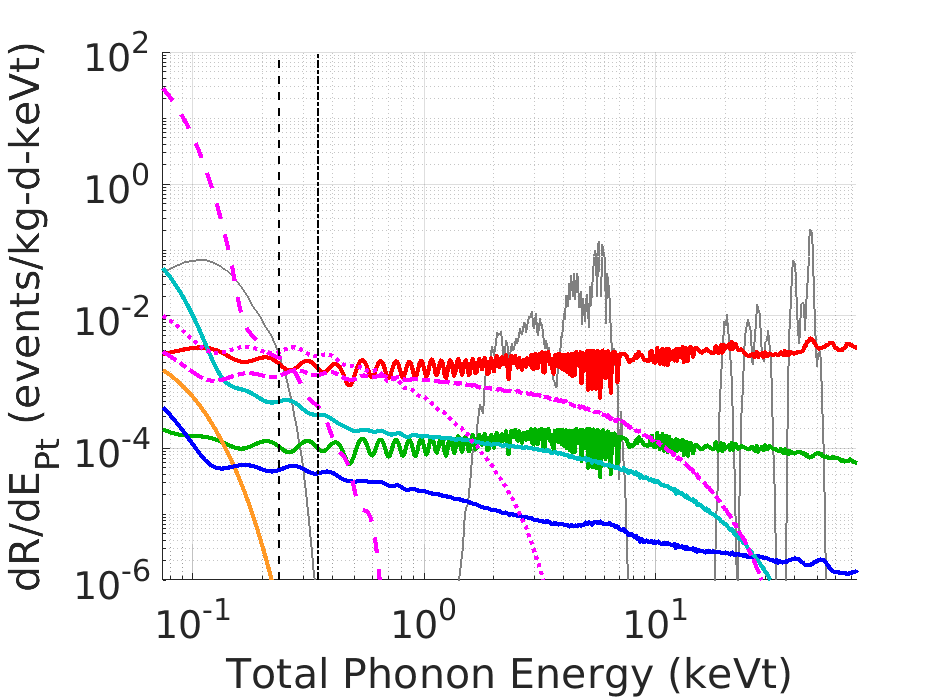}
\hfill
\includegraphics*[width=0.46\textwidth,viewport=0 0 420 322]{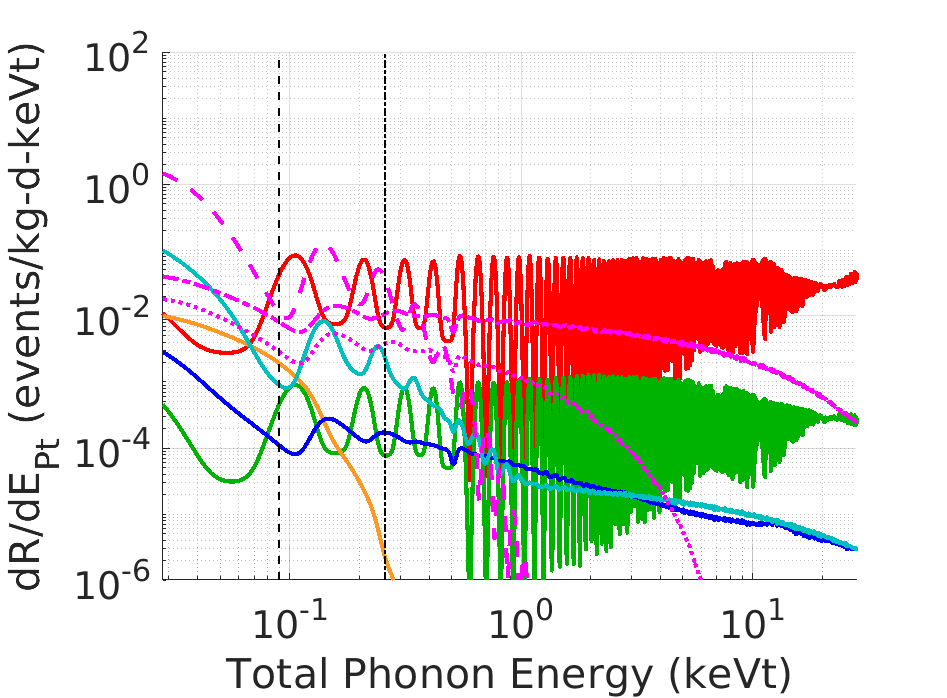}
\end{center}
\caption[Expected sensitivity of SuperCDMS SNOLAB for nucleon-coupled dark matter]{\textbf{Expected sensitivity of SuperCDMS SNOLAB for nucleon-coupled dark matter.}  (Top left) Optimum interval exclusion sensitivity at 90\%~CL from \cite{sensitivity2016} (dashed) and based on current estimates of detector performance (Table~\ref{tab:ExpOperations}) and backgrounds (solid).  (Top right) Optimum interval (dashed) and profile-likelihood ratio (solid) exclusion sensitivity at 90\%~CL based on current estimates of detector performance and backgrounds.  In all cases: (red-brown)~Ge~HV; (blue)~Si~HV; (mustard)~Ge~iZIP; (cyan)~Si~iZIP; (magenta long dashed and shaded)~``neutrino fog'' based on~\cite{Billard:2013qya,Ruppin:2014prd}; (magenta short dashed)~``single neutrino'' sensitivity, where one neutrino event can be expected on average; (grey shaded)~current exclusion limits.
(Middle and bottom)~Background spectra in phonon energy after all analysis cuts described in \S\ref{sec:forecast_procedure_detail} for Ge (left) and Si (right) and iZIPs (middle) and HV (bottom).  Black vertical dashed line:~$7\,\sigpt$ analysis threshold. Red:~bulk ERs due to Compton and cosmogenic backgrounds (\hthreenosp, \sittnosp); the bumps correspond to Compton features. Green:~surface ERs. Mustard:~surface NRs due to \pbtsnosp. Blue:~bulk NRs due to neutrons. Cyan:~bulk NRs due to solar neutrino \cevnsnosp; the two bumps correspond, in order of increasing energy, to pp and \beseven neutrinos. Magenta:~candidate DM signals for $\mathit{\mdm \approx}$~1.6, 5, and 16~\GeV for iZIP and 0.5, 1.6, and 5~\GeV for HV for cross sections at the exclusion sensitivity.}
%\hrule
\label{fig:snolab_reach_nrdm}
\end{figure}

Figure~\ref{fig:snolab_reach_nrdm} shows the expected nu\-cle\-on-coupled dark matter (i.e., nuclear recoil) sensitivity of SuperCDMS SNOLAB based on this forecasting procedure as well as the post-cuts background spectra used to generate the sensitivity estimates.  For completeness, we also show a comparison to previously published expected sensitivity estimates from \cite{sensitivity2016}.  As noted above, we presented optimum interval (OI) estimates in that work while we present PLR-based estimates going forward.  Therefore, we provide two incremental comparisons: one comparing current OI sensitivities to those from~\cite{sensitivity2016}, and one comparing current OI and PLR sensitivities.  The former plot thus reflects changes in experimental design, background estimates, and expected detector performance as well as increased sophistication of the forecasting procedure since \cite{sensitivity2016} while the OI-PLR comparison for SuperCDMS SNOLAB shows the expected gain possible from incorporating knowledge of backgrounds.  It is also the  baseline for comparison for the future forecasts that will be presented in \S\ref{sec:experience_forecasts}.

% should add a reference for leakage, but have to find the right one
We can be more specific about the major changes between \cite{sensitivity2016} and the current estimates.  The most important are that, for HV detectors, which dominate the sensitivity, we now account for ionization leakage (\cite{Agnese_2018}; see \S\ref{sec:ionization_leakage_upgrades} and \S\ref{sec:ionization_leakage})\footnote{We  note that HV leakage is not yet measured in full-size detectors but rather is scaled from smaller devices by surface area.}, which increases the HV detector analysis threshold, and we also account for the fact that ionization production is discretized and therefore fluctuates statistically (as opposed to a continuous yield function mapping recoil energy to ionization energy with no fluctuations).  The impact on mass reach and sensitivity for HV detectors is noticeable but not significant.  (The improved Si HV reach at low mass, discussed below, is unrelated to these changes).  Other changes that affected both HV and iZIP detectors include: change of the cryostat geometry from 31 towers to 7 towers (\S\ref{sec:SNOBOX}); updates to the background model, including, in particular, an increase in the neutron background from the mu-metal magnetic shield\footnote{Assay results indicate that the U/Th chain in the mu-metal is not in secular equilibrium.  The later part of the chain, which mainly sources gammas, is unchanged, but contamination by the earlier part of the chain is significantly higher than previously estimated, yielding a NR background contribution comparable to that of solar neutrinos.  The effect ends up being modest because it is more important for Si while Ge dominates the sensitivity.\label{footnote:mumetal}} and approximate addition of \cft and \artn backgrounds (\S\ref{sec:Backgrounds}); increased realism of the assumed ionization yield function (\S\ref{sec:ionization_yield}) based on recent measurements~\cite{supercdms_impact}; corrections to the ionization collection model; an improved energy resolution scaling model; and use of film parameters from witness fabrication wafers\footnote{All the SuperCDMS SNOLAB detectors have been fabricated at this point, though their \Tctxt values were not available at the time these forecasts were made.  Hence the reliance on witness wafers.}.   These latter changes dominate the impact on the iZIP detector reach.  There is modest degradation at higher masses, where event-by-event NR discrimination based on ionization yield is effective.  The ionization yield model changes actually improve the Si iZIP reach modestly at lower masses, where yield-based discrimination is used statistically.  The ionization yield model change is also the cause for the improved Si HV reach at low mass. 

The difference between OI and PLR sensitivities is quite significant, with the PLR science reach more than compensating for the degradations noted above and significantly extending the cross-section reach overall.  We believe that our background modeling is sufficiently sophisticated and accurate to warrant the use of the PLR estimates.

Most importantly, the relatively modest changes in sensitivity between prior work and these estimates confirm the robustness of the expectation that SuperCDMS SNOLAB will test multiple square decades of new parameter space for dark matter in the 0.5--5~\GeV mass range, complementing experiments using other techniques in this mass range.

\paragraph{New Forecasts for Electron-Coupled Dark Matter}

\begin{figure}
\begin{center}
\textbf{SuperCDMS SNOLAB \eV Electron-Coupled DPDM and ALPDM} \\
\vspace{3pt}
%\vspace{-12pt} \green{221003 up-to-date} \\ 
%\red{Harrison, could you change log spectra to be 0.05-5 keV, and also plot linear spectra going up to 3 keV, ticks at 0, 1, 2, 3 keV?  Keep log option too, though} \\
\begin{tabular}{p{0.2in}cp{2.1in}c}
& Dark Photon & & Axion-Like Particle  \\
\end{tabular}\\
\includegraphics*[width=0.48\textwidth,viewport=0 0 576 410]{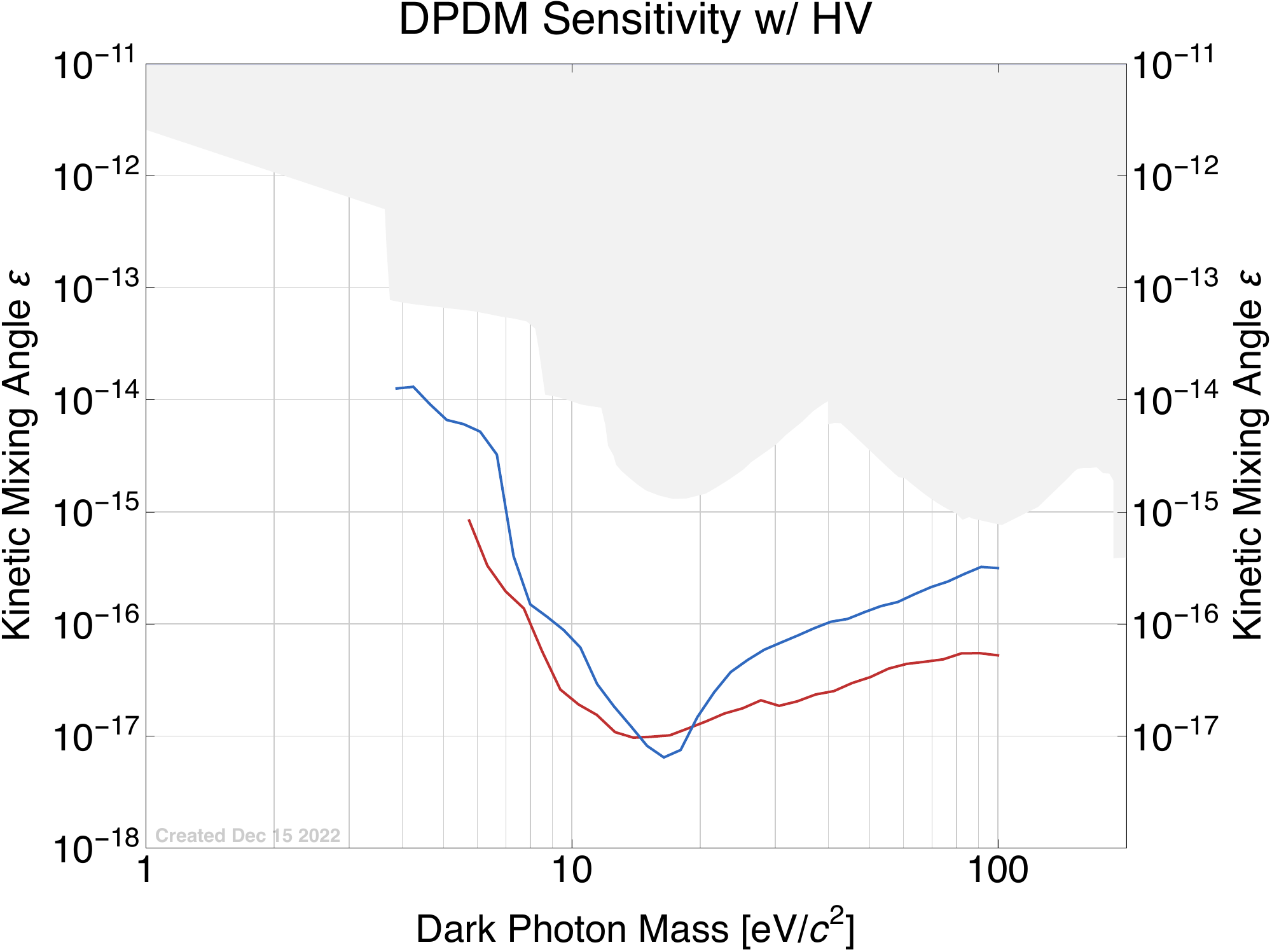} 
\hfill
\includegraphics*[width=0.48\textwidth,viewport=0 0 576 410]{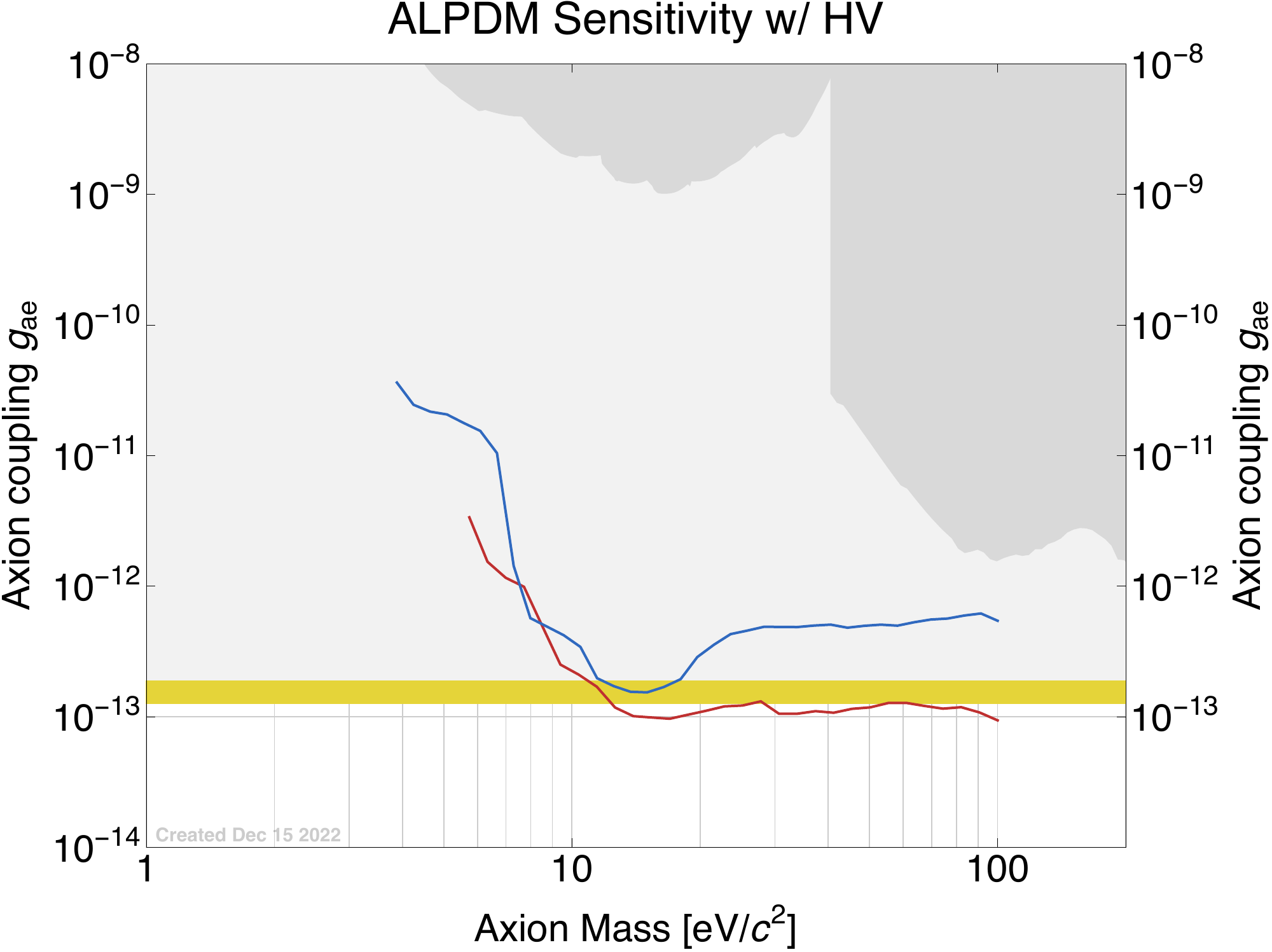} \\
\vspace{6pt}
\noindent
\hspace{0.45in}Ge\hspace{3.35in}Si\hfill \\
\includegraphics*[width=0.46\textwidth,viewport=0 0 420 322]{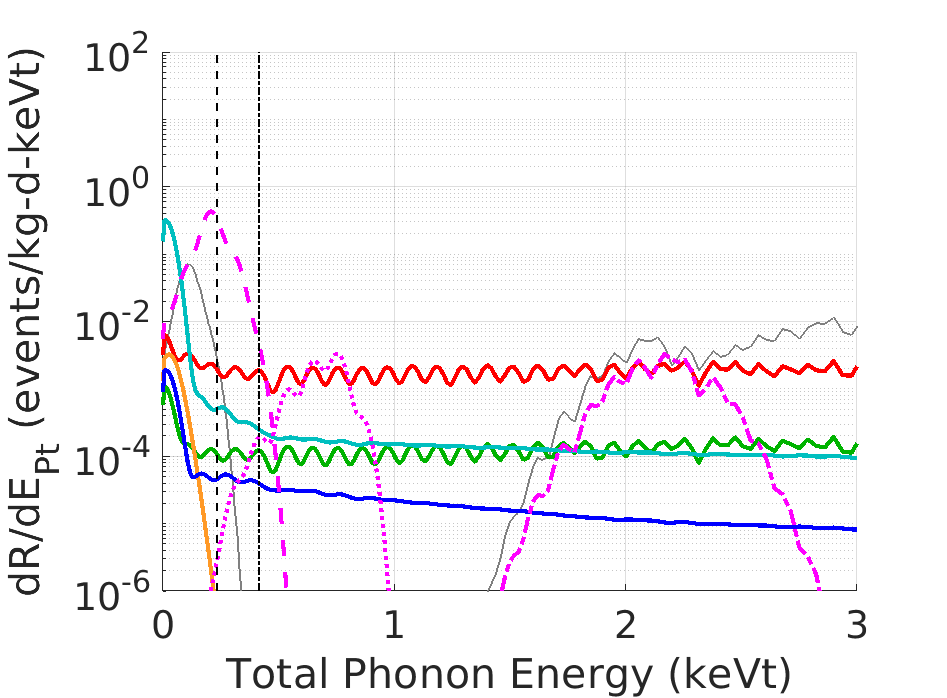}
\hfill
\includegraphics*[width=0.46\textwidth,viewport=0 0 420 322]{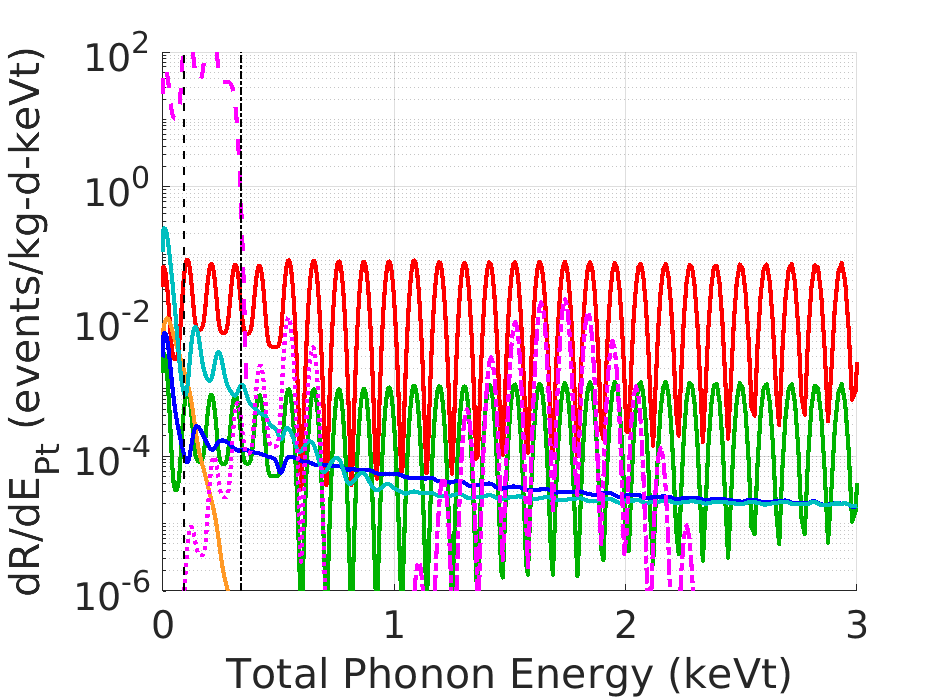}
% keep linear plots
%\\
%\includegraphics*[width=0.46\textwidth,viewport=0 0 420 322]{spectra/snolab/DP_HV_bkgSNOLAB_detSNOLAB_cons_Ge_postCuts.png}
%\hfill
%\includegraphics*[width=0.46\textwidth,viewport=0 0 420 322]{spectra/snolab/DP_HV_bkgSNOLAB_detSNOLAB_cons_Si_postCuts.png}
\end{center}
%\vspace{-6pt}
\caption[Expected sensitivity of SuperCDMS SNOLAB for dark photon and axion-like-particle dark matter (DPDM and ALPDM).]{\textbf{Expected sensitivity of SuperCDMS SNOLAB for dark photon and axion-like-particle dark matter (DPDM and ALPDM).}   All forecasts are based on current estimates of detector performance and backgrounds.  No comparisons to previous forecasts are shown because none have been published (in particular, none were presented in \cite{sensitivity2016}). (Top) Median expected PLR-based 90\%~CL exclusion sensitivity for HV detectors (Si: blue; Ge: red) for DPDM~(left) and ALPDM~(right).   Light shaded regions show the outline of all current limits (astrophysical and from terrestrial detectors).  Dark shaded regions highlight current exclusion limits from  searches for ALPDM absorption in terrestrial targets.  The yellow band indicates the ALP stellar cooling hint~\cite{stellarcooling2016}. (Bottom)~Background spectrum in phonon energy after all analysis cuts described in \S\ref{sec:forecast_procedure_detail} for Ge~(left) and Si~(right), HV detectors only.  Legend as in Figure~\ref{fig:snolab_reach_nrdm}.  Magenta:~candidate DM signals for $\mathit{\mdm \approx}$~6, 20, and 60 \eV and coupling at the exclusion sensitivity.}
%\hrule
\label{fig:snolab_reach_dpdm_alpdm}
\end{figure}

\begin{figure}
\begin{center}
\textbf{SuperCDMS SNOLAB  \eV Electron-Coupled LDM} \\
\vspace{3pt}
%\vspace{-12pt} \green{220924 up-to-date}  \\ 
%\red{Harrison: Use linear plots, change x-range to 0 to 1.5 keV, tick labels every 0.5 keV} \\
\begin{tabular}{p{0.1in}cp{2.5in}c}
& $F(q) = 1$ & & $F(q) = 1/q^2$  \\
\end{tabular}\\
\includegraphics*[width=0.48\textwidth,viewport=0 0 576 410]{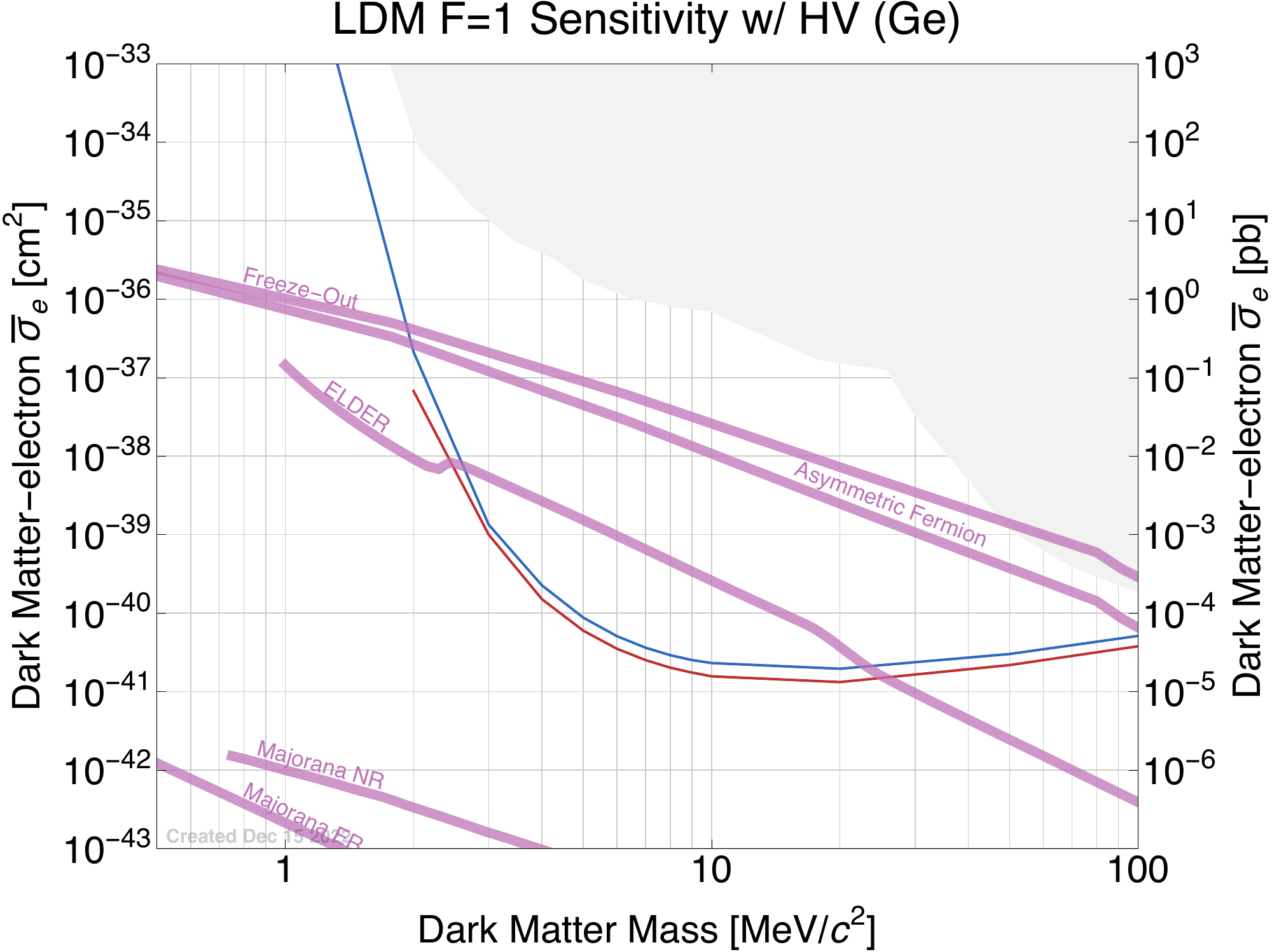} 
\hfill
\includegraphics*[width=0.48\textwidth,viewport=0 0 576 410]{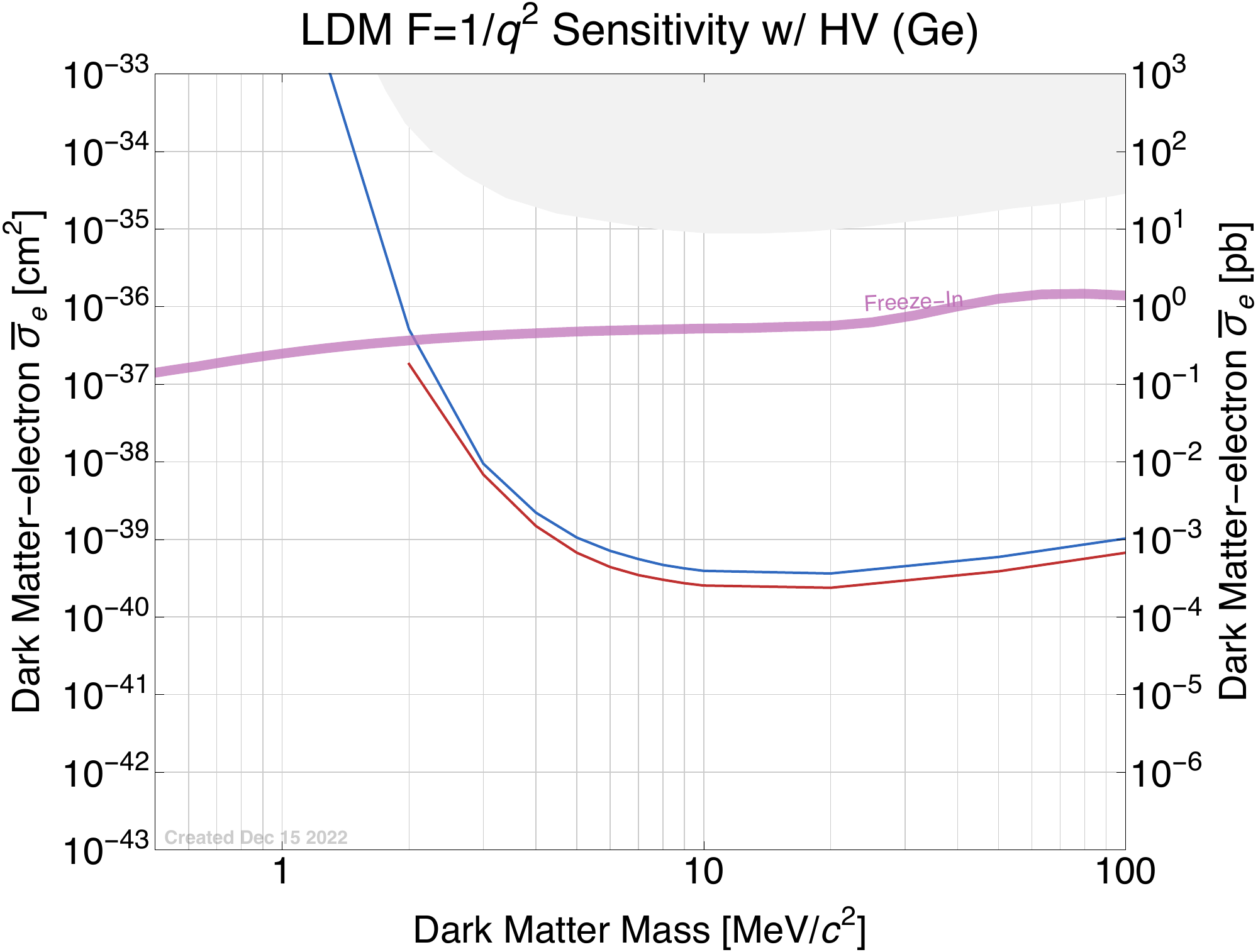}\\ 
\vspace{12pt}
\begin{tabular}{p{0.55in}cp{2.35in}c}
& Ge, $F(q) = 1$ & & Ge, $F(q) = 1/q^2$  \\
\end{tabular} \\
\includegraphics*[width=0.46\textwidth,viewport=0 0 430 322]{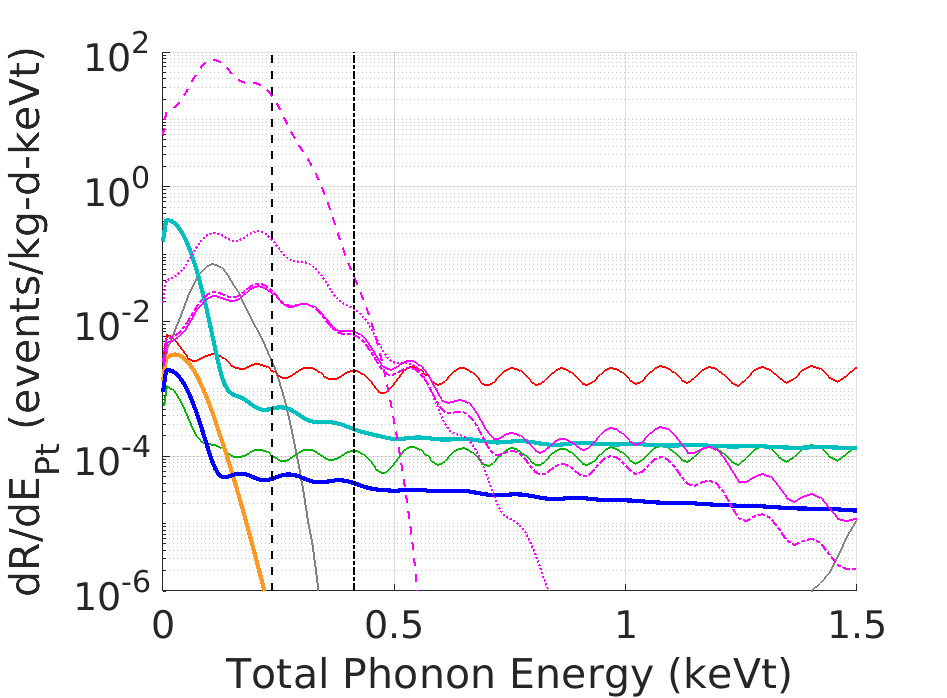}
\hfill
\includegraphics*[width=0.46\textwidth,viewport=0 0 430 322]{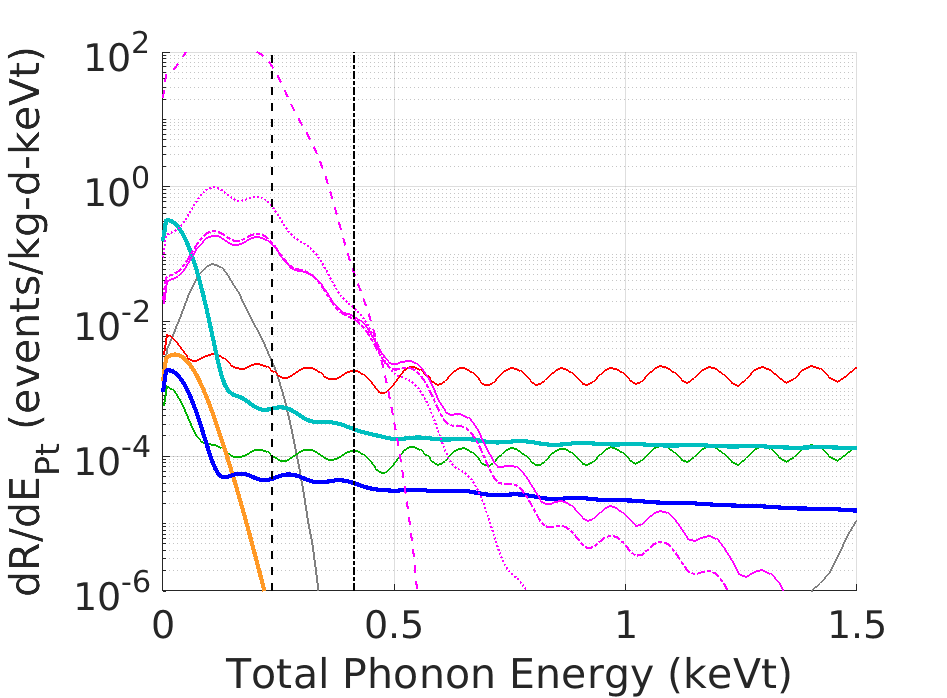}
\\
\vspace{6pt}
\begin{tabular}{p{0.55in}cp{2.35in}c}
& Si, $F(q) = 1$ & & Si, $F(q) = 1/q^2$  \\
\end{tabular}\\
\includegraphics*[width=0.46\textwidth,viewport=0 0 430 322]{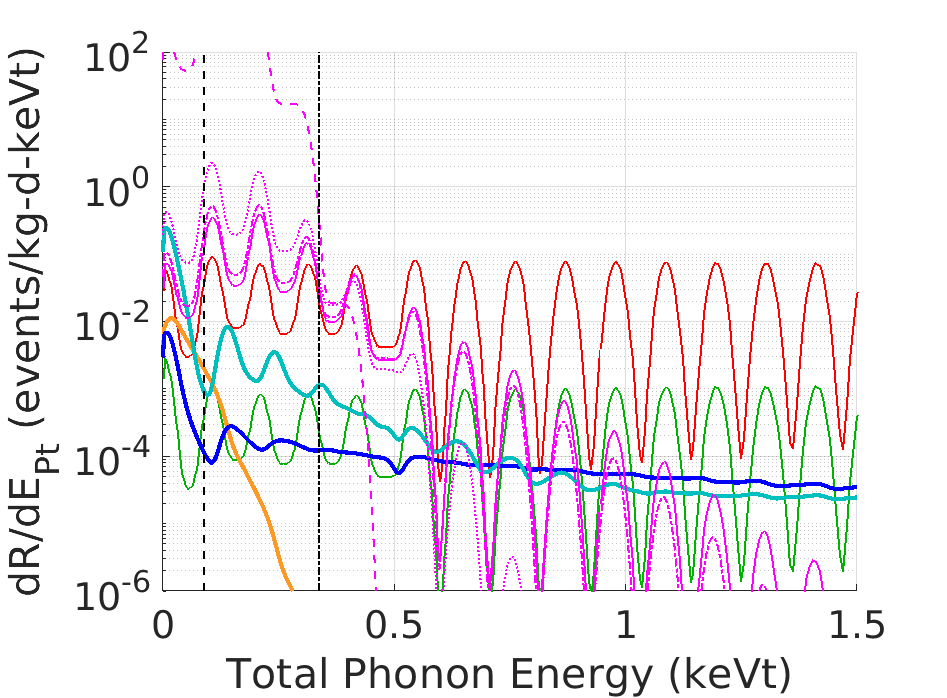}
\hfill
\includegraphics*[width=0.46\textwidth,viewport=0 0 430 322]{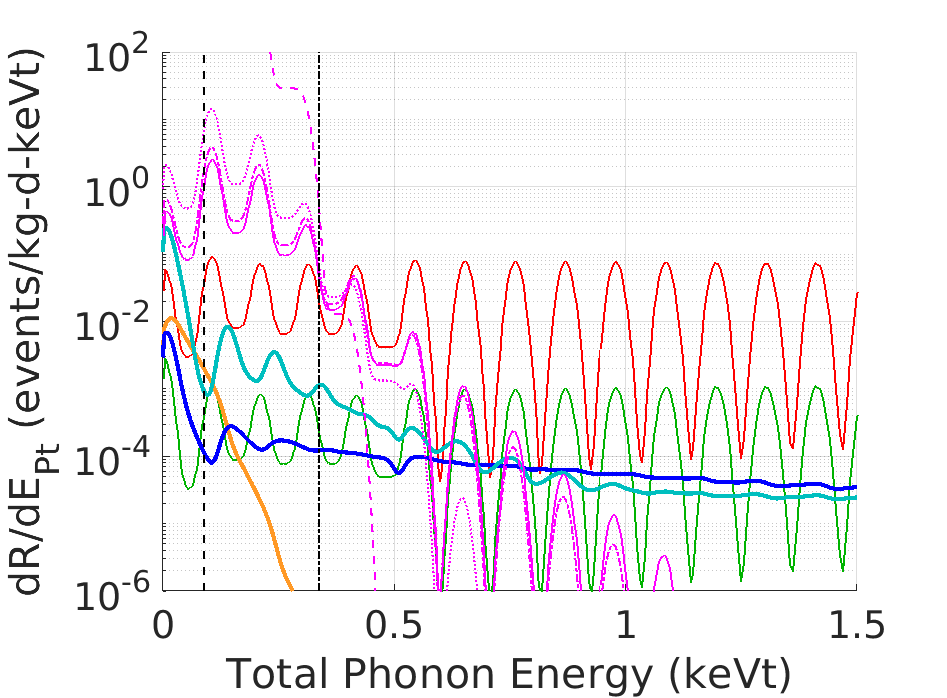}
%\vspace{-6pt}
\end{center}
\caption[Expected sensitivity of SuperCDMS SNOLAB for dark-photon-coupled light dark matter (LDM).]{\textbf{Expected sensitivity of SuperCDMS SNOLAB for dark-photon-coupled light dark matter (LDM).}   All forecasts are based on current estimates of detector performance and backgrounds.  No comparisons to previous forecasts are shown because none have been published (in particular, none were presented in \cite{sensitivity2016}). (Top) Median expected PLR-based 90\%~CL exclusion sensitivity for HV detectors~(Si: blue; Ge: red) for heavy mediator~(form factor $\mathit{F(q) = 1}$, left) or a light mediator~(form factor $\mathit{F(q) = 1/q^2}$, right).   Specific sharp targets~\cite{cvdm2017} are shown as thick magenta lines, assuming $\mathit{M_{A'} = 3 \mdm}$ where $A'$ is the dark photon and $\chi$ is the dark matter. (Middle and bottom)~Background spectrum in phonon energy after all analysis cuts described in \S\ref{sec:forecast_procedure_detail} for Ge~(middle) and Si~(bottom) for heavy~(left) and light~(right) mediators, HV detectors only.  Legend as in Figure~\ref{fig:snolab_reach_nrdm}.  Magenta:~candidate DM signals for $\mathit{\mdm \approx}$~1, 3, 10, and 30 \MeV and cross section at the exclusion sensitivity.}
%\hrule
\label{fig:snolab_reach_ldm}
\end{figure}

We also show in Figures~\ref{fig:snolab_reach_dpdm_alpdm} and~\ref{fig:snolab_reach_ldm}, for the first time, expected sensitivity estimates for HV detector searches for various electron-coupled dark matter models: dark photon, axion-like-particle, and dark-photon-coupled light dark matter (DPDM, ALPDM, and LDM).  We only show PLR estimates because OI expectations have not been presented in the past and, as noted above, we will make forecasts for PLR sensitivity for future scenarios.  We offer the following commentary on these forecasts:

\subparagraph*{Dark Photon Absorption} Perhaps most important for dark photon searches is that the signature is a spectral line, broadened of course but quite different from the falling exponential characteristic of nucleon or electron scattering.  The main difference between Si and Ge reach is threshold and exposure.  Both estimates incorporate the dielectric loss function for the material based, to the extent possible, on measured photoelectric cross sections.  In-medium effects are also incorporated. Significant new reach is possible at all masses accessible (down to a few \eVnosp).  The greatest sensitivity is in the 10--30~\eV\ range, covering $\ge 2$ orders of magnitude in kinetic mixing parameter ($\ge 4$ orders in rate) beyond current limits.

\subparagraph*{Axion-Like Particle Absorption} Reach for ALPDM is related to that for DPDM since the spectral shape is similar, though the mass-dependent conversion to coupling parameter differs.  ALPDM searches with SuperCDMS SNOLAB will extend the mass and coupling-parameter reach of terrestrial particle searches significantly, though much of this parameter space has already been ruled out by astrophysical constraints and solar axion searches.  Of particular note, however, is that the stellar cooling hint~\cite{stellarcooling2016} may be testable by Ge HV detectors above about 12~\eVnosp.

\subparagraph*{Dark-Photon-Coupled Light Dark Matter} SuperCDMS SNOLAB HV detectors will test new parameter space at all masses to which it is sensitive, down to about 1~\MeVnosp.  The greatest new reach will be near 10~\MeVnosp, probing nearly five orders of magnitude in cross section beyond current limits.  SuperCDMS SNOLAB will test significant portions of the sharp-target parameter space for Freeze-out, Asymmetric Fermion, SIMP, ELDER, and Freeze-in models as outlined in~\cite{cvdm2017}.

\clearpage

\subsection{Overview of Future Opportunities}
\label{sec:future_overview}

It is known that thermally produced, fermionic or bosonic DM can be no less massive than a few keV\footnote{The phase-space of galaxies below the escape velocity would be insufficient for fermions having lower mass and thus higher speed (the phase space density is fixed in momentum units)\nocite{}, and cosmological observations of the Ly-$\alpha$ forest would observe suppressed structure were the DM lighter\nocite{}.}.  The momentum of such particles in our galactic halo places a meV lower limit on energy depositions via scattering for such DM.  Bosonic DM remains viable below the few \keV mass scale because it can be produced non-thermally (as well as thermally), and absorption of such lower-mass bosonic DM would be the detection mechanism\footnote{Scattering would require sensitivity to even smaller energy depositions.}.  The same \meV regime, however, is where bosonic DM transitions to semiclassical wave-like, rather than particle-like, behavior\footnote{The bosonic DM occupation number reaches unity for eV masses, but a substantially higher occupation number is needed for particle detection to become ineffective.}, rendering conventional particle detection ineffective.  The meV--keV energy deposition regime is thus the clear experimental target for low-mass particle DM searches.  
% even though pt threshold is 100-200 eVt, HV detector get us to tens of eVnr:
% ionization yield ~ 0.1 at 100 eVr, but x100 in NTL gain, so 100 eVnr -> 1 keVt.
% 100-200 eVt is therefore in the tens of eVnr.

SuperCDMS SNOLAB is an important foray into this meV--keV regime for DM searches not just in sensitivity but also because it validates scaling laws for detector performance down to an order of magnitude lower energies.  These scaling laws are critical to forecasting future improvements and DM mass reach.  The facility itself provides a vast improvement in background levels over the previous generation experiment, SuperCDMS Soudan.
% 221016 S. Golwala: Don't need this anymore since don't need background improvements
%, and it is amenable to further significant background reductions via well-understood approaches.  

These two inputs undergird ``experience-based,'' quantitative forecasts for Ge and Si targets going forward, which we know now can explore down to eV-scale energy depositions.  They also provide a solid foundation for the extension of the technology to the meV--eV regime on these same substrates as well as on new substrates with sensitivity to dark matter models inaccessible with Ge and Si.  In this section, we first discuss some general considerations about how detector response to energy deposition will change as we enter the regime below tens of eV (\S\ref{sec:novel_energy_regimes}), followed by summaries of the opportunities of both experience-based upgrades  (\S\ref{sec:experience_upgrades}) and of extension to the sub-eV energy regime and new substrates (\S\ref{sec:novel_directions}).  We then provide detailed forecasts for the experience-based upgrades (\S\ref{sec:experience_forecasts}) and more discussion of ongoing efforts in novel directions (\S\ref{sec:novel_forecasts}).

%Our experience-based upgrades (\S\ref{sec:experience_upgrades}, \S\ref{sec:experience_forecasts}) offer the potential to extend this reach by 2-3 decades to $\mathcal{O}$(0.1--1)~eV thresholds with our current Si and Ge targets but leave an equal number of decades unprobed.

\subsubsection{New Regimes in Energy Deposition below Tens of~eV}
\label{sec:novel_energy_regimes}

The physics of how energy deposition changes in the energy regime below tens of eV impacts the types of detectors and materials we should consider and how various particle backgrounds manifest in our detectors.

\paragraph{Energy Deposition via Interactions with Nuclei}
\label{sec:nucleon_coupling}

\begin{figure}
\begin{center}
\begin{tabular}{p{2.65in}p{3.65in}}
\includegraphics*[height=2.5in,viewport=160 510 430 760]{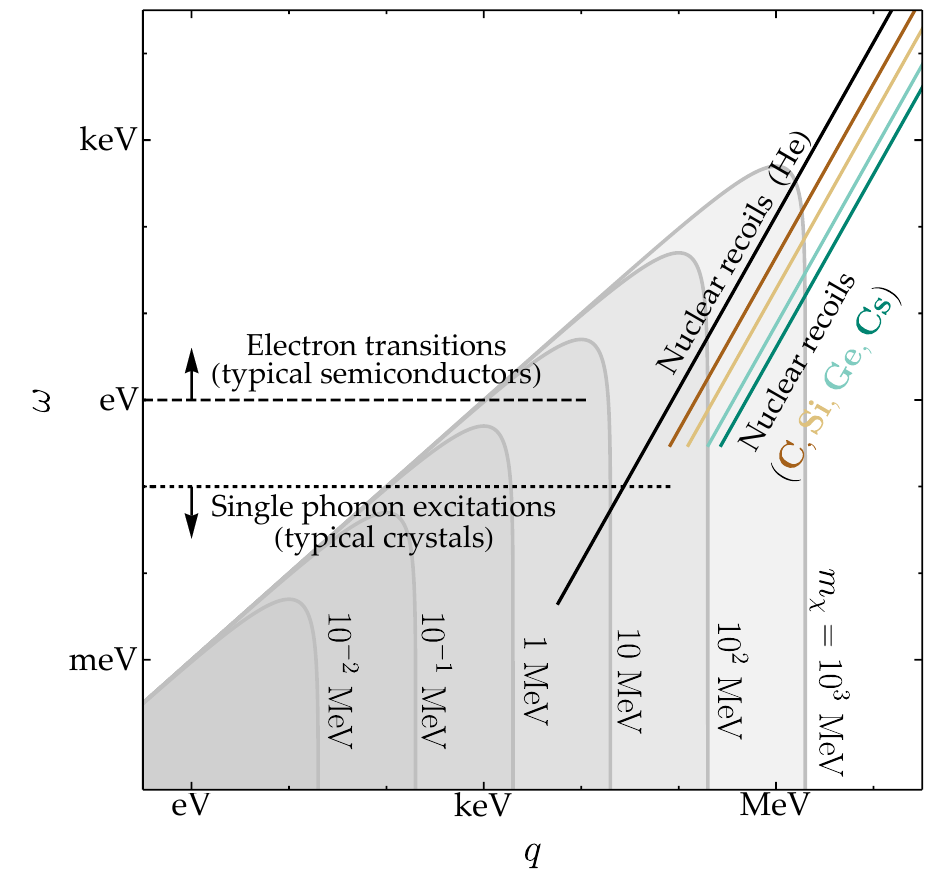} 
& \includegraphics*[height=2.4in,viewport=320 555 560 735]{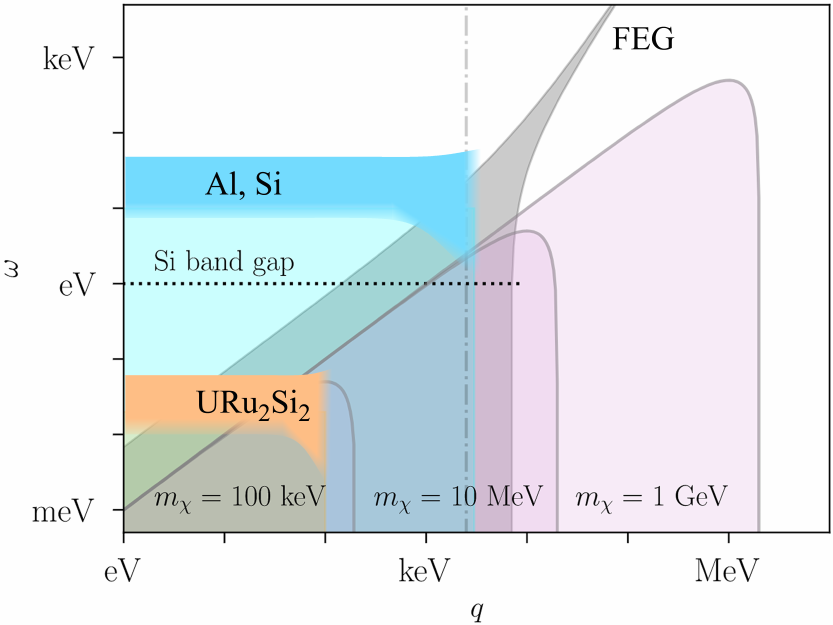} \\
\end{tabular}
\end{center}\vspace{-6pt}
\caption[Connection between DM kinematics and detector excitations]{\textbf{Connection between DM kinematics and detector excitations.}  The inverted parabolas are the space of energy-momentum transfer accessible for the indicated DM particle mass and $\mathit{v = 10^{-3}c}$.  (The $q$ axis is obtained by multiplying momentum by $c$.)  (Left)~Lines indicate the locus of nuclear recoils for various materials, terminating where scattering can no longer be treated as involving a single nucleus (about an order of magnitude below the energy needed to displace a nucleus from its ionic site).  Regions occupied by electronic transitions in semiconductors and single-phonon excitations are indicated.  (Right)~The blue and orange regions indicate the support of the plasmon part of the dielectric loss function for typical (Al, Si) and heavy-fermion (URu$_2$Si$_2$) materials.  The former has a tail extending into the DM region while the latter strongly overlaps it, indicating plasmon excitation is feasible.  The grey ``FEG'' region is for a free-electron gas and is indicative of scattering with unbound electrons.  The Si bandgap is indicated, consistent with the left figure.  The dashed-dot line is the edge of the silicon Brillouin zone, below which single-phonon-creation becomes viable.  Figures from~\cite{trickle2020,hochberg_dielectric_function_prl2021}.}
\label{fig:excitations}
\hrule
\vspace{6pt}
\end{figure}

Dark matter that interacts with nuclei includes~(see, e.g.,~\cite{cvdm2017}): WIMP-like Majorana fermion DM interacting via scalar or vector bosons; scalar or Dirac fermion DM interacting via a kinetically mixed dark photon with proton number; and, scalar or Dirac fermion DM interacting via a hadrophilic scalar mediator with nucleon number.\footnote{Bosonic DM cannot be absorbed by nuclei, as there is no final state that conserves energy and momentum.}  For such DM candidates, we must address two new regimes of recoil energy deposition, as illustrated in Figure~\ref{fig:excitations}.  

\subparagraph*{eV to Few Tens of eV Nucleon-Coupling-Mediated Depositions} In the first regime, below recoil energies of tens of eV, ionization yield for nuclear recoils vanishes almost completely because the momentum transfer from a DM particle is insufficient to displace the nucleus from its ionic site\footnote{The Frenkel defect energy; see, e.g.,~\cite{supercdms_frenkel2018}.}.  Instead, the energy will be deposited completely as phonons.\footnote{The fundamental cross section is unchanged because the interaction is still  with a single nucleus.  The primary channel available for energy transfer is phonons.  Some energy may transfer to electron recoils via the Migdal effect\nocite{}, but that coupling is highly suppressed relative to the ionic motion's coupling to phonons.}  In this regime, NR discrimination via ionization yield remains possible in principle because ERs still produce ionization, but the lack of ionization for NRs limits rejection of noise events and environmental backgrounds (\S\ref{sec:env_backgrounds_summary}).  Spectral discrimination --- here, the steepness of the DM recoil spectra compared to those of particle backgrounds --- becomes more useful in this regime.  The interactions do remain energetic enough, well above the energies of individual phonons, that the material's phonon spectrum does not affect the DM cross section.

\subparagraph*{Sub-eV Nucleon-Coupling-Mediated Depositions} As we consider energy depositions below $\sim$1~eV, the picture of an interaction with a single nucleus fails because the wavelength of the corresponding momentum transfer, $\lambda_q = h/\sqrt{2\,m_N\,\er}$, becomes larger than the spatial extent of the ion wavefunction in the crystal lattice's harmonic potential, $\lambda_\text{ion} = h/\sqrt{2\,m_N\,\ephonon}$~\cite{trickle2020}.  That is, the regime of validity of the standard nuclear interaction treatment is $\er \gg \ephonon$.  In the extreme other limit, $\er \sim \ephonon$, the DM couples directly to the target material's phonon spectrum via the nucleon coupling.   The range in between single-phonon coupling and the standard nuclear interaction, $\ephonon < \er \lesssim 10\,\ephonon$, is where direct multi-phonon creation must be considered, work that is currently underway but not yet published~\cite{trickle2020}. 

Si and Ge have maximum phonon energies of tens of meV (acoustic and optical), so the standard treatment is valid down to hundreds of meV, approximately the lower limit of the experience-based upgrades considered in \S\ref{sec:experience_upgrades} and \S\ref{sec:experience_forecasts}.  Some materials with high sound speeds have phonon spectra reaching 100--200~meV, opening up the direct phonon-creation channel as the conventional nuclear recoil channel closes in Ge and Si.  In addition, creation of ionization becomes impossible even for ERs below the bandgap energy ($\approx$1~eV in Si and Ge, larger in the high-sound-speed materials), so the utility of ionization measurement for ER rejection vanishes at that energy for most materials, rendering phonon-only measurements the default modality.  We will consider such new targets in \S\ref{sec:novel_directions}.

ER rejection can in principle be extended to lower energy via use of lower-bandgap targets, such as InSb, Dirac materials, and magnetically ordered materials.  We will explore this option in \S\ref{sec:novel_directions} also.  

\paragraph{Energy Deposition via Couplings to Electrons}
\label{sec:electron_coupling}

For dark matter that interacts with electrons~(see, e.g.,~\cite{cvdm2017}) --- complex scalar and Dirac fermion DM scattering via a kinetically mixed dark photon, bosonic DM that can be absorbed --- similar phenomena with regard to binding and multi-site interaction occur as the energy deposition decreases.  The phenomenology is in theory richer because electrons, unlike nuclei, can be bound in a  variety of ways, ranging from keV binding energies in inner atomic shells of high-$Z$ materials to meV binding energies in superconducting Cooper pairs.  Figure~\ref{fig:excitations} illustrates the kinematic match between these different electron bound states and dark matter.

\subparagraph*{$\boldsymbol{\gtrsim}$ eV Electron-Coupled Depositions} In the first regime, down to the bandgap or electron binding energy of the particular material (typically $\gtrsim 1$~eV, though various materials may offer smaller gaps (\S\ref{sec:novel_directions})), DM interactions can create electron-hole pairs or liberate electrons in a manner similar to higher energies, though the production may now be visibly discretized (and the statistics of that discretization, such as Fano factors, must be accounted for).  The material's dielectric loss function, Im$(q)$, plays a role in modulating the coupling strength and thus creates mass-dependent structure in dark-photon absorption sensitivity curves.  This structure is integrated over for scattering of low-mass fermionic DM and not present for axion-like-particle DM.

\subparagraph*{$\boldsymbol{\lesssim}$ eV Electron-Coupled Depositions} Below the \textit{typical} bandgap or ionization energy, it is not kinematically possible to liberate an electron or electron-hole pair.  One approach is to consider materials with smaller bandgaps (\S\ref{sec:novel_directions}).  Reach down to the few keV thermal limit, however, is most readily available via the use of polar crystals, which have optical phonons at tens of meV that couple well to the dark photon\footnote{Due to the electric dipole moment of the unit cell in such materials, which gives them a strong coupling to dark photons via kinetic mixing.} (as the DM or as mediators).\footnote{Superconductors may offer reach down to even smaller energy depositions via their meV binding energies, but there is some debate in the literature on this topic (e.g., \cite{hochberg_pyle_zurek_jhep2016} vs.\ \cite{hochberg_dielectric_function_prl2021}), yielding different conclusions about how screening degrades the sensitivity of superconductors and whether it also applies to light scalar and non-kinetically mixed vector mediators. (Recent work using an effective theory (EFT) framework \cite{Mitridate:2021} concludes that dark matter absorption rates for vector (dark photon) and pseudoscalar (axion-like particle) dark matter can be calculated directly from the dielectric function as in \cite{hochberg_dielectric_function_prl2021}, but that, for scalar DM, the EFT framework is required, and gives different results than previous scalar DM scattering rate calculations~\cite{Hochberg:2016ultralight,Gelmini:2020,hochberg_dielectric_function_prl2021}.)
% above footnote is text from Noah in internal report
Also, using superconductors to sense smaller energy depositions is less well motivated, as fermionic candidates do not deposit such small energies, leaving only DPDM and ALPDM in the narrow mass range between the meV particle/wave boundary and the tens of meV energies of optical phonons.}

\subsubsection{Experience-Based Upgrades to Reach eV Energy Deposition}
\label{sec:experience_upgrades}

With that general discussion in hand, we summarize here experience-based upgrades covering the first regime discussed in the prior section, eV-scale energy depositions (and a bit below).  The detector and background models for SuperCDMS SNOLAB are highly developed, the former based on extensive prototyping and latter on exhaustive assay work and detailed simulations, and these models provide the basis for detailed sensitivity forecasts in \S\ref{sec:experience_forecasts}.  While such upgrades may seem evolutionary, we will see that they are both ambitious and capable of substantial new science reach. 

\paragraph{Background Improvements} We have four different avenues for improvement in backgrounds:
\label{sec:background_upgrades_summary}

\vspace{-12pt}

\subparagraph*{Radiogenic activity} A handful of specific components or materials dominate the non-Rn radiogenic background.  Replacements for these now exist or can be expected to be identified.

\subparagraph*{Cosmogenic activation of non-detector components} While efforts have been made to limit cosmogenic exposure of the copper in the apparatus, the large mass implies any above-ground exposure will result in a non-negligible background contribution.  Underground electroforming and machining is a known means to vastly reduce this background.

\subparagraph*{Radon exposure} Similarly, while efforts have been made to limit radon exposure of detectors and line-of-sight components, additional measures can be taken.  Some are relatively modest changes to detector fabrication procedures that were identified too late to impact SuperCDMS SNOLAB.  Others are wholesale changes in the environment in which the work would be done, such as use of radon-suppressed environments for the majority of steps in detector fabrication and testing.

\subparagraph*{Cosmogenic activation of detector material} The sensitivity of the SuperCDMS SNOLAB will be limited by bulk ER backgrounds due to \hthree and \sitt created cosmogenically in the detector material, as noted in \S\ref{sec:volumesources}.  (For Ge, spectral lines due to activation are also important.)  \sitt can potentially be reduced by modified sourcing or isotopic enrichment.  Reduction of \hthree requires the full detector life-cycle to be underground: from zone refinement and crystal growth through crystal shaping and polishing, detector fabrication, and detector testing.  Reduction of Ge activation isotopes requires the Ge material be underground for an extended amount of time, which also requires a fully underground detector life-cycle to prevent re-activation.

\paragraph{Cryostat Size} 
\label{sec:cryostat_upgrades_summary} As noted in \S\ref{sec:SNOBOX}, the cryogenic system was originally designed for a larger cryostat capable of accommodating 31 Towers, though the cryostat itself was downsized to 7 Towers.  Replacing the cryostat with one having the original design capacity, without changing the existing refrigerator, is feasible.

\paragraph{Detector Improvements} For detectors, there are three different areas in which improvements are expected:
\label{sec:detector_upgrades_summary}

\subparagraph*{Phonon Energy Resolution} For phonon energy resolution, three orders of magnitude of improvement is considered within reach.  Such resolution improvements would be obtained by three different changes: reducing the detector size, reducing the \Tctxt of the TES films used to ultimately transduce the phonon signal to an electrical signal, and by improving the transmission/trapping probability of the interface between the Al films that absorb phonons and the W TES films.  Simple, validated scaling laws predict the expected improvements.  

In addition to simply reducing energy threshold for all detectors, improved phonon resolution has specific additional effects for HV detectors.  The first is to make \textbf{spectral discrimination of NRs from ERs} possible.  In HV detectors, events with recoil energy \ertxt reside at $\neh \evb + \er$ where $\neh$ is the number of electron-hole pairs produced.  For ERs, $\neh \approx \er/\epseh$ (recall, \epsehtxt is the typical energy needed to create an electron-hole pair).  For NRs, $\neh \approx \er\,y(\er)/\epseh$ where $y(\er)$ is the NR ionization yield at \ertxtnosp, taking on value 0.3--0.5 at high energies but dropping to 0.1 (and eventually vanishing) for sub-keV energies.  When $\sigpt \ll \evb$, ERs appear as spectral peaks at $\neh\, (\evb + \epseh)$ while NRs appear in the region between the $\neh$ and $\neh+1$ ER peaks.  The $\sigpt \ll \evb$ resolution condition holds for all upgrade scenarios we consider.  The second effect is to mitigate ionization leakage (see below) by improving the separation of ERs and NRs from leakage spectral peaks at $\neh\, \evb$ and by improving the rejection of pileup of individual leakage events via the enhanced timing resolution ensuing from enhanced phonon resolution.  We will discuss the latter two effects in more detail in \S\ref{sec:detector_upgrades}.

\subparagraph*{Ionization Energy Resolution} Improvements in ionization resolution would extend ionization-yield-based nuclear recoil discrimination to lower energies and thus lower DM masses, but the prospects are far less promising: a factor of 3 from changes to the readout circuit and a factor of 3 from reduction in detector size, or one order of magnitude overall.  Larger improvements are highly uncertain because the physical path to significantly reducing the noise of the front-end high-electron-mobility transistor (HEMT) in the ionization readout circuit is not clear. 

Such ionization resolution improvements may in fact be obviated by phonon resolution improvements combined with the new ``piZIP'' (phonon iZIP) phonon-sensor configuration.  This concept uses the electric field configuration of the iZIP but seeks to measure ionization production via separation of the NTL and primary phonon components, which is possible because the former are generated primarily near the electrodes where the electric field is high.  If a fraction of the phonons \fdisctxt carry this NR/ER discrimination capacity, than the equivalent ionization resolution is 
\begin{align}
\sigqpizip & = \frac{1}{\fdisc}\, \frac{\epseh}{e\,V_b}\, \sigpt
\end{align}
where \sigpttxt is the phonon energy resolution.  With $\fdisc \approx 0.2$ expected and $\epseh < \evb$, the substantial improvements in phonon resolution may yield \sigqpiziptxt that outstrips the expected improvements in conventional ionization readout resolution.  An explicit demonstration of the piZIP detector concept would substantially enhance confidence in this option.

\subparagraph*{Ionization Measurement Nonidealities} For HV detectors, which rely on a phonon-based measurement of the ionization produced by an event using NTL gain, two non-idealities may be amenable to improvements.  The first, ionization leakage, consists of the tunneling-driven injection of individual charge carriers from the electrodes into the bulk, creating events that are the equivalent of a single electron-hole pair.  (These events lack recoil energy, so there is some ability to distinguish them on the basis of energy, noted earlier.)  Pileup of such events can yield multi-electron-hole-pair events.  Impact ionization and charge trapping can cause such events to appear at energies corresponding to a non-integral number of electron-hole pairs.  The leakage phenomenon itself may be reduced by blocking layers, while improved phonon energy resolution can improve spectral discrimination and reduce pileup.  The potential to reduce impact ionization and charge trapping is less well understood, but these non-idealities are not as limiting as leakage and leakage pileup.

\paragraph{Low-Energy Particle and Environmental Backgrounds: a New Challenge} 
\label{sec:low_energy_particle_backgrounds_summary}
\label{sec:env_backgrounds_summary}

New, low-energy backgrounds just now being explored may become visible in SuperCDMS SNOLAB and future upgrades as we explore the eV regime for the first time with this experimental apparatus.  Experience gained from SuperCDMS SNOLAB will guide the development of the mitigations necessary to ensure the experience-based upgrade projections provided in \S\ref{sec:experience_forecasts} can be realized.  

\subparagraph*{Low-Energy Particle Backgrounds} Recent low-mass dark matter searches (e.g., \cite{SENSEI2019, Amaral:2020ryn}) have measured backgrounds in the eV regime for the first time.  This work has encouraged an exploration of low-energy backgrounds previously neglected, including coherent photonuclear scattering~\cite{robinson2017}, secondary photon and phonon generation by atomic processes excited by high-energy particles~\cite{du_essig_secondary_photon_bgnd_2022}, and direct phonon creation by coherent photon-electron Rayleigh scattering as well as other processes~\cite{berghaus_essig_coherent_photon_scattering_phonons_2022}.  Standard particle background simulation packages such as GEANT4 do not include these processes yet, and work is ongoing to incorporate them as add-ons.  It will be necessary to complete the list of such backgrounds considered, calculate their rates, and incorporate them into simulations.  Our implementation of coherent photonuclear scattering is discussed in \S\ref{sec:low_energy_particle_backgrounds}, while the other two backgrounds are currently neglected.  Full consideration of such backgrounds may motivate changes in the experimental apparatus, such as modifying the detector mounting hardware or implementation of an active veto, to ensure the background levels assumed here can be achieved.

\subparagraph*{Low-Energy Environmental Backgrounds} Historically, particle backgrounds (\S\ref{sec:background_upgrades_summary}) have been the primary concern for rare-event experiments like DM searches, with so-called ``environmental backgrounds'' --- spurious events or noise floors caused by physical or electrical activity in the local environment --- deemed to be non-fundamental and amenable to experiment-specific debugging.  It is already clear~\cite{stress_induced_2022}, however, from prototypes that approach the sub-eV deposition regime that these environmental backgrounds will become a more serious obstacle because they may begin to arise from fundamental design aspects of the experiment in much the same way as particle backgrounds.  These backgrounds must now be treated more systematically than in the past.  Given the still immature understanding of how to predict such backgrounds, we do not include them here.  Experience from SuperCDMS SNOLAB may motivate changes in the detector mounting hardware to address such backgrounds.

\subsubsection{Novel Directions}
\label{sec:novel_directions}

We discuss here a set of ideas that would enable mass and/or cross-section reach beyond that possible with the experience-based upgrades.  These ideas are not sufficiently well-developed to enable sensitivity projections of the type we can make for experience-based upgrades, but they are \rnd directions that will hopefully be realized on a longer timescale.  Specific activities on these fronts will be described in \S\ref{sec:novel_forecasts}.

\paragraph{New Channels and New Targets for New Regimes in Energy Deposition} 
\label{sec:novel_targets_summary}

\begin{figure}
\begin{center}
\begin{tabular}{p{3.25in}p{3.25in}}
\includegraphics*[width=3.2in,viewport=60 445 565 745]{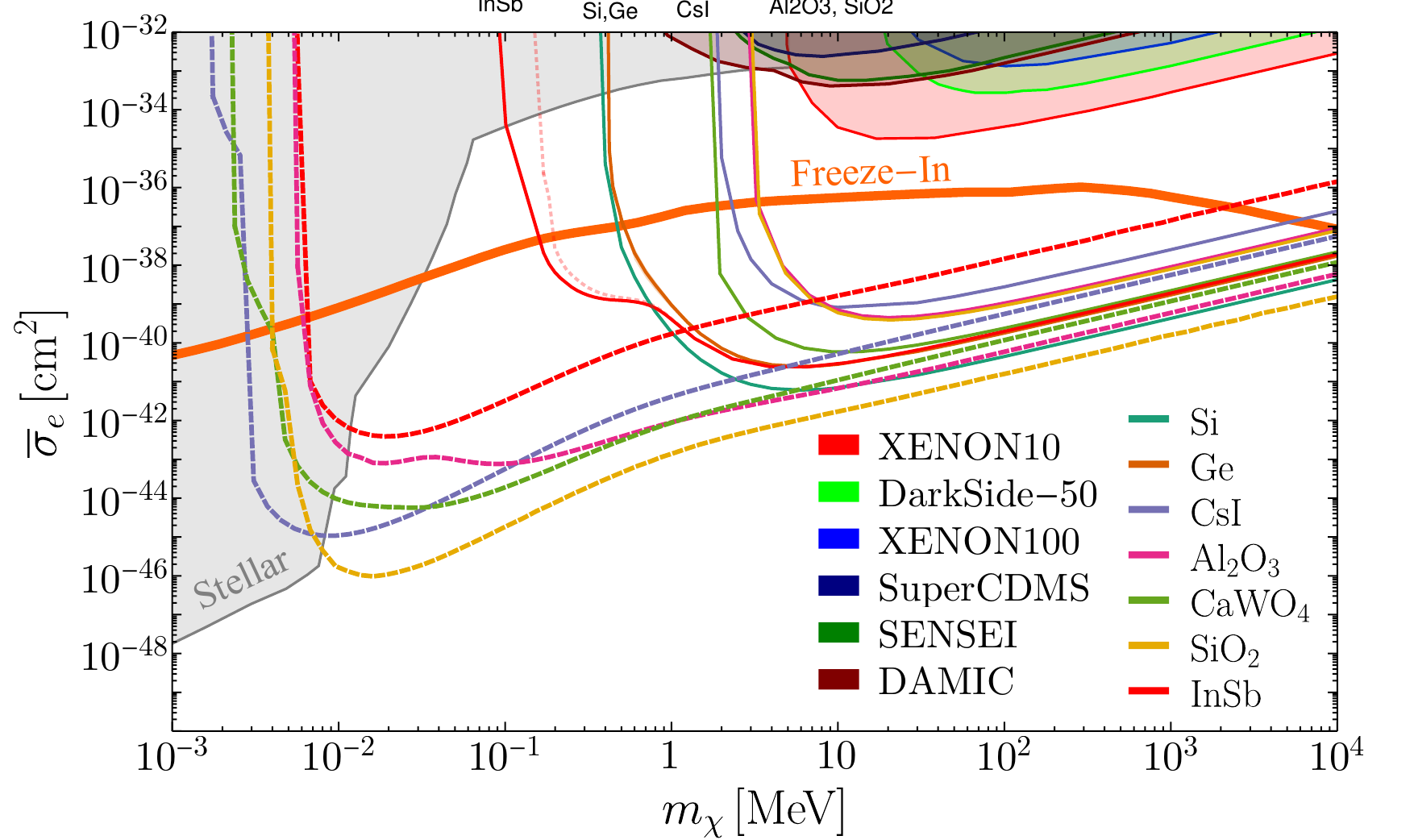} 
& \includegraphics*[width=3.2in,viewport=60 445 565 745]{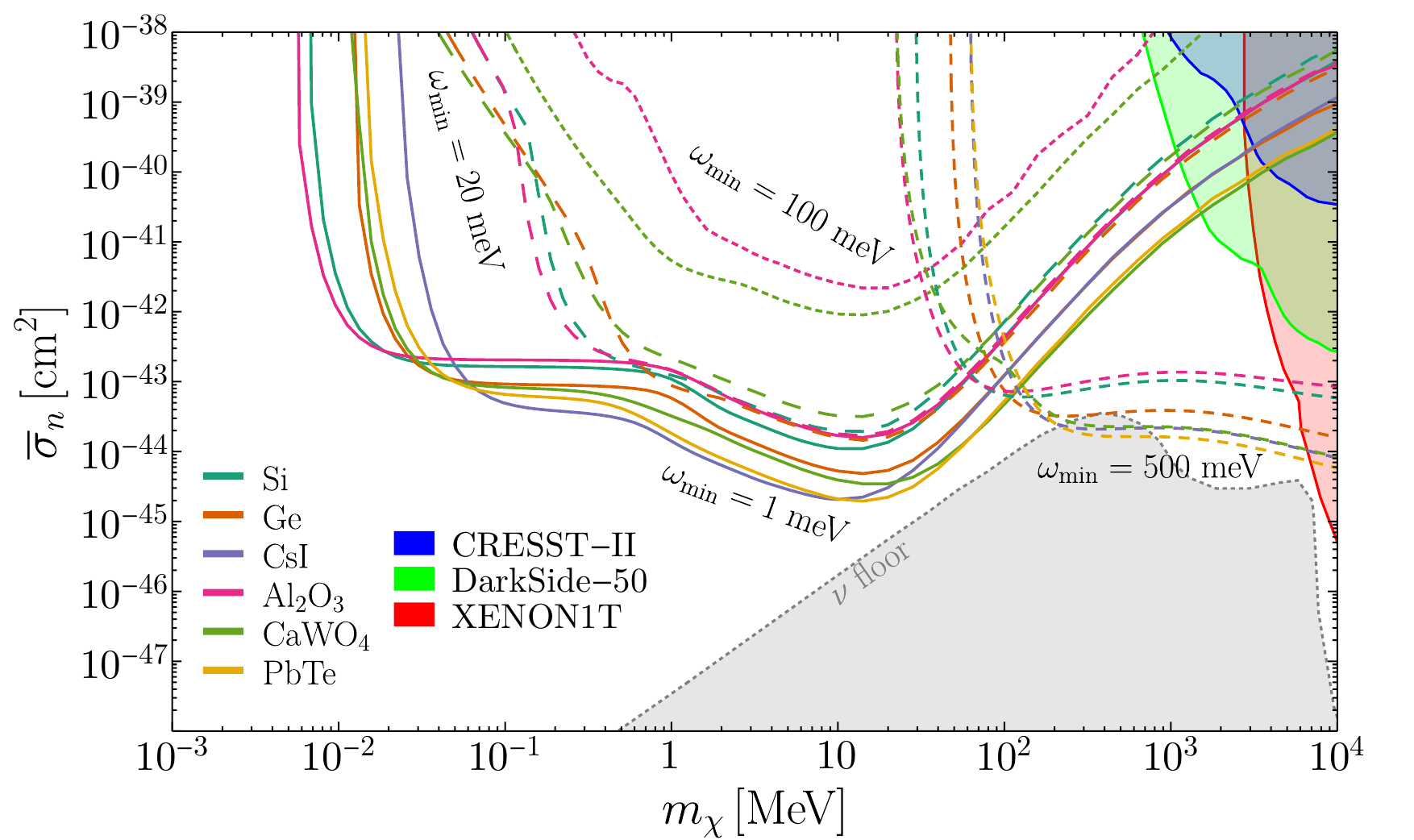} \\ 
\end{tabular}
\end{center}
\vspace{-6pt}
\caption[Target material sensitivity comparison]{\textbf{Target material sensitivity comparison.}  Expected sensitivity of various materials for 1~kg-yr exposure for (left)~nucleon scattering mediated by a heavy ($\mathit{\gtrsim 400}$~\MeVnosp) scalar that couples primarily to hadrons and (right)~electron scattering mediated by a light (low-mass; $\mathit{F(q) = 1/q^2}$) dark photon.  Nucleon scattering uses conventional nuclear recoils ($\mathit{\omega_\text{min} = 500}$~meV) and direct creation of acoustic phonons ($\mathit{\omega_\text{min} = 1}$, 20, and 100~meV).  Current constraints are the shaded, colored regions.  The ``$\nu$ floor'' region indicates the neutrino fog.  Electron scattering uses detection of electron-hole pairs (curves terminating near 1~\MeVnosp) or optical phonons (curves terminating near 10~\keVnosp).  ``Freeze-In'' refers to one DM creation scenario.  Current constraints from stellar astrophysics (``Stellar'') and direct searches (shaded, colored regions at upper right) are shown. Figures from~\cite{griffin2020}.}
\label{fig:targets}
\end{figure}

\subparagraph*{Sub-eV Nucleon-Coupling-Mediated Depositions} 
As noted above, the picture of coupling to a single nucleus fails below a fraction of an eV for Ge and Si (i.e., around 10\ephonontxtnosp) and one must consider direct phonon production.  The theoretical situation is not clear yet because, as noted in \S\ref{sec:novel_energy_regimes}, only single-phonon production has been calculated while multi-phonon production calculations are still in process.  

It is certainly true that, if one only considers single-phonon production, Si and Ge do not become viable options again until one reaches thresholds of \ephonontxtnosp~$\sim$ tens of meV, their maximum phonon energies.  Other materials with higher sound speeds and thus higher maximum phonon energies become attractive alternatives; see Figure~\ref{fig:targets}.  Selection on this criterion and on heritage for thin-film photolithographic fabrication motivates a focus on sapphire (\sapphirenosp), quartz (\quartznosp), and \cawofournosp.  To these choices, we add the high-sound-speed materials diamond (C) and SiC because of their likelihood of consistency with photolithographic fabrication.

Two caveats on new materials must, however, be appreciated.  First, as thresholds decrease further, to the tens of meV regime, Figure~\ref{fig:targets} (left) shows that, at the meV thresholds required to reach the few keV thermal mass limit for fermionic nucleon-coupled DM, there is little distinction between Ge/Si and other materials.  Second, without multi-phonon calculations, it is not yet clear that this channel does \textit{not} provide Ge and Si continued sensitivity in the decade of energy between interaction with single nuclei and single-phonon production.  Given the significant investment involved in fabrication development and establishing radiopurity for new targets, it may be premature to conclude that wholesale transition to other materials is necessary \textit{for nucleon-coupled DM}.

Another avenue is to extend ER rejection to substantially lower energies via lower-gap materials.   The semiconductor InSb has a bandgap of $\approx$0.1~eV, about 10$\times$ smaller than Ge and Si.  Dirac materials such as ZrTe$_5$ (30~meV) are available, and smaller gaps are potentially realizable as new Dirac materials are developed.  Magnetically ordered materials may provide meV gaps~\cite{magnetic_order}. One caveat is that a smaller electron-hole-pair creation energy does not change the fact that nuclear recoils simply do not occur below the tens of eV energy deposition scale, and thus ionization production would only be used to reject ERs.  Another caveat is that, below tens of eV energy deposition, \cevns of solar neutrinos (\S\ref{sec:volumesources}) dominates over ER backgrounds, reducing the utility of such ER rejection.

\subparagraph*{Sub-eV Electron-Coupling-Mediated Depositions} 

Down to the $\approx$eV bandgaps of Ge and Si, they remain quite competitive for dark-photon-mediated scattering of light dark matter via electron-hole pair creation (\S\ref{sec:ldmhv0v}), with rates comparable to other targets.  They start to lose sensitivity below a few eV for dark photon and axion-like particle absorption (\S\ref{sec:dpdmhv0v}, \S\ref{sec:alpdmhv0v}), 
%due \red{presumably} to structure in the dielectric loss function \red{(what is the cause for ALPs?)}
% Belina comments:
% In vacuum both curves, the one for DPs and the one for ALPs, would have the same shape. But for DPs the effective mixing angle of the DP and the SM photon in matter (here Si or Ge) varies compared to the vacuum mixing parameter. The mixing in matter requires the consideration of the polarization tensor and with it the complex part of the conductivity, i.e. the response of the medium to the electro-magnetic wave/field. This is the actual BSM interaction in this case, i.e. the DP -> SM photon mixing. The SM photon -> electron coupling is SM stuff.
%The case for the ALPs coupling (somewhat) directly to electrons lies different and the gaee coupling remains unchanged also in matter. Hence the difference between DP and ALPs curves, at least at lowest masses and as long as in-medium effects are taken into account. Note that the effective kinetic mixing in matter and the vacuum kinetic mixing only differ below e.g. 20 eV or so for Si and becomes stronger the lower in mass you go. So you notice it the most at a few eV.
% SG -- all that said, this would suggest the ALP sensitivity would be flat down to the bandgap, so there must be some other effect, presumably just due to energy-momentum conservation as you approach the bandgap.  Could be a final-state phase-space issue.
motivating consideration of other materials for electron-hole-pair creation such as InSb, Dirac materials, and magnetically ordered materials.

A different approach, based on optical phonon creation in polar materials, may, however, be more promising, at least for dark photons.  Figure~\ref{fig:targets} (right) illustrates that optical phonon production in a variety of materials can provide access to DP-coupled fermionic LDM down to the few keV thermal limit.  Such materials also presumably promise DPDM reach to the bottom of their optical phonon spectra, which can be ten to tens of meV.  Three of these materials --- \sapphirenosp, \quartznosp, and \cawofour --- are known to be amenable to photolithography, with \sapphire and \quartz being most common.  \sapphire and \cawofour have in fact been in use by CRESST and other experiments for decades and thus there is enormous experience with radiopurity, too.  

A reasonable strategy would be to focus on polar materials first until the few keV fermionic thermal limit is reached, and then, if meV-gap Dirac or other materials are available at that point, turn attention to them to extend the reach for bosonic absorption to the meV particle/wave boundary.  The maturity of Dirac materials as potential fabrication substrates and low-radioactivity targets may evolve significantly by that point.  

\subparagraph*{Spin-Dependent Coupings Down to meV Scales}

Magnons are the electron-spin equivalent of phonons: they are a wave in the spin ordering of magnetically ordered materials, corresponding to precession of the electron magnetic moment about the nominal magnetic moment axis.  They have dispersion relations and energies similar to that of phonons.  Their main utility would be for the detection of DM (or neutrinos) interacting via a spin-dependent coupling (the non-relativistic limit of magnetic dipole, anapole, and pseudo-mediated couplings), which may not couple well to other excitations~\cite{trickle_magnons_2020}.  While there have been calculations, there has not yet been any experimental exploration of such materials.  They must have magnetic order, which generally requires heavier elements with $d$- and $f$-shell electrons, and so radiopurity would be a significant concern.  It is also not yet clear what sensor would be compatible with them.  As the literature on these materials evolves, further exploration of their potential for DM detection may be warranted.

\paragraph{Addressing Ionization Leakage with New Materials} 

In our experience-based upgrade scenarios, we assume improvements of an order of magnitude in ionization leakage in each of the \detB and \detC scenarios.  Should this optimism not be borne out, other approaches to reducing HV detector ionization leakage may be necessary.  

Some of the same materials discussed in the context of high sound speeds and optical phonon production --- mainly single-crystal diamond~\cite{diamond2019}, but perhaps also SiC~\cite{griffin_SiC_2021} --- have been proposed for HV operation because of their larger bandgaps, $\approx$2.5--5.5~eV, and potential for or demonstrated high charge collection efficiency.  The large bandgap suggests that impurity levels will be quite deep, perhaps mitigating ionization leakage as well as impact ionization and charge trapping.  While these higher bandgaps may limit mass reach for electron-coupled DM, that limit is likely no higher than that imposed by ionization leakage.  In particular, even the \detC HV scenario, in which ionization leakage is reduced by 100$\times$ from current values (Table~\ref{tbl:detector_upgrade_scenarios}), has an effective threshold of roughly 4 (3)~eV in Si (Ge), just above two electron-hole pairs (Figures~\ref{fig:dpdmhv_reach} and \ref{fig:ldmhv_reach}).  The bandgap in diamond and SiC are similar so, if they can be operated with much lower leakage current, they may provide comparable mass reach.  Futhermore, if leakage is not just negligible but is nonexistent, it may be possible to employ a more aggressive analysis mode, using the regions in between the electron-hole pair peaks all the way down to one pair or possibly even below one pair.  This capability would enhance NR/ER spectral separation, improving the reach for nucleon-coupled DM with HV detectors in the 0.5--5~\GeV mass range where ER backgrounds are most important.  (Below 0.5~\GeVnosp, neutrino backgrounds dominate, limiting the utility of this approach.)

\paragraph{Diurnal Modulation}

Diurnal modulation is a ``smoking gun'' for DM detection, a signal that can only be produced by DM.  Even in the absence of a signal, it may provide a path through the ``neutrino fog'' by exploiting the different directions of neutrinos from the Sun and incoming DM.  The energy regimes to be explored with SuperCDMS SNOLAB and beyond will begin to open up the potential for diurnal modulation via the anisotropic response of target materials, which has two regimes: anisotropy in response to the direction of motion of a nuclear recoil, and anisotropy in the response to the momentum transfer of the scattering event.  Such modulations may be as large or larger than the mean signal, motivating early consideration of these channels. 

\subparagraph*{Nuclear Recoil Response Anisotropy} As explained in \S\ref{sec:nucleon_coupling}, there is a threshold displacement energy for the ejection of an ion from its site of order tens of eV.  This threshold displacement energy strongly depends on the direction of the recoiling atom and target material~\cite{Nor18, Nor18b, Vaj77}.  This effect alone may be sufficient to produce measurable diurnal modulation in a DM signal whose typical energy deposition is in the vicinity of the displacement energy.  Two additional effects may also be relevant.  First, there is experimental and theoretical evidence indicating that the threshold for ionization production by a recoiling ion displays a nonlinear dependence on velocity at low energies due to electronic band structure effects~\cite{Val03,Mar09,Pri12}.  Second, the final state of the recoiling ion is likely interstitial, creating electronic states in the bandgap~\cite{Lim16} that can serve to decrease the energy required to create an electron-hole pair~\cite{Hor16}.  These two effects together may render the ionization yield dependent on the direction of the recoiling ion and thus the initial DM's direction of motion.  There has been some modeling of these effects~\cite{Sebastian_neutrinofloor}, suggesting they may yield tens of \% modulation, and more sophisticated modeling of the underlying condensed-matter physics is the next step.

\subparagraph*{Collective Excitation Response Anisotropy} As DM energy depositions become so small that they create discrete numbers of excitations --- an electron-hole pair at eV energies, a few or single phonons or magnons at lower energies --- diurnal modulation becomes possible simply because everything in a crystal lattice --- electronic band structure, dispersion relations of phonons or other collective excitations --- is anisotropic.  A number of papers have explored these effects~\cite{directional_sapphire_2018, griffin2020, directional_dirac_2020, directional_dirac_2021, griffin_SiC_2021}, some using the same high-sound-speed and polar materials considered earlier, and they conclude that modulations of order unity are possible.  This work has employed sophisticated condensed-matter physics models of the materials from the start, so serious analysis of the literature on the topic may tip the balance in favor of new target materials.

\clearpage

\section{Sensitivity Expectations for Experienced-Based Upgrades}
\label{sec:experience_forecasts}

We began our study of potential future directions by developing a matrix of upgrade scenarios.  As noted in \S\ref{sec:future_overview}, the two axes are backgrounds and detector improvements.  Each axis had three scenarios, corresponding to increasing levels of cost or risk and decreasing levels of maturity.  Risk and maturity refer to ``development risk'' --- how certain can one be that the specific approach is capable of achieving the desired goal? --- not ``implementation risk'' --- how certain can one be of achieving the goal in practice?  Cost, on the other hand, refers to a combination of development cost and implementation cost.  Implementation risk can generally be traded against implementation cost, but there is usually a floor for implementation cost for the construction of a new Tower.

The first scenario in each case was defined to consist of upgrades based on developments already or very close at hand but too late to be implemented for SuperCDMS SNOLAB, termed \bkgon and \detAnosp.  In the case of backgrounds, the next two scenarios (\bkgtw and \bkgthnosp) consisted of upgrades that depended on well-understood techniques for reducing backgrounds but that required some implementation development and would be more costly to realize.  In the case of detector upgrades, the next two scenarios (\detB and \detCnosp) consisted of plausible but not yet demonstrated evolution in detector parameters that would yield substantially improved detector performance in energy resolution, ER rejection/NR discrimination threshold, and/or ionization measurement non-idealities.

It became clear from this analysis that more advanced background reduction measures, while feasible, did not provide much scientific gain over the \bkgon scenario, certainly not enough to warrant their potential cost.  This outcome reflects \textbf{a huge strength of the SuperCDMS approach: improvements in detector performance can be used to circumvent backgrounds}.  The quantitative forecasts presented in \S\ref{sec:nrdm_0v}, \S\ref{sec:nrdm_neutrino_fog}, and \S\ref{sec:erdm} will show that \textbf{a vast amount of parameter space for both nucleon-coupled and electron-coupled dark matter can be explored with minimal background improvements}, the \bkgon scenario.

In terms of specific detector improvement options, the analysis shows that \textbf{phonon energy resolution improvements combined with detector size reductions are the best strategy for obtaining new reach to low masses for both nucleon-coupled and electron-coupled dark matter and for reaching the neutrino fog in the 0.5--5 \GeV mass range.}

We focus the following discussion on these most promising upgrade directions.  We begin with a somewhat generic discussion of the potential upgrades (\S\ref{sec:upgrades}) and then describe how various specific science goals (\S\ref{sec:forecasts}) make use of specific subsets of those upgrades.  The forecasting procedure was described in \S\ref{sec:forecast_procedure}.

\subsection{Upgrade Elements}
\label{sec:upgrades}

\subsubsection{Background Improvements}
\label{sec:background_upgrades}

%\red{For now, this is being written under the assumption of the \bkgon scenario.  SG thinks that we should run a \bkgz scenario: are we so dominated by \hthree and \sitt that reducing anything else is pointless?  Anticipated reductions in \bkgon discussed below.  This \bkgz scenario should re-grab bgnds files to incorporate split-chain mu-metal.}

%\red{Leave \bkgon discussion unpolished for now in case we just cut it.}

The \bkgon scenario seeks to address the two backgrounds that are both significant and most easily amenable to improvement:
\begin{itemize}
\item U/Th in Kapton and Cirlex\footnote{Cirlex is a thick version of Kapton made by laminating Kapton with an adhesive.} in flex cables and detector clamps together contribute 18\% of the bulk ER background.
% 2022/09/27 S. Golwala: Forecasts have shown we don't need to improve this
%\item \red{As noted earlier$^\text{\ref{footnote:mumetal}}$, the neutron background from the mu-metal shield is significant, about 35--40\% of the total NR background and comparable to the solar neutrino \cevns background.} 
\item Radon daughter plateout on detector surfaces dominates surface backgrounds.
\end{itemize}
The scenario consists of the following specific improvements:
\begin{itemize} 
    \item Remake the flex cables with lower activity kapton identified in \cite{arnquist_kapton}, which would  reduce \utte (\thtttnosp) contamination by a factor of 100 (6).
    \item Replace the Cirlex detector clamps with much lower contamination alternative materials, eliminating this background.
% 2022/09/27 S. Golwala: Forecasts have shown we don't need to improve this
%    \item \red{Reduce the neutron background from the magnetic shield by sourcing new mu-metal material or replacing the shield with Helmholtz bucking coils to cancel Earth's field at the detectors.  It is not yet clear this is necessary --- it may be that the recoil spectrum due to neutrons is sufficiently hard to be subdominant to neutrinos in the energy range of most interest --- but the forecasts have been done without this large neutron background and so further investigation will be needed.}
    \item Reduce \pbtt on the detector faces from 25~n\bqcmsq to 10~n\bqcmsq via 
       \begin{itemize}
       \item fabrication of a single detector (rather than two) at a time;
       \item crystal polishing in a lower radon environment (10~\bqmcunosp) than the shallow underground but non-radon-abated environment currently used\footnote{The crystal polishing slurry provides good protection against radon daughter plateout, hence the choice to not undertake this measure for SuperCDMS SNOLAB.}; 
       \item and, detector packaging and tower assembly in an environment similar to the SuperCDMS low-radon cleanroom ($< 0.1$~\bqmcunosp) at SNOLAB, which would improve upon levels at the Stanford Radon Suppression Facility (RSF;  $\sim$10~\bqmcunosp), where packaging currently occurs. 
    \end{itemize}
    \item Reduce \pbtt on the detector sidewalls from 50~n\bqcmsq to 1~n\bqcmsq via post-fabrication etching.
\end{itemize}
Currently, estimates of naturally occurring \sitt activity are based on~\cite{Aguilar-Arevalo:2015lvd} (\S\ref{sec:Backgrounds}).
While lower levels of \sitt were later observed in a more precise measurement, 11.5$\pm2.4$/kg/day~\cite{damic_32si_2021} vs.\ $80^{110}_{-65}$/kg/day~\cite{Aguilar-Arevalo:2015lvd}, 
and lower-activity silicon is in principle available for rare-event searches~\cite{plaga:1991}, the difference may be source-to-source variation that is not easily controllable~\cite{orrell_si32}, so we retain the higher, conservative \sitt level.  All other backgrounds are also left unchanged relative to~\cite{sensitivity2016} except for updates listed specifically in\S\ref{sec:Backgrounds}.

As discussed in \S\ref{sec:background_upgrades_summary}, we considered an extensive suite of more advanced background scenarios including lower rates of \sitt via sourcing or isotopic separation, reductions in cosmogenic activation of both detector and cryostat materials, and further reductions in materials contamination.  The \bkgth case, in particular, involved isotopic purification of silicon, extended storage of germanium underground to permit decay of cosmogenic germanium isotopes, an all-underground and radon-abated detector life-cycle from before crystal growth onward, and underground electroforming and machining of all copper parts.  As noted above, we found that, with the expected detector performance improvements discussed in \S\ref{sec:alt_detector_sizes} and \S\ref{sec:detector_upgrades}, none of these more advanced backgrounds scenarios provided substantial additional science reach beyond that possible with the \bkgon scenario, while they would incur significant (perhaps prohibitive) new costs.

\subsubsection{Alternative Detector Sizes}
\label{sec:alt_detector_sizes}

Reducing the detector size from the current $\diameter 10$~cm~$\times$~3.3~cm detectors, with volume 260~\cmcu and mass 0.6/1.4~kg (Si/Ge), offers two avenues to improved detector performance.  Doing so reduces the phonon sensor area and thus the total volume of TES film.  The resolution in energy units scales as the square root of the TES volume.  Thus, any detector size reduction improves phonon energy resolution, albeit with a reduction of mass per detector.  Reducing detector size in a manner that preserves aspect ratio can also reduce the detector capacitance and thereby the ionization energy resolution (\S\ref{sec:ionization_resolution_upgrades}).  

We consider three ``canonical'' detector sizes, differing by approximately a factor of 25 in mass/volume: \textbf{SNOLAB-sized} as already indicated, 
``\textbf{10~cm}$^\mathbf{3}$'' ($3\times 3\times 1.2$~\cmcunosp), and ``\textbf{1~cm}$^\mathbf{3}$'' ($1\times 1\times 0.4$~\cmcunosp).  The expected energy resolutions will be described below and in Table~\ref{tbl:detector_upgrade_scenarios}.  The number of channels per detector for these smaller detectors depends on the type of detector.  We describe the channel counts in \S\ref{sec:live_time_channel_counts} and list the detector masses, the total numbers of detectors, and the raw exposures in Table~\ref{tbl:sizes_exposures}.

\subsubsection{Alternative Detector Types}
\label{sec:alt_detector_types}

Two new detector types, modest evolutions of the current types, provide interesting new scientific reach.  We introduced in \S\ref{sec:detector_upgrades_summary} the ``\textbf{phonon iZIP (piZIP)}'' concept, which is a competitive option for reaching the neutrino fog in the 0.5--5~\GeV mass range (\S\ref{sec:nrdmpizip}).  Another concept, ideal for sub-\GeV nucleon-coupled dark matter, is the ``\textbf{0V}'' detector.  At these masses, NR/ER rejection becomes less important because the DM spectral shape becomes much steeper than most backgrounds except \cevns, separation of DM and \cevns does not benefit from NR/ER discrimination, and ionization production for NRs vanishes below tens of eV recoil energy, corresoponding to roughly 1--2~\GeV DM mass.  A simplified detector design, with no ionization electrodes or electric field applied, and only one phonon sensor per 1~\cmcu or 10~\cmcu detector, becomes the means to reach lower masses while maximizing channel count and thus target mass and exposure.

\subsubsection{Detector Performance}
\label{sec:detector_upgrades}

For improvements in detector performance, we consider the impact of changing detector size and we define three scenarios corresponding to varying levels of development maturity --- \detAnosp, \detBnosp, and \detCnosp.  In the case of detector performance improvements, there is not an obvious scaling of cost with scenario.  The development cost of more distant scenarios will obviously be higher, though it is not clear by how much.  The implementation cost of the different scenarios are all roughly the same --- the cost of fabricating new Towers.  Changes in detector size  will also incur a cost in adapting the current Tower design.  Implementing 20$\times$ mass upgrades (\S\ref{sec:live_time_channel_counts}) for small detector sizes will incur substantially larger redesign and fabrication costs.  %\red{We will estimate costs in \S\ref{sec:upgrade_costs}.}

\paragraph{Phonon Energy Resolution}
\label{sec:phonon_resolution_upgrades}

\subparagraph*{Baseline Phonon Energy Resolution}

The baseline phonon energy resolution is the resolution contribution due only to fundamental and electronics noises and is valid at low energy.  It determines energy threshold and thus mass reach.

\begin{description}

\item[Detector Size] As noted above, reducing detector size improves energy resolution.  We re-optimize the phonon sensor design as the detector size changes, so the improvement is not linear in a specific dimensional parameter.  There are also two types of phonon sensor design optimization: driven only by energy resolution (iZIP and 0V), or balancing energy resolution and position information (HV and piZIP), though the differences are small.

\end{description}

\noindent Beyond changes in detector size, we define specific changes/improvement in phonon sensor parameters.  The SuperCDMS phonon sensors~\cite{irwin1,irwin2} consist of overlapping films of  aluminum (Al) and tungsten (W).  Athermal phonons incident on the large (100-$\mu$m-scale) Al films break Cooper pairs, creating quasiparticles (similar to free electrons).  They diffuse about, trapping into the much smaller (1-2 $\mu$m wide) W transition-edge-sensor (TES) films, which transduce the quasiparticle energy into an electrical signal.  The improvements projected for the three scenarios are as follows:

\begin{description}

\item[TES $\boldsymbol{\Tc}$:] We  assume reductions of the TES \Tctxt to 40, 30, and 20~mK for the three scenarios \detAnosp, \detBnosp, and \detCnosp, respectively.  Each yields substantial resolution improvement because, all other quantities held fixed, the energy resolution scales as $\Tc^3$.  The first value has already been achieved on Si but not Ge.\footnote{This low \Tctxt seems to be reliably achieved when the TES film is deposited directly on the crystalline Si substrate.  This process has been implemented for the SuperCDMS SNOLAB Si detectors.  Etch selectivity necessitates that, for Ge substrates, a layer of amorphous silicon be deposited before the W film.   $\Tc = 40$~mK on Ge+a-Si has been demonstrated, but reliability is currently poor.  Detector size reduction alone may improve reliability, as the thermal behavior of kg substrates during deposition differs from that of the inexpensive thin substrates on which the recipes are tuned.  See also the comment that follows about use of Xe sputter gas.}  The latter two values are expected to be feasible because the SuperCDMS W films are a mixture of two crystalline phases of W, $\alpha$ and $\beta$, with \Tctxt values of 15~mK and $\sim$3~K.  In fact, in development work for the TESSERACT experiment~\cite{tesseract}, building on~\cite{bouziane}, the SuperCDMS group at Texas A\&M University has demonstrated \Tctxt~=~19~mK on Si by use of Xe instead of Ar sputter gas.  This result must be shown to be reliable and robust and transferrable to the full detector fabrication process.

\item[W/Al transmission/trapping:] This quantity is the probability per attempt that a quasiparticle created in the Al film and incident on the W/Al film interface passes through the interface and is trapped in the TES, depositing its energy there.  The current value is estimated to be $10^{-4}$: remarkably low, compensated by the large number of attempts each quasiparticle makes on the interface before being lost to recombination with another quasiparticle.  %\marginpar{\red{\footnotesize Can we be offer more of a justification for W/Al transmission improvement?}} 
The \detA scenario assumes no improvement, while the \detB and \detC scenarios assume improvements to $10^{-3.5}$ and $10^{-3}$ respectively.  While there is no specific demonstration in hand of such an improvement, it is expected that fabrication process changes can enhance transmission and design changes can enhance trapping.

\end{description}

\noindent The quantitative impact of these improvements depends on the detector size and design optimization (energy vs.~position).  Table~\ref{tbl:detector_upgrade_scenarios} summarizes the baseline resolutions calculated for each detector size and upgrade scenario.

\subparagraph*{Systematic Phonon Energy Resolution}

At higher energies, the phonon energy resolution becomes limited by the fact that phonon pulse shapes are position-dependent and this position dependence cannot be fully corrected in energy estimation.  This becomes important for HV detector spectral discrimination of NRs from ERs and of ERs from leakage.  For example, at the first electron-hole pair peak for $\vb = 100$~V, a baseline resolution of 1~eV corresponds to fractional energy resolution of 0.01.  (For non-HV detectors, this effect only impacts the sharpness of spectral lines due to cosmogenic activation, which do not play an important role in our sensitivity estimates.)

We incorporate this ``systematic'' contribution to energy resolution by including in quadrature a contribution $\fsig \ept$ where \fsigtxt is the \textit{fractional} energy resolution limit set by this systematic.  Table~\ref{tbl:detector_upgrade_scenarios} lists our assumptions for \fsigtxtnosp.  $\fsig = 0.01$ is regularly achieved currently, so we believe the indicated improvements are optimistic but within reach.

\paragraph{Ionization Energy Resolution}
\label{sec:ionization_resolution_upgrades}

\begin{description}

\item[Detector Size] The ionization readout noise scales as the ratio of two capacitances: the capacitance at the input to the HEMT and the transimpedance amplifier feedback capacitance.  For SuperCDMS SNOLAB, the former is dominated by the $\approx$200~pF detector capacitance.  This quantity scales as $A/d$ where $A$ and $d$ are the detector area and thickness, so the smaller detector sizes provide proportional reductions in $A/d$.  For the reduction in detector size between SNOLAB-sized and 10~\cmcunosp, the improvement is a factor of 3.\footnote{There appears to be no value in considering ionization readout for 1~\cmcu detectors: the vast improvement in energy resolution for this detector size provides new sensitivity at energy depositions for which NRs produce no ionization signal.}

\end{description}

\noindent There are two ways to improve the ionization readout circuit itself:

\begin{description}

\item[Elimination of Feedback Resistor] The ionization readout circuit is a transimpedance integrating amplifier with a parallel capacitor-resistor feedback network.  The resistor resides at 4~K\footnote{This choice was made to enable a circuit design that mitigates microphonic noise due to a nonzero voltage on the gate wire, but which is inconsistent with placing the feedback resistor at a lower temperature.} and thus its Johnson noise is a significant contributor.  This resistor can in principle be replaced with a HEMT switch (infinite ``off-state'' impedance) to discharge the capacitor on a periodic basis, eliminating this Johnson noise.  This concept is fairly mature, with prototype circuits tested with a detector~\cite{hemt_amp_phipps}, but it was not ready in time to be incorporated into SuperCDMS SNOLAB.  Based on a calculation of the Johnson noise contribution, we assume a factor of 3 improvement from this upgrade in the \detB scenario.

\item[Reduction of HEMT $\mathbf{1/f}$ Noise] The ionization resolution is also limited by the $1/f$ noise of the HEMT amplifiers, which counters some of the improvement one would obtain from increasing integration time were the noise purely white.  The noise is believed to be due to two-level systems at the HEMT gate~\cite{jin_hemt2012, jin_hemt2014}.  There is not a well-defined path to improving this noise, but it is reasonable to expect it is modestly amenable to \rndnosp, so we assume a further factor of 3 improvement in the lowest-maturity scenario, \detC (in addition to eliminating the feedback resistor).
\end{description}

\paragraph{Ionization Leakage at High Voltage}
\label{sec:ionization_leakage_upgrades}

The SuperCDMS HV detectors suffer from tunneling of charge carriers from the electrodes, a standard tunneling ionization phenomenon enhanced by the large voltage applied to these detectors~\cite{Agnese_2018}.  The resulting leakage current is a random Poisson process, generating drift of a charge carrier across the full bias voltage and thus appearing at an energy of \evbtxtnosp, very close to the energy of an event in which a single electron-hole pair was created by a particle interaction.  Such events can pile up, yielding two or more leakage events too close in time to distinguish as independent.  The result is an exponentially falling background at \nleaktxtnosp~\evbtxt for \nleaktxtnosp~=~2, 3, ....  Additionally, impact ionization (II) and charge trapping (CT), described below, distribute events from these peaks into the regions in between. 

The ability of HV detectors to distinguish spectrally between ERs and NRs can mitigate the impact of leakage because leakage events have no recoil energy: they appear at precisely \nleaktxtnosp \evbtxt (when not affected by II or CT) while particle interactions appear at 
\nehtxtnosp\evbtxtnosp~+~\ertxtnosp.  ERs thus manifest as spectral peaks separated by \ertxtnosp~=~\nehtxtnosp\epsehtxt from leakage spectral peaks.  With fine enough phonon energy resolution, this separation is visible, recovering sensitivity to DPDM, ALPDM, or LDM events that would otherwise be hidden under the leakage peaks.  The separation is even greater for NRs given the prior discussion of NR/ER spectral discrimination.  Of course, in all cases, II and CT of leakage events cause fill-in of the inter-peak regions and thus degrade sensitivity.  \S\ref{sec:ionization_leakage} discusses how we limit the use of spectral discrimination in these energy regions highly impacted by leakage, II, and CT.

We consider improvements of four types relevant to ionization leakage:

\begin{description}

\item[Ionization leakage] Reducing surface injection would be the most effective means to reduce the impact of ionization leakage.  The use of an insulating blocking layer, such as deposited amorphous \quartznosp, is deemed to have the best chance of success and is being explored.\footnote{Such insulating layers were never considered prior to the advent of HV detectors because they would likely degrade ionization collection at low voltage by creating a barrier to ionization collection.  %There was work in the 1990s to use higher-bandgap materials such as amorphous Si in order to block diffusion of wrong-sign carriers while still permitting correct-sign carriers through.   
It is expected that, with high voltage applied, the potential function will be so tilted that tunneling of correct-sign carriers from the bulk through the barrier into the electrode will be permitted while injection from the electrode back into the bulk will be blocked.}  The lack of a quantitative microphysical model for surface injection renders it difficult to make precise predictions, so we assume one order of magnitude improvement in leakage improvement from each upgrade stage: the current 1.25~mHz/\cmsq for \detA and improvements to 0.125 and 0.0125~mHz/\cmsq for \detB and \detC.  We assume this leakage scales as the area of the detector faces instrumented with electrodes, which is 2~\cmsq for the 1~\cmcu detectors for which the rate has been measured to be 2.5~mHz.

\begin{table}[t!]
\begin{center}
\vspace{-12pt}
%\green{221008 up-to-date}  \\
\includegraphics*[width=\textwidth,page=1,viewport=71 240 541 721]{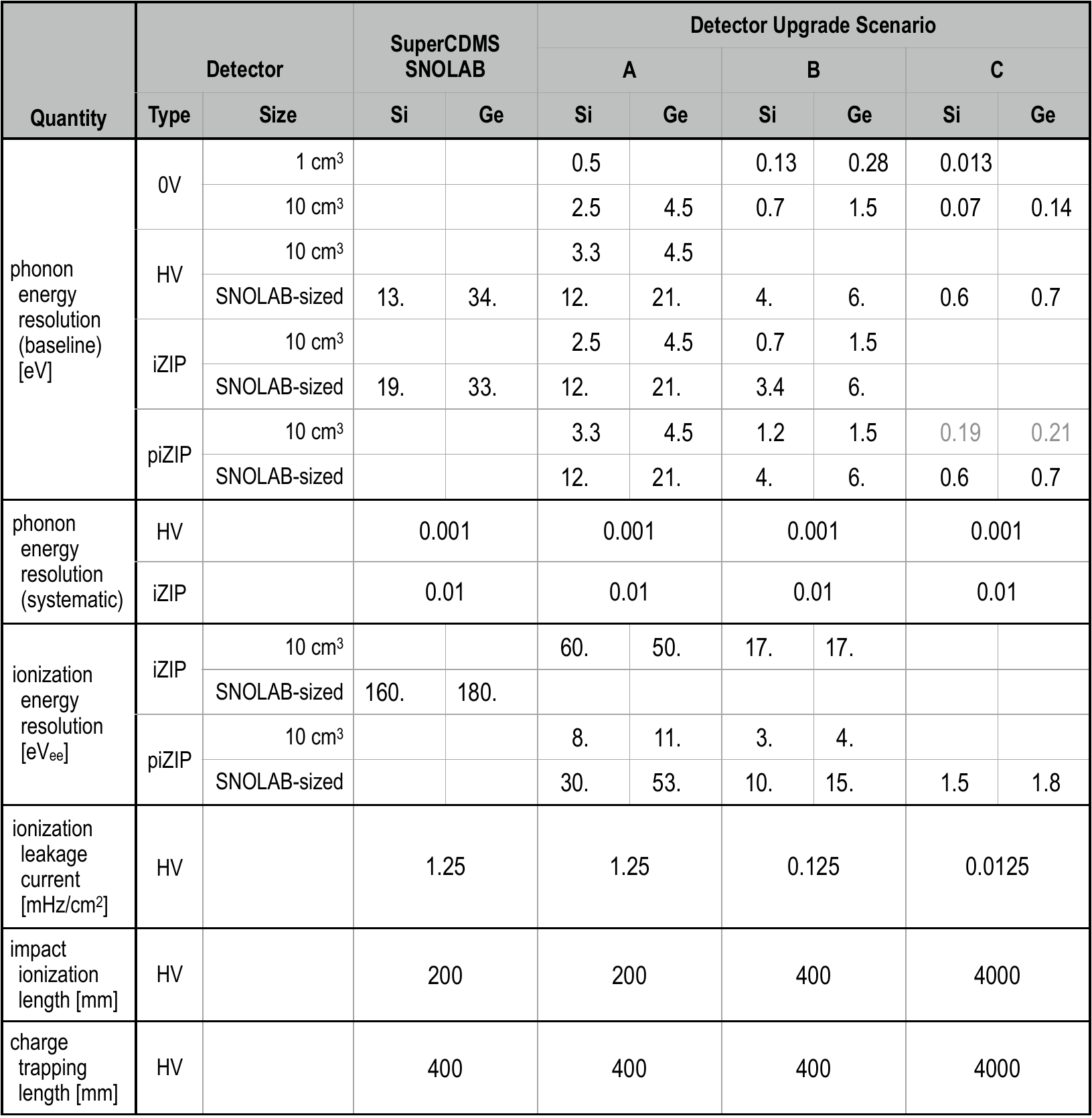}
\end{center}
\vspace{-6pt}
% See https://confluence.slac.stanford.edu/display/CDMS/Long-Term+Planning+Sensitivities#LongTermPlanningSensitivities-ImprovedDetectorPerformanceParameters for discussion of fudge factor applied in addition to nominal Tc and W/Al interface scaling
\caption[Detector upgrade scenarios]{\textbf{Detector upgrade scenarios.}  We provide the quantitative improvements for each detector type and size and for each scenario (\detAnosp, \detBnosp, \detCnosp).  These improvements are described in detail in \S\ref{sec:detector_upgrades}.  The detector sizes are explained in \S\ref{sec:alt_detector_sizes}.  The \detC 10~\cmcu piZIP scenarios are greyed out because the continuous ionization model used for the piZIP becomes invalid for such good effective ionization resolution.    We also list the SuperCDMS SNOLAB values for reference (duplicated from Table~\ref{tbl:sizes_exposures}).  In all cases, we assume a phonon energy threshold of $7\sigpt$ (\S\ref{sec:threshold}).}
\label{tbl:detector_upgrade_scenarios}
\hrule
\vspace{6pt}
\end{table}

\item[Impact ionization (II) and charge trapping (CT)] Si and Ge host donor and acceptor impurities, yielding bound states for, respectively, electrons and holes with $\approx$10~meV binding energies.  Such states can act as trapping sites for drifting charges.  These traps are mitigated by flooding the detectors with free charge (from a radioactive source or a LED) after cooldown and re-flooding them regularly for shorter periods of time during normal operation.  The filled traps, however, become potential sites for ``impact ionization,'' wherein a drifting charge collides with such a site and transfers enough energy to liberate the charge that neutralized the site.  The liberated charge then drifts through only a portion of the detector bias voltage, adding a fraction of \evbtxt in phonon energy and increasing its energy from its nominal value by this fraction of \evbtxtnosp.  Conversely, any unneutralized sites, either present in spite of neutralization or created by prior impact ionization, become potential sites for ``charge trapping,'' whereby a drifting charge becomes bound before traversing the entire \vbtxt voltage drop and thus reducing the NTL energy from its nominal value.  In both cases, ionization leakage events become a background for any DM searches that benefit from the spectral separation of true recoil events from leakage events.  Current values for the probability that a given charge will impact ionize or become trapped are 0.02 and 0.01 in 1~\cmcu detectors.  We convert these probabilities to lengths so they may be applied for any detector thickness, yielding lengths of 200~mm and 400~mm, respectively.  We assume these values for the \detA scenario, assume a modest improvement for \detB so that both lengths are 400~mm, and assume an improvement to 4000~mm for \detCnosp.
%\red{(over the 4 mm HVeV drift?)} 
% right now, the above probabilities are just fixed
% since we only consider SNOLAB-sized HV detectors, there may be an overall scaling error of around
% 10 (from 4 mm HVeV to 33 mm), but at least everything here is internally consistent

\item[Phonon energy resolution] As noted above, both ERs and NRs appear at different energies in the HV detector energy spectrum than do leakage events, modulo impact ionization and charge trapping.  Thus, while the aforementioned improvements in phonon energy resolution will not reduce HV detector threshold, which is in practice limited by ionization leakage, they will have the secondary positive impact of improving spectral separation of leakage, ERs, and NRs, enhancing reach for both electron-coupled and nucleon-coupled DM.

\item[Leakage pile-up] The ability to identify two or more piled-up leakage events as separate events scales with the phonon timing resolution, \sigttxtnosp, which itself scales with the fractional phonon energy resolution, \sigpttxtnosp/\epttxtnosp.  Thus, the aforementioned improvements in phonon energy resolution also improve pile-up rejection of leakage events, improving the leakage background in the $\nleak > 1$ leakage peaks.

\end{description}

\subsubsection{Cryostat Size}
\label{sec:alt_cryostat_size}

We will show that the existing, 7-tower cryostat (\S\ref{sec:SNOBOX}) is capable of enormous scientific reach over multiple generations of detector upgrades for all three detector sizes considered.  When that reach is exhausted, only exposure-limited SNOLAB-sized detector scenarios would benefit from the original 31-tower cryostat with the existing dilution refrigerator, which is feasible if the scientific reach of enhanced exposure for that detector size is compelling.  Other detector sizes considered are small enough that a payload capable of fully exploiting the existing cryostat's \textit{readout-channel capacity} can fit in the existing cryostat's volume.  To extend scientific reach beyond exposure-limited scenarios for these smaller detector sizes would require an \textit{entirely new refrigerator and cryostat} capable of supporting a larger channel count.  See \S\ref{sec:live_time_channel_counts} for details.

\subsection{Experimental Configuration: Live Time and Baseline Detector Counts}
\label{sec:live_time_channel_counts}

For our sensitivity estimates, we make the same live-time assumptions as for SuperCDMS SNOLAB: 4 years of data-taking, an 80\% data-taking duty cycle, and a uniform 95\% efficiency for data-quality cuts.  The exposure will vary by detector mass and channel count.

% don't need this
%\subsubsection{Data Quality Cuts} Based on past experience, we assume that data-quality cuts have an energy-independent efficiency of 0.95.  Since actual pulse timestreams are not generated in our forecasting model, the specific cuts are left unspecified.  Historically, though, they typically include cuts against specific times of known high noise or other disruption, cuts against a detector when its individual behavior is anomalous, and cuts of individual events when pre-pulse conditions such as the trace baseline or noise are anomalous.  

The configuration of SuperCDMS SNOLAB under construction (\S\ref{sec:supercdms_snolab}) has physical space for up to six Towers holding six detectors each, with 12 phonon channels and 4 ionization channels available per detector.  The cryogenic system was designed to have sufficient cooling power for the readout electronics load and thermal conduction of 30 Towers\footnote{31 Towers, actually, but we round for simplicity.}, so we consider the equivalent of \textit{30 Towers of readout channels} (not 30 Towers of SNOLAB-sized detectors\footnote{As noted in \S\ref{sec:alt_cryostat_size}, such an upgrade of the cryostat to the original designed size is feasible but only useful for SNOLAB-sized detector scenarios that are exposure-limited.\label{footnote:cryostat_size}}) to be an upgrade that is feasible with the current cryostat, though it would necessitate a redesign of the Towers to accommodate many more readout channels.

The above heritage leads us to consider the following three options for numbers of detectors and readout channels, where an option with a larger number of detectors is considered when the corresponding option with fewer detectors is seen to be exposure-limited:

\begin{description}

\item[nominal] Two Towers with the equivalent of 144 phonon channels (+ 48 ionization channels for iZIPs), corresponding to 12 SNOLAB-sized iZIP or HV detectors.  (The phonon/ionization channel balance will depend on the upgrade option.)  The intent in considering only two Towers is to provide the flexibility to run multiple detector types in parallel using the six-Tower capacity of the cryostat.

\item[3$\times$ mass] Six Towers with the equivalent of 432 phonon channels (+ 144 ionization channels for iZIPs; same potential for adjusting phonon/ionization channel balance), corresponding to 36 SNOLAB-sized iZIP or HV detectors.  

\item[20$\times$ mass] 2880 \textit{total} channels, corresponding approximately to the wiring equivalent of 30 Towers.   Without a cryostat upgrade (\S\ref{sec:alt_cryostat_size}), this option is only feasible for the alternative detector sizes discussed in \S\ref{sec:alt_detector_sizes}.  We elide the details of detector sizes and channel counts that determine why this is a 20$\times$ rather than 30 towers/2 towers = 15$\times$ increase in exposure.

\end{description}
The numbers of detectors of each type, and their exposures for the nominal scenario, are listed in Table~\ref{tbl:sizes_exposures}.  The exposures for the larger mass scenarios scale proportionally with the payload mass.

\subsection{Sensitivity Forecasts}
\label{sec:forecasts}

Consideration of the full matrix of potential upgrades led us to identify five distinct science goals that these upgrades can address:
\begin{description}
\item[SG-1] Nucleon couplings of sub-\GeVnosp-scale (0.05--0.5~\GeVnosp) dark matter (\S\ref{sec:nrdm_0v})
\item[SG-2] Nucleon couplings of \GeVnosp-scale (0.5--5~\GeVnosp) dark matter down to the neutrino fog (\S\ref{sec:nrdm_neutrino_fog})
\item[SG-3] Electron couplings of kinetically mixed \eVnosp-scale (1--100~\eVnosp) dark photon dark matter (\S\ref{sec:dpdmhv0v})
\item[SG-4] Electron couplings of \eVnosp-scale (1--100~\eVnosp) axion and axion-like particle dark matter (\S\ref{sec:alpdmhv0v})
\item[SG-5] Dark-photon-mediated couplings of \MeVnosp-scale (1--100~\MeVnosp) light dark matter (\S\ref{sec:ldmhv0v})
\end{description}
In this section, we detail our reach for each of these science goals.  Each one relies on a somewhat different detector portfolio.  \textbf{Many of the science goals are amenable to more than one kind of detector, motivating our multi-pronged detector upgrade program, which will provide robustness against challenges in detector development and, potentially, multiple detection channels for a specific dark matter candidate.}  Table~\ref{tbl:science_goal_detector_mapping} shows how the various detector types and sizes map onto the various science goals.

We illustrate our sensitivity for each science goal with a mass vs.~coupling strength plot that shows the median expected 90\%~CL exclusion limit over a set of 100 simulated experimental realizations as described in \S\ref{sec:forecast_procedure}.  Each plot represents only one detector type (0V, HV, iZIP) but includes multiple upgrade scenarios and may include both Ge and Si and multiple detector sizes.

Apropos of the possibility of a DM detection, we also illustrate ``discovery potential'' by showing allowed regions for a representative set of DM models (mass and coupling strength) that could be detected.  Specifically, based on the 90\% exclusion sensitivity curve for a particular scenario, we estimate, for a representative set of masses, the coupling strengths that would be detected at approximately 3$\sigma$.  For each such representative model, we then simulate experimental realizations with signal injected on the basis of that model.  For each realization, we obtain a best-fit mass and coupling strength and 3$\sigma$ ($\Delta \chi^2 = 9$; 99.7\% CL) allowed region.  We plot the injected signal model as well as the allowed region for the realization whose best-fit mass and coupling strength is closest to the injected signal model.  The contour is also a reasonable approximation to the size of the region that the best-fit points of 99.7\% of realizations will occupy.  The contour thus characterizes the constraining power and also the realization-to-realization fluctuations.  While we must limit the number of such discovery potential contours shown for legibility, even this limited set is broadly representative of constraining power.  

Some of these allowed regions reside in the expected SuperCDMS SNOLAB 90\%~CL exclusion portions of parameter space.  This overlap indicates that hints of these signals could be seen in SuperCDMS SNOLAB and that these future upgrades would increase the significance of such hints to $3\sigma$.

We complement these exclusion sensitivity and discovery potential plots with energy spectra in phonon energy displaying the individual backgrounds --- bulk and surface electron and nuclear recoils due to radiocontamination, bulk nuclear recoils due to solar neutrino \cevns\ --- and DM spectra for a representative set of masses, normalized in coupling strength to match the 90\%~CL exclusion sensitivity.  These plots provide intuition for how the DM and background spectra determine the exclusion sensitivity curves.  To provide the greatest amount of information, we make the plot for the relevant detector scenario with the combination of best energy resolution and highest background levels.  One can reasonably extrapolate from the given plot to coarser resolution, higher threshold, and lower background levels.

\begin{table}[t!]
%\hrule
\vspace{-9pt}
\begin{center}
{\small
\begin{tabular}{|rr|c|cc|c|cc|}
\hline
%\multicolumn{2}{|c|}{} &            & \multicolumn{2}{c|}{} &                & \multicolumn{2}{c|}{} \\
\multicolumn{2}{|c|}{detector} &            & \multicolumn{2}{c|}{mass [gm]} & number of      & \multicolumn{2}{c|}{raw exposure [kg$\cdot$yr]} \\
\multicolumn{1}{|c}{size} & \multicolumn{1}{c|}{type} & dimensions & Ge & Si                  & detectors & Ge & Si       \\ \hline\hline
SNOLAB- & HV/iZIP & $\diameter 10$~cm~$\times$~3.3~cm & 1400    & 610   &  12 & 54  & 23 \\  
sized   & piZIP &                                     &        &        &   6 & 27  & 12 \\ \hline
10~\cmcu & HV    & $3\times 3\times 1.2$~\cmcu       &    57  &    25  &  36 & 6.6 & 2.9 \\
         & iZIP  &                                   &        &        &  24 & 4.4 & 1.9 \\
         & piZIP &                                   &        &        &  12 & 2.2 & 1.0 \\
         & 0V    &                                   &        &        & 144 & 26  & 12  \\ \hline
1~\cmcu & 0V     & $1 \times 1 \times 0.4$~\cmcu     &    2.1 &   0.9  & 144 & 1.0 & 0.42 \\ \hline
\end{tabular}
}
\end{center}
\caption[Masses and exposures for nominal exposure option]{\textbf{Masses and exposures for nominal exposure option (two Towers).}  Different detector types of the same size can have different channel counts, resulting in differing numbers of detectors that can be accommodated by two Towers.  As for SuperCDMS SNOLAB, we assume 4 years of data-taking and 80\% duty cycle to obtain the raw exposure.}
\hrule
\vspace{6pt}
\label{tbl:sizes_exposures}
\end{table}

\subsubsection{SG-1: Sub-GeV/$\mathbf{c^2}$ Nucleon Couplings: ``1 \cmcunosp'' and ``10 \cmcunosp'' Detectors}
\label{sec:nrdm_0v}

\textit{It is extremely promising to consider smaller detectors with vastly improved phonon energy resolution, operating at 0V without event-by-event discrimination of DM from backgrounds, because they offer access to nucleon-coupled dark matter at masses for which the recoil energy spectrum is so steep that it can be spectrally distinguished from conventional particle and even \cevns backgrounds.}

% this is a funny statement since it compares 0V to HV resolution, but it neglects the fact that
% HV resolution get divided by the NTL gain and then the threshold suffers from leakage. 
The \detA scenario for 1~\cmcu 0V detectors already provides 1.5 orders of magnitude improvement in phonon energy resolution over the SuperCDMS SNOLAB HV detectors via detector size reduction, and the \detB and \detC scenarios each provide an additional half to one order of magnitude.  Mass reach for nucleon couplings scales approximately as the square root of recoil energy threshold.

\begin{table}[t!]
\begin{center}
%\green{220928 up to date}\\
\includegraphics*[width=\textwidth,page=3,viewport=70 500 541 721]{tables/LTPSnowmassTable221001.pdf}
\end{center}
\vspace{-9pt}
\caption[Relevance of various detector types, sizes, and upgrade scenarios to science goals]{\textbf{Relevance of various detector types, sizes, and upgrade scenarios to science goals}. The numbers indicate the science goals.  If a science goal is boldfaced and underlined, then a ``sharp target'' benchmark model or models can be tested.  The color shading, starting from light green and going to green-yellow-red, indicates finer gradings in maturity.  For example, the 1~\cmcunosp~Si 0V \detA option is light green because such detectors have been made and the necessary resolution nearly achieved, while the 10~\cmcunosp~Ge 0V Ge piZIP option is shaded with more yellow-red because neither a piZIP nor a 10~\cmcu detector has been fabricated.}
\hrule
\vspace{6pt}
\label{tbl:science_goal_detector_mapping}
\end{table}

\begin{figure}[t!]
\begin{center}
\textbf{SG-1: Sub-\GeV Nucleon-Coupled DM with 0V } \\
\vspace{6pt}
%\green{221003 up to date}\\
%\green{TS: 0927 discovery regions updated}\\
%\red{Harrison: change horizonal axis to eV - Done} \\
%
%\red{\textbf{Sub-GeV NRDM with 0V Sensitivity plot showing:}} \\
%{\footnotesize
%\begin{tabular}{l} \\
%\textbf{Exclusion Limits:} \\
%\bkgon \detA 0V 1~\cmcu Si  \\
%\bkgon \detB 0V 1~\cmcu Si  \\
%\bkgon \detC 0V 1~\cmcu Si  \\
%\bkgon \detC 0V 1~\cmcu Si 20$\times$ \\
%\bkgon \detA 0V 10~\cmcu Si \\
%\bkgon \detB 0V 10~\cmcu Si \\
%\bkgon \detC 0V 10~\cmcu Si \\
%\bkgon \detC 0V 10~\cmcu Si 20$\times$ \\
%\bkgon \detA 0V 10~\cmcu Ge \\
%\bkgon \detB 0V 10~\cmcu Ge \\
%\bkgon \detC 0V 10~\cmcu Ge \\
%\bkgon \detC 0V 10~\cmcu Ge 20$\times$ mass\\
%\end{tabular}} 
%\red{In process plots.  Need to change scale to be 0.05-10 GeVnosp.} \\
%\includegraphics[width=0.33\textwidth]{limit_plots/nrdm_0v/0V1cm3_Si_plrlimit.png}
%\includegraphics[width=0.33\textwidth]{limit_plots/nrdm_0v/0V10cm3_Si_plrlimit.png} \\
%\includegraphics[width=0.33\textwidth]{limit_plots/nrdm_0v/0V10cm3_Ge_plrlimit.png} 
%\includegraphics[width=0.33\textwidth]{limit_plots/nrdm_0v/0V1cm3_Si_plrlimit3x.png} \\
\includegraphics*[width=0.48\textwidth,viewport=0 0 576 410]{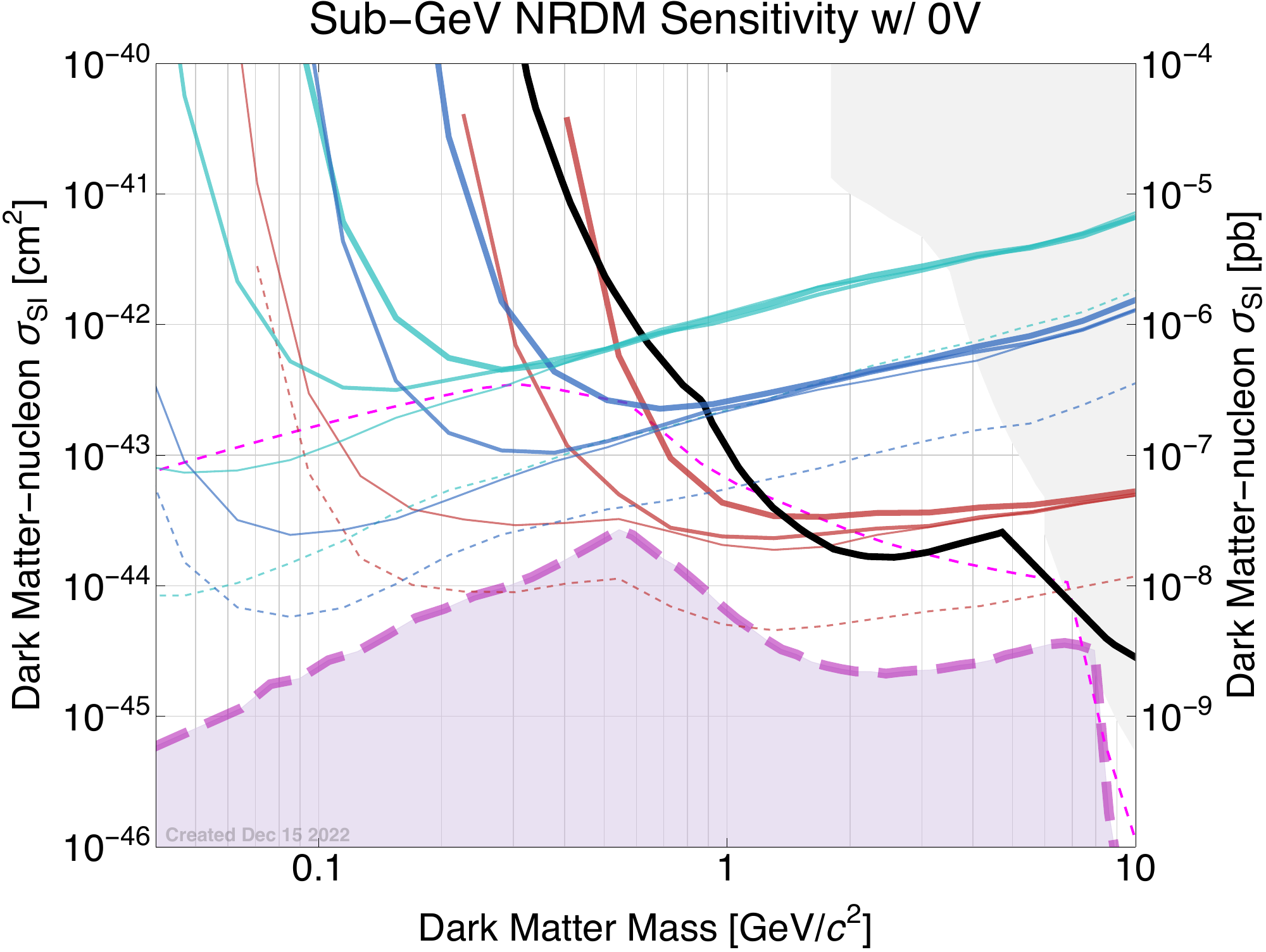}
\hfill
\includegraphics*[width=0.48\textwidth,viewport=0 0 576 410]{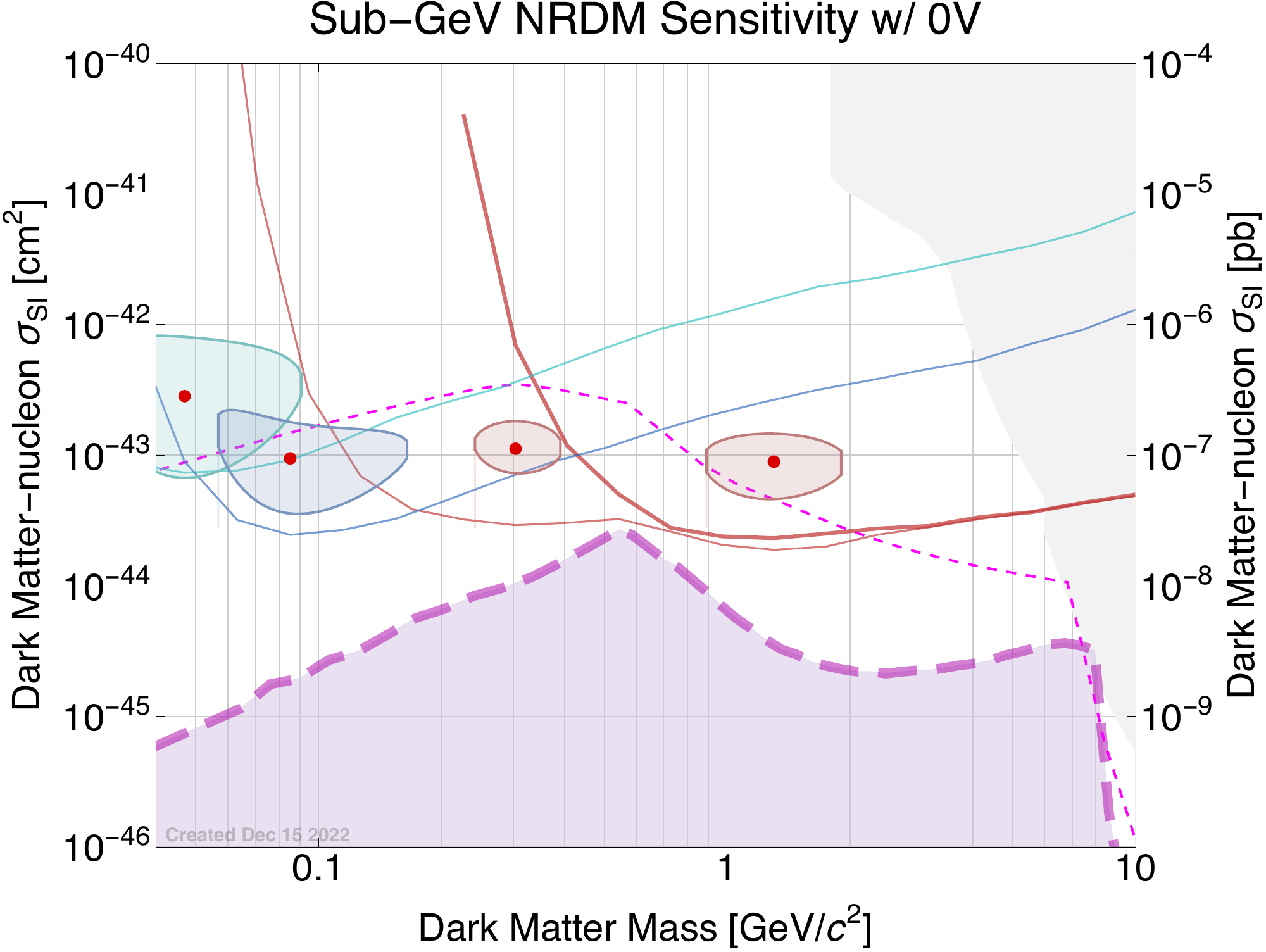}
% \includegraphics*[width=0.46\textwidth,viewport=0 0 808 618]{limit_plots/nrdm_0v/NRDM_0V1cm3_0V10cm3_GeSi_discovery.png}
%\parbox[b]{3in}{\red{\footnotesize
%\begin{tabular}{p{3in}} \\
%\textbf{Discovery Potential Plot:} \\
%For the following cases, pick the mass at which the exclusion limit sensitivity is greatest, multiply the cross section at the exclusion sensitivity limit by 2.57 (90\%~CL to 99.7\% CL exclusion limit rate factor), and use that as the mass/cross section for an entry in a separate discovery potential plot that also shows the exclusion regions.  \\
%\bkgon \detC 0V 1~\cmcu Si  \\
%\bkgon \detC 0V 10~\cmcu Si  \\
%\bkgon \detA 0V 10~\cmcu Ge \\
%\bkgon \detB 0V 10~\cmcu Ge \\
%\bkgon \detC 0V 10~\cmcu Ge \\
%\end{tabular}} \\}  
\\
\medskip
\includegraphics*[width=0.46\textwidth,viewport=0 0 420 322]{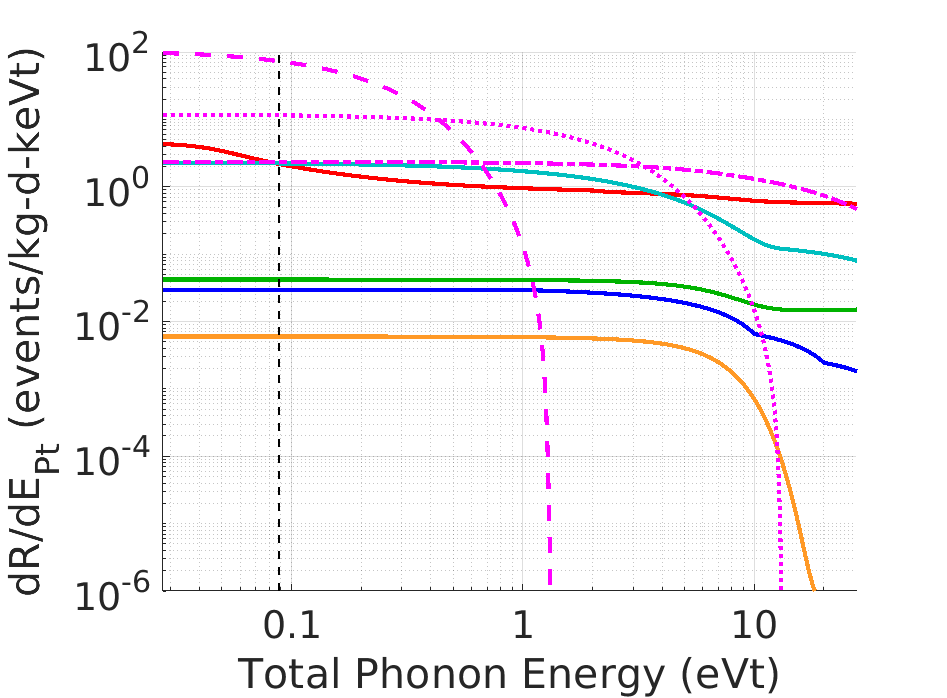}
\\
\end{center}
\caption[Sensitivity to sub-\GeV nucleon-coupled DM with 1~\cmcu and 10~\cmcu 0V detectors]{\textbf{Sensitivity to sub-\GeV nucleon-coupled DM with 1~\cmcu and 10~\cmcu 0V detectors.} (Top left) Median expected PLR-based 90\%~CL exclusion sensitivity for 1~\cmcunosp~Si,  10~\cmcunosp~Si, and 10~\cmcunosp~Ge 0V detectors for the three phonon-resolution improvements (\detAnosp, \detBnosp, and \detCnosp) and ready-to-implement improvements in backgrounds (\bkgon scenario) along with the sensitivity of SuperCDMS SNOLAB (envelope of all detector types, thick black).  Colors (cyan, blue, red-brown) correspond to detector type (1~\cmcunosp~Si, 10~\cmcunosp~Si, 10~\cmcunosp~Ge) while decreasing line thickness corresponds to more advanced phonon-resolution improvement scenario.   For each detector type and for the \detC improvement scenario, forecasts are also shown for the 20$\times$ mass payload (dotted).  The magenta short- and long-dashed lines indicate the ``single-neutrino'' sensitivity and the neutrino fog for Si~\cite{Billard:2013qya}.  The corresponding lines for Ge would be at higher cross section.  The grey region shows parameter space currently excluded at 90\%~CL.  The sensitivity curves for a given detector type and payload mass asymptote to a single line at high mass because they all become exposure-limited in the same way while showing progressively greater reach at low mass.  The improvement from increasing mass by 20$\times$ becomes progressively smaller in going from Si~1~\cmcu to Si~10~\cmcu to Ge~10~\cmcunosp, with the first case being almost exposure-limited while the last case is limited by statistical uncertainty on background subtraction via PLR.  Sensitivity can extend beyond the neutrino fog line because \cite{Billard:2013qya} assumes a specific set of detector parameters, differing from those used here.  
(Top right)~Discovery potential for the \detCnosp~1~\cmcunosp~Si, \detCnosp~10~\cmcunosp~Si, and \detBnosp/\detCnosp~10~\cmcunosp~Ge cases, along with their corresponding exclusion limits.  For each case, the red point indicates the model for which DM signal is injected and the contour the 99.7\% CL (3$\sigma$) allowed region as described in the introductory part of \S\ref{sec:forecasts}.
(Bottom)~Example spectrum plot showing the background spectra for the 1~\cmcu Si 0V \bkgonnosp, \detC case and candidate nucleon-coupled DM signal spectra \underline{after} all analysis cuts described in \S\ref{sec:forecast_procedure_detail}.  Legend as in Figure~\ref{fig:snolab_reach_nrdm}.  The candidate DM signals are for $\mathit{\mdm \approx}$~0.16, 0.5, and 1.6~\GeVnosp.  The two bumps in the \cevns spectrum correspond, in order of increasing energy, to $pp$ and \beseven neutrinos (the $pep$, CNO, and \beight neutrinos are not visible in this phonon energy range).  The corresponding 10~\cmcu Si and Ge cases are similar but with poorer phonon resolution, and the corresponding 10~\cmcu Ge case additionally has a lower bulk ER background and a neutrino spectrum reduced by about 2.5$\times$ in recoil energy.}
\label{fig:nrdm0v_reach}
\hrule
\end{figure}

This enhanced mass reach requires \textbf{no particle background improvements beyond \bkgonnosp}.  All the ER backgrounds are fairly flat at these low energies (even coherent photonuclear scattering), while the DM recoil energy spectrum is steeply rising, both resulting in small numbers of ER background events in the DM region of interest (ROI) and making the two relatively easily distinguished by PLR.  The dominant background ends up being \cevns of $pp$ solar neutrinos with nuclei.  This background is not amenable to the yield-based or spectral discrimination power of iZIP and HV detectors, respectively, so the simple 0V phonon-sensor-only detector is the optimal approach.  We display the sensitivity of two Towers of 1~\cmcu Si 0V detectors for the various phonon-resolution improvements in Figure~\ref{fig:nrdm0v_reach}. 

Because of this exposure-limited behavior, it is sensible to consider alternatives.  One option is 10~\cmcu 0V detectors, which offer more exposure per readout channel in exchange for less mass reach.  Another option is germanium.  Due to differing energy resolutions, Si mass reach is about half an order of magnitude greater than for Ge, while the point of maximum cross section sensitivity is 2--3$\times$ lower for Ge.  With all three types of detectors, one can obtain excellent complementarity in coverage of parameter space.  Figure~\ref{fig:nrdm0v_reach} illustrates this complementarity by displaying, for the \bkgon scenario, the reach of two Towers each of 1~\cmcu Si, 10~\cmcu Si, and 10~\cmcu Ge 0V detectors for all three detector upgrade scenarios.  It is clear that Si 1~\cmcu gives the greatest immediate gain in mass reach and should be the first option pursued, but, in the longer-term \detB and \detC scenarios, the 10~\cmcu detector offers the best balance between mass and cross-section reach by following the neutrino fog contour.  In the \detC scenario, the optimal combination appears to be two Towers of each of these types of detectors (six Towers total), combining mass reach to below 0.05~\GeV with neutrino fog sensitivity above 0.2~\GeVnosp.

All three of these detector types can benefit from more exposure, again because the background spectra are so flat, so we show the potential of 20$\times$ mass cases for all three.  (These could not all be installed simultaneously.)  The benefit of the greater mass increases in going from Ge~10~\cmcu to Si~10~\cmcu to Si~1~\cmcunosp.  The optimal combination may be a mix of smaller mass increases for each of the three so they can all be operated at the same time.

We should discuss the validity of the standard assumption of interaction of DM with individual nuclei for the most advanced scenarios.  This assumption begins to fail for the Si 0V~1~\cmcunosp~\detC scenario.  A $7\,\sigpt^\text{Si} = 0.1$~eV threshold for energy deposition is solidly in the multi-phonon $\ephonon < \er < 10\,\ephonon$ regime.  Figure~3 of \cite{griffin2020} illustrates the end of the standard nuclear recoil regime with its $(\hbar) \omega_{min} = 0.5$~eV curves.  Their threshold corresponds to approximately $10\ephonon$ for Si, so their projection of a response cutoff in the 30--60~\MeV regime motivates the 50~\MeV cutoff for our plots.  In Ge, a $7\,\sigpt^\text{Ge} = 0.5$~eV threshold is about $15\,\ephonon$.  (The standard nuclear recoil regime extends a bit lower in Ge because the phonon energies are lower than in Si).  We therefore believe our Ge projections are reasonably accurate.   Understanding the full potential of the Si 0V~1~\cmcunosp~\detC detectors below 50~\MeV will await the DM-multi-phonon calculations.

\subsubsection{SG-2: Reaching the Neutrino Fog, 0.5--5 GeV/$\mathbf{c^2}$}
\label{sec:nrdm_neutrino_fog}

\textit{We have three different potential paths to reach the ``neutrino fog'' in the 0.5--5 \GeV energy range, where it is determined by \besevennosp, CNO, and \beight solar neutrinos.  This is an important target for the DM community, along with the neutrino fog at higher masses due to atmospheric and diffuse supernova neutrinos.}

All three paths take the approach of \textit{rejecting} the dominant bulk and surface electron recoil backgrounds rather than trying to subtract them.  Reaching the neutrino fog via background subtraction alone would require the enormous effort and cost of the \bkgth scenario.

\paragraph{SG-2A: Reaching the Neutrino Fog with HV Detectors}
\label{sec:nrdmhv}

As has been noted above, improving the energy resolution of HV detectors makes it possible to spectrally separate ER backgrounds from nuclear recoils (both signal and backgrounds).  This regime will not quite be reached by SuperCDMS SNOLAB, so operating in this regime in the future provides significant new sensitivity.  In parallel, reducing ionization leakage reduces the energy down to which this effect can be used (assuming the conservative approach to spectral discrimination in the presence of leakage as discussed in \S\ref{sec:ionization_leakage_upgrades} and \S\ref{sec:ionization_leakage}).  These two improvements work together to make it possible to reach the neutrino fog over most of the 0.5--5~\GeV region with HV detectors.  In particular, two Towers of Ge SNOLAB-sized HV detectors can reach this goal under the \detB scenario with only the \bkgon scenario improvements in backgrounds.  Six Towers of Si SNOLAB-sized HV detectors are sensitive to lower mass than Ge for \detBnosp, albeit with poorer cross-section reach above 1~\GeVnosp.  The Ge and Si \detC scenarios are not shown because they offers no additional reach in the 0.5--5~\GeV mass range, though \detC does improve sensitivity above 5~\GeV to approach the 5--10~\GeV neutrino fog because the improved systematic energy resolution contribution (Table~\ref{tbl:detector_upgrade_scenarios}) maintains spectral NR-ER discrimination to higher phonon energy and thus DM mass.  While the \detA scenario for SNOLAB-sized HV detectors is an insufficient improvement over SuperCDMS SNOLAB to provide interesting reach, a 20$\times$ mass payload of 10~\cmcu Ge HV detectors with only the \detA improvements does also reach the neutrino fog.  (Note that 1~\cmcu HV detectors would provide the necessary detector performance in the \detA scenario but would be exposure-limited at even 20$\times$ payload mass.)  \textbf{None of these scenarios require background improvements beyond the \bkgon scenario,} though the \bkgth scenario would improve Ge HV sensitivity above 5~\GeV by reducing ER backgrounds.  

\begin{figure}[t!]
\begin{center}
\textbf{SG-2A: \GeV Nucleon-Coupled DM with HV} \\
\includegraphics*[width=0.48\textwidth,viewport=0 0 576 410]{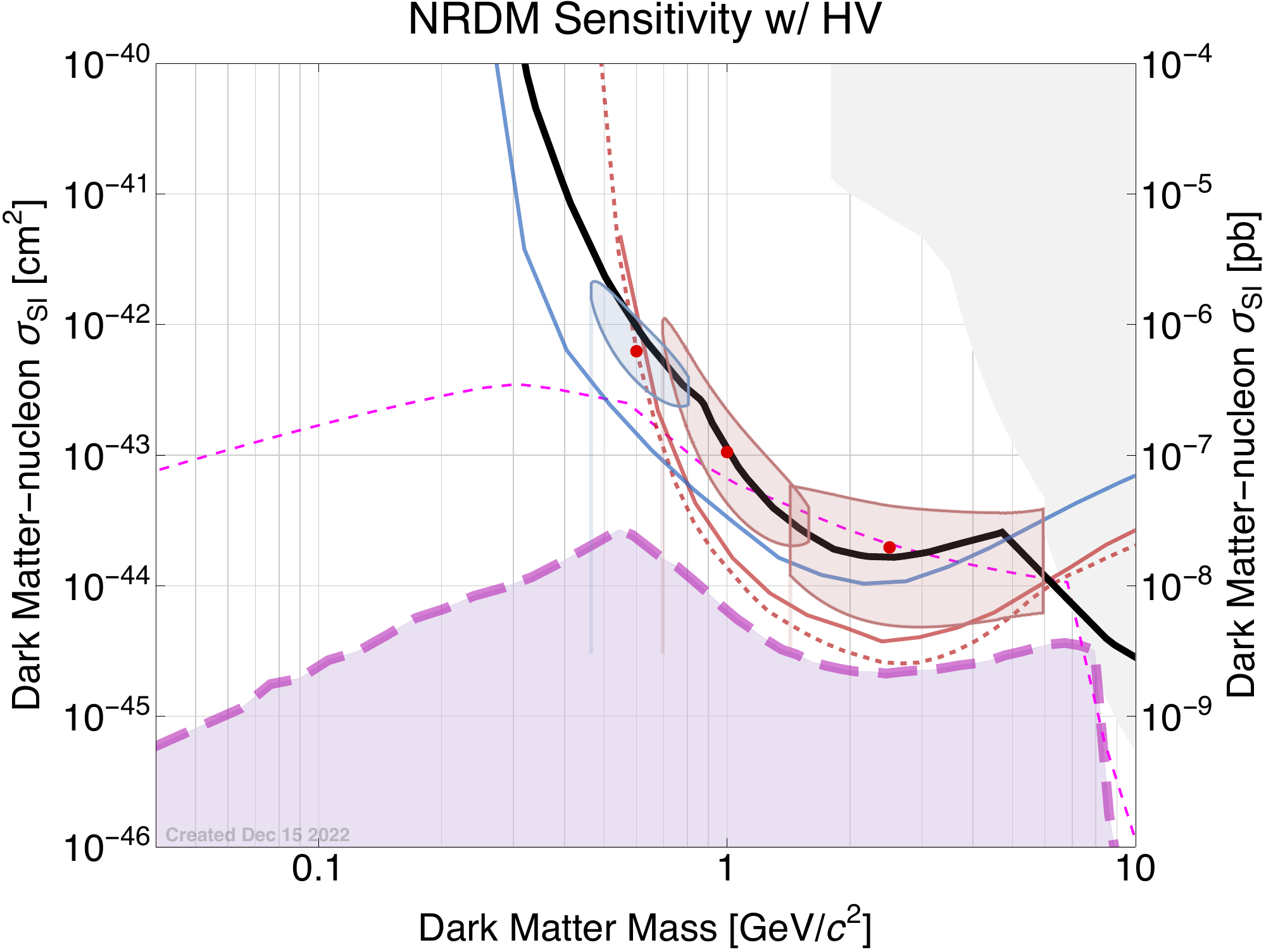} %\\
\hfill
% Mar 2022
%\includegraphics*[width=0.46\textwidth,viewport=0 0 420 322]{spectra/HV10cm3_bkgA_detA_cons_Si_postCuts_linear_v4.png} \\
% 220930 
% changes: 
% lower ionization leakage cutoff due to fix to leakage timing pileup
% fix to implementation of IMPACT NR yield function, which makes DM signal shallower (pushed up to higher yield) and makes surface NR signal steeper (easier to reject surface NRs because they also get higher yield)
\includegraphics*[width=0.46\textwidth,viewport=0 0 420 322]{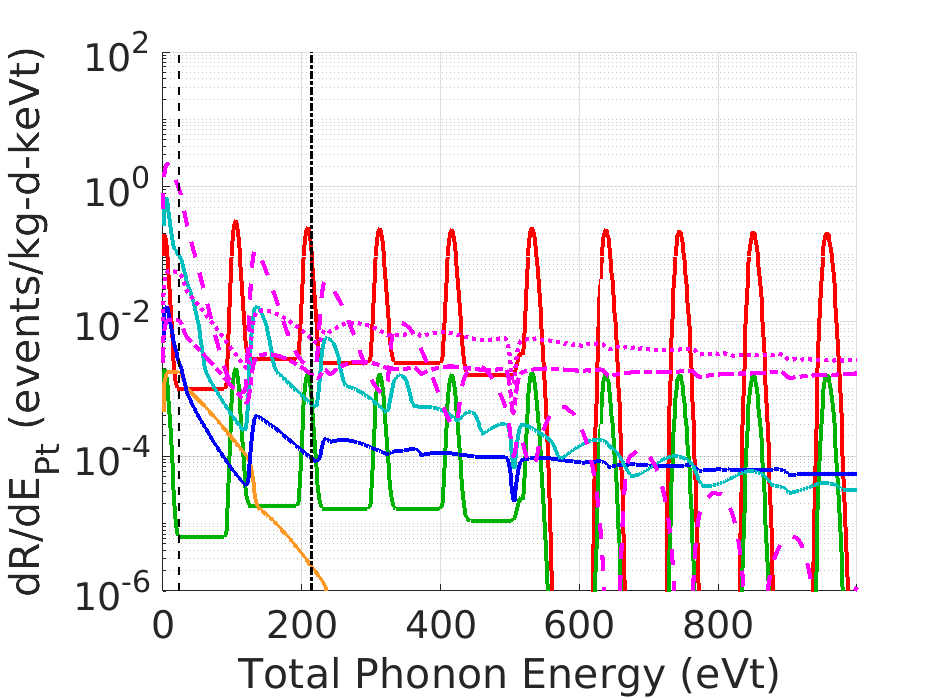} \\
\end{center}
\caption[Sensitivity to \GeV nucleon-coupled DM with SNOLAB-sized HV detectors]{\textbf{Sensitivity to \GeV nucleon-coupled DM with SNOLAB-sized HV detectors.} (Left)~Median expected PLR-based 90\%~CL exclusion sensitivity, assuming ready-to-implement improvements in backgrounds (\bkgon scenario), for SNOLAB-sized Ge (two Towers) and Si (six Towers) HV detectors for the \detB improvement scenario (solid, red-brown and blue) and 20$\times$ mass payload of 10~\cmcu Ge HV detectors for the \detA improvement scenario (dashed, red-brown) along with the sensitivity of SuperCDMS SNOLAB (envelope of all detector types, thick black).  The single-neutrino and neutrino fog sensitivity and grey shaded region are as in Figure~\ref{fig:nrdm0v_reach}.  %The 10~\cmcu HV detectors achieve the energy resolution and ionization leakage needed to reach the neutrino fog in the more mature \detA improvement scenario, but the much larger number of detectors required would necessitate a significant Tower redesign.  SNOLAB-sized HV detectors achieve similar sensitivity with fewer individual detectors but require the more advanced \detB detector improvements.
In both cases, Si detectors of the same types provide additional reach to lower mass than Ge though poorer cross section reach at masses for which Ge is sensitive.
We also show discovery potential for the \detBnosp~SNOLAB-sized Ge and Si cases for candidate DM masses, with legend as in Figure~\ref{fig:nrdm0v_reach}.
(Right)~Example spectrum plot showing the background spectra for the 10~\cmcu Si HV \bkgonnosp, \detA case and three candidate \GeV nucleon-coupled DM signal spectra \underline{after} all analysis cuts described in \S\ref{sec:forecast_procedure_detail}.  The SNOLAB-sized Si HV \detB case is similar because the resolutions are similar, and the corresponding Ge cases are also similar but with lower bulk ER background.  Legend as in Figure~\ref{fig:snolab_reach_nrdm} with the addition of the threshold imposed by ionization leakage (\S\ref{sec:threshold}; black dashed-dotted vertical line).  The convolution of the solar neutrino interaction spectrum with the HV response function makes interpretation in terms of the individual reaction chains less clear than for the other detector types.  The candidate DM signals are for $\mathit{\mdm \approx}$~0.5, 1.6, and 5~\GeVnosp.}
\label{fig:nrdmhv_reach}
\hrule
\end{figure}

\paragraph{SG-2B: Reaching the Neutrino Fog with iZIP Detectors}
\label{sec:nrdmizip}

\begin{figure}[t!]
\begin{center}
\textbf{SG-2B: \GeV Nucleon-Coupled DM with iZIP} \\
\includegraphics*[width=0.48\textwidth,viewport=0 0 576 410]{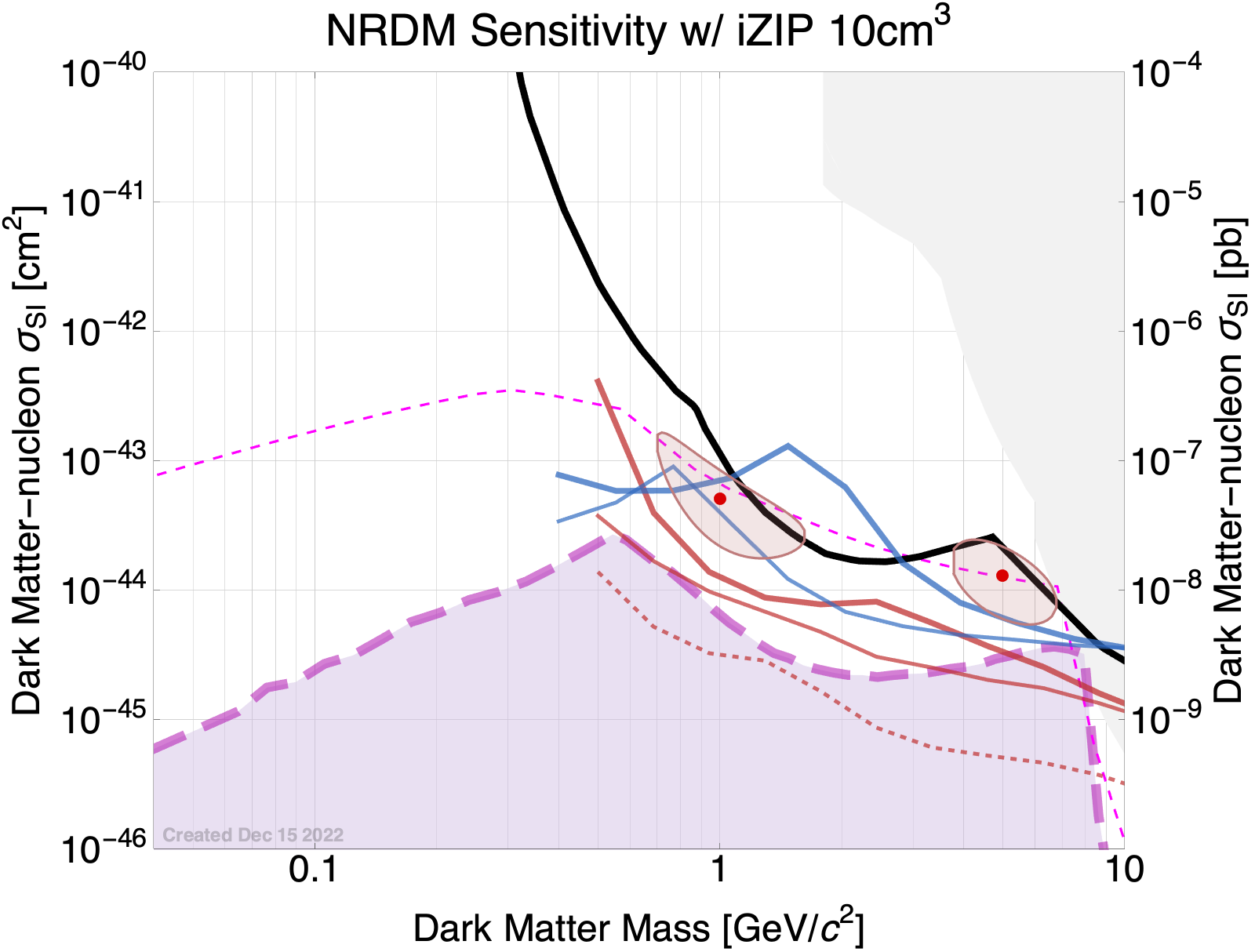}
\hfill
%\parbox[b]{3in}{\red{\footnotesize
%\begin{tabular}{p{3in}} \\
%\textbf{Tarek:} I think color code for Si and Ge is the opposite as in the 0V and %piZIP cases.  Could you correct?  If not, could you correct the caption below? 
%\textcolor{blue}{TS: Fixed the caption}\\
%  \\
% \textbf{Discovery Potential Plot:} \\
% Three regions at 1, 2, and 4 GeV for \bkgon \detA iZIP 10~\cmcu Ge \\\vspace{60pt} \\
%\end{tabular}} \\} 
\medskip
% Mar 2022
%\includegraphics*[width=0.46\textwidth,viewport=0 0 420 322]{spectra/iZIP10cm3_bkgA_detB_Si_postCuts_ER_v4.png}\\
% 220930
% changes: DM spectrum harder because of fix to implementation of IMPACT yield function
\includegraphics*[width=0.46\textwidth,viewport=0 0 420 322]{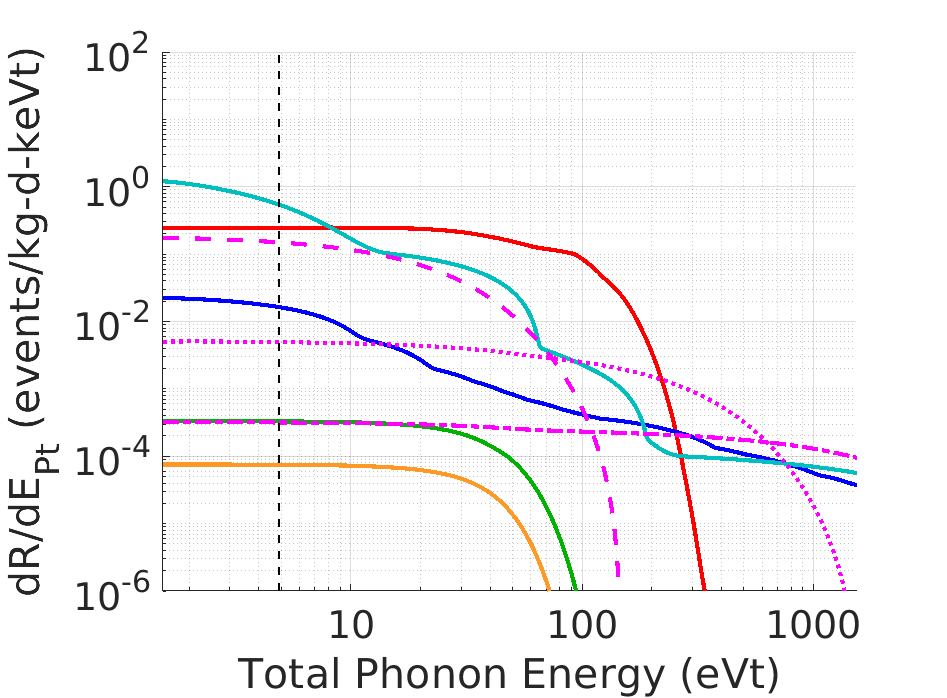}\\
\end{center}
\caption[Sensitivity to \GeV nucleon-coupled DM with 10~\cmcunosp~iZIP detectors]{\textbf{Sensitivity to \GeV nucleon-coupled DM with 10~\cmcunosp~iZIP detectors.} (Left)~Median expected PLR-based 90\%~CL exclusion sensitivity, assuming ready-to-implement improvements in backgrounds (\bkgon scenario), for two Towers of 10~\cmcunosp~Ge (red-brown) and six Towers of 10~\cmcunosp~Si iZIP (blue) detectors for the \detA (thick) and \detB (thin) improvement scenarios along with the sensitivity of SuperCDMS SNOLAB (envelope of all detector types, thick black).   We also include a 20$\times$ exposure case for the Ge iZIP \detB scenario to illustrate the ultimate reach of the existing SuperCDMS SNOLAB cryostat with iZIPs.  The single-neutrino and neutrino fog sensitivity and grey shaded region are as in Figure~\ref{fig:nrdm0v_reach}.  The SNOLAB-sized Ge and Si iZIP \detB forecasted sensitivities are approximately the same as the corresponding 10~\cmcu iZIP \detA sensitivities.  We show six-Tower forecasts for Si detectors because two-Tower forecasts are exposure-limited.  We also show discovery potential for the \detAnosp~10~\cmcunosp~Ge case for candidate DM masses, with legend as in Figure~\ref{fig:nrdm0v_reach}. 
(Right)~Example spectrum plot showing the background spectra for the 10~\cmcu Si iZIP \bkgonnosp, \detB case and candidate nucleon-coupled DM signal spectra \underline{after} all analysis cuts described in \S\ref{sec:forecast_procedure_detail}.  Legend as in Figure~\ref{fig:snolab_reach_nrdm}. The candidate DM signals are for $\mathit{\mdm \approx}$~0.5, 1.6, and 5~\GeVnosp.  The four bumps in the \cevns spectrum correspond, in order of increasing energy, to $pp$, \besevennosp, $pep$ and CNO, and \beight neutrinos.  The corresponding Ge cases are similar but with poorer phonon resolution, lower bulk ER background, significant cosmogenic spectral lines, and a neutrino spectrum reduced by about 2.5$\times$ in recoil energy.  The bulk ER spectrum rolls off as ionization-yield-based ER rejection becomes effective, while surface ER and NR curves roll off due to fiducial volume cuts becoming effective.  The 10~\cmcu Ge iZIP \bkgonnosp, \detB case is more sensitive because the ER background is lower (no \sittnosp). }
\label{fig:nrdmizip_reach}
\hrule
\end{figure}

The SuperCDMS SNOLAB iZIP detectors lose sensitivity below 5~\GeV because the ionization measurement signal-to-noise becomes too small to perform event-by-event NR discrimination.  Improving the ionization resolution reduces the threshold for NR discrimination and thus extends such discrimination to lower DM mass.  \textbf{No background improvements beyond the \bkgon scenario are necessary} because the dominant backgrounds are bulk ERs due to Compton backgrounds and cosmogenics, all of which are amenable to NR discrimination.  We use two of the three paths described in \S\ref{sec:ionization_resolution_upgrades} for improving ionization resolution in preparing forecasts, shown in Figure~\ref{fig:nrdmizip_reach}.

The first and most straightforward option is simply to reduce the detector size and capacitance without significant advances in detector performance otherwise (\detAnosp).  We find that this option, two Towers of 10~\cmcu Ge iZIPs with the \bkgonnosp, \detA upgrade scenario, provides reach between  single-neutrino sensitivity and the neutrino fog.  This already yields appreciable progress over SuperCDMS SNOLAB.  For Si, similar reach requires both the \detB scenario and six Towers of such detectors.  The maturity of reduced TES \Tctxt on Si is greater than on Ge (\S\ref{sec:phonon_resolution_upgrades}), so this option may not be as mismatched as it first seems.

To fully reach the neutrino fog requires further improvement in ionization resolution: the upward bumps in sensitivity seen at 2.5~\GeV for Ge and 1.5~\GeV for Si are due to the similarity between the spectra for these DM masses and the shape of Compton and cosmogenic bulk ER backgrounds as modulated by the rolloff of NR rejection in the \detA scenario.  The \detB detector upgrade scenario provides the necessary improvement for Ge, resulting in sensitivity reaching the neutrino fog for the same two-Tower Ge 10~\cmcu payload down to 0.5~\GeVnosp.  (Achieving comparable sensitivity with Si requires the \detC scenario, for which the required ionization resolution improvement is far less certain (\S\ref{sec:ionization_resolution_upgrades})).  We note that the flattening of the sensitivity curves below 1~\GeVnosp, in contrast to the fast rise at low mass seen for less advanced \detX scenarios, occurs because the phonon energy threshold is sufficiently low that the iZIPs mimic 0V detectors, discriminating NRDM from bulk ERs on the basis of the steepness of the spectrum rather than ionization yield: Figure~\ref{fig:nrdmizip_reach} shows the DM recoil spectrum for these low masses is entirely in the regime for which yield-based NR discrimination is ineffective.  

To determine the ultimate reach with iZIPs in the existing SuperCDMS SNOLAB cryostat, we consider the Ge 10~\cmcunosp~\detB scenario with 20$\times$ exposure.  This large exposure combined with the low-energy yield-based discrimination probes significantly into the neutrino fog.

It is more difficult to achieve these sensitivities with SNOLAB-sized iZIPs because of their higher detector capacitances, and, in fact, they are unnecessary: the 10~\cmcu detectors provide sufficient exposure to reach the neutrino fog, and go well into it.  That said, the \detB scenario with SNOLAB-sized Ge iZIPs achieves comparable sensitivity to the \detA scenario for 10~\cmcu Ge iZIPs.  Obviously, both the detector size reduction and the \detB improvement in ionization readout noise via elimination of the feedback resistor should be pursued, deploying whichever one is successful first and then eventually deploying both (10~\cmcunosp, \detBnosp) to obtain the greatest reach.

We do not pursue \detC iZIP upgrade scenarios for three reasons.  First, as was noted above, it becomes clear that, below 0.5--1~\GeVnosp, the spectral discrimination of the 0V approach is far more effective than yield-based discrimination, which, even in the \detC scenario for 10~\cmcu detectors, is ineffective below about 1~\GeVnosp.  Second, the \detC ionization resolution scenario requires improving the intrinsic noise of the HEMT ionization amplifiers, for which the path is currently unknown.  Lastly, for DM below 0.1--0.2~\GeVnosp, the recoil energies are so low, a few \evrnosp, that the ER bgnds at that energy no longer create ionization and the utility of the ionization measurement is completely lost.\footnote{At masses half an order of magnitude larger, corresponding to recoil energies an order of magnitude larger, tens of \evrnosp, the ionization yield for NRs likely vanishes (though the detailed manner is uncertain due to lack of ionization yield measurements at these energies), but ER rejection remains valuable if the ionization production of ERs can still be detected.  Direct ionization measurement cannot reach down to such low energies --- in even the \detC scenario, the ionization \textit{threshold} is around 50~\evr --- but the piZIP design (\S\ref{sec:nrdmpizip}) can maintain this ER rejection down to the few \evr energies at which it is possible.}

\paragraph{SG-2C: Reaching the Neutrino Fog with piZIP Detectors}
\label{sec:nrdmpizip}

\begin{figure}[t!]
\begin{center}
\textbf{SG-2C: \GeV Nucleon-Coupled DM with piZIP} \\
\includegraphics*[width=0.48\textwidth,viewport=0 0 576 410]{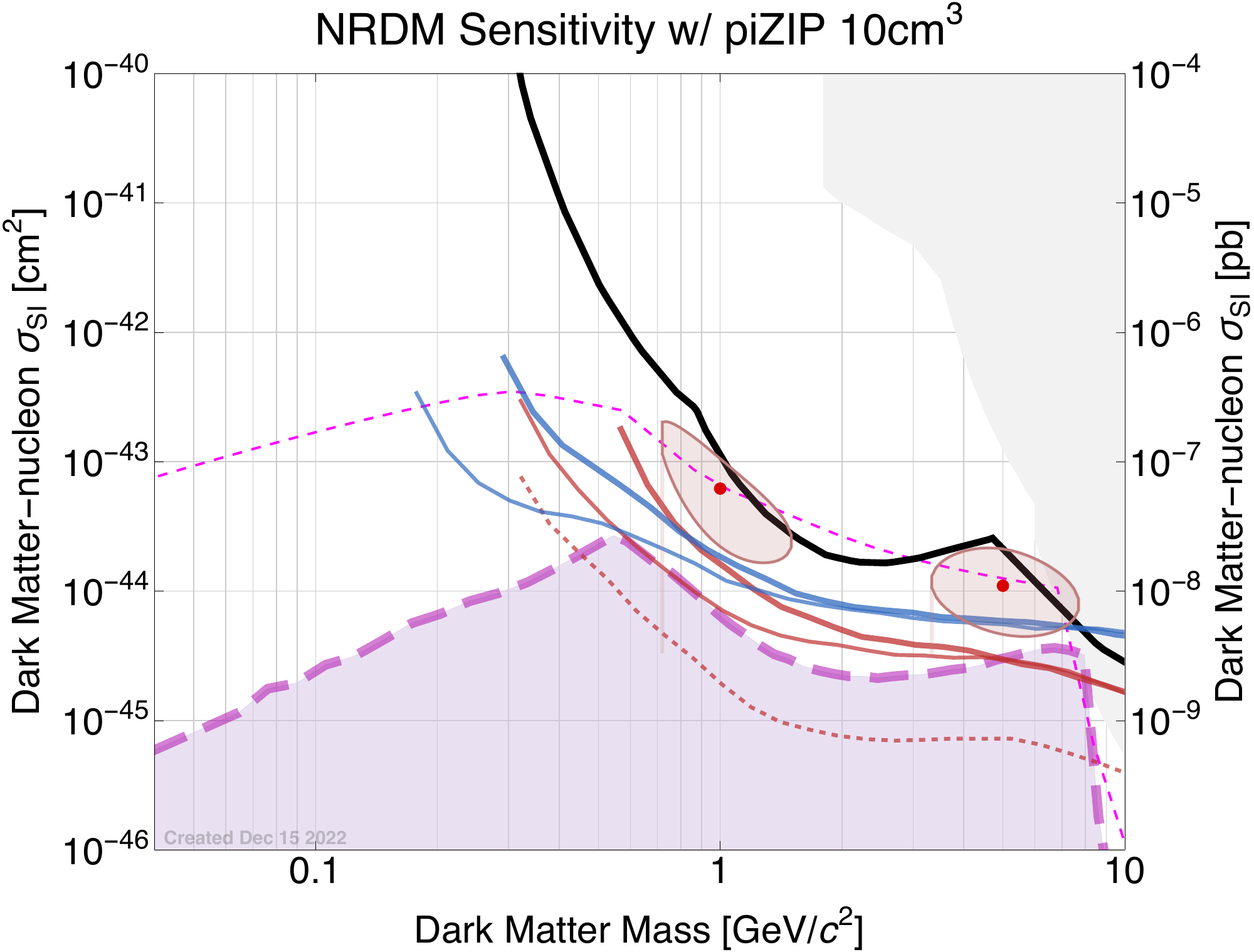}
\hfill
% \parbox[b]{3in}{\red{\footnotesize
% \begin{tabular}{p{3in}} \\
% \textbf{Discovery Potential Plot:} \\
% Three regions at 1, 2, and 4 GeV for \bkgon \detA piZIP 10~\cmcu Ge \\
% \vspace{60pt} \\
% \end{tabular}} \\} 
% \\
% \medskip
% Mar 2022
%\includegraphics*[width=0.46\textwidth,viewport=0 0 420 322]{spectra/piZIP10cm3_bkgA_detB_Si_postCuts_ER_v2.png}
% 220930
% changes: nothing much, really.  Funny bump in spectra has appeared at 30 eV, asking for explanation
\includegraphics*[width=0.46\textwidth,viewport=0 0 420 322]{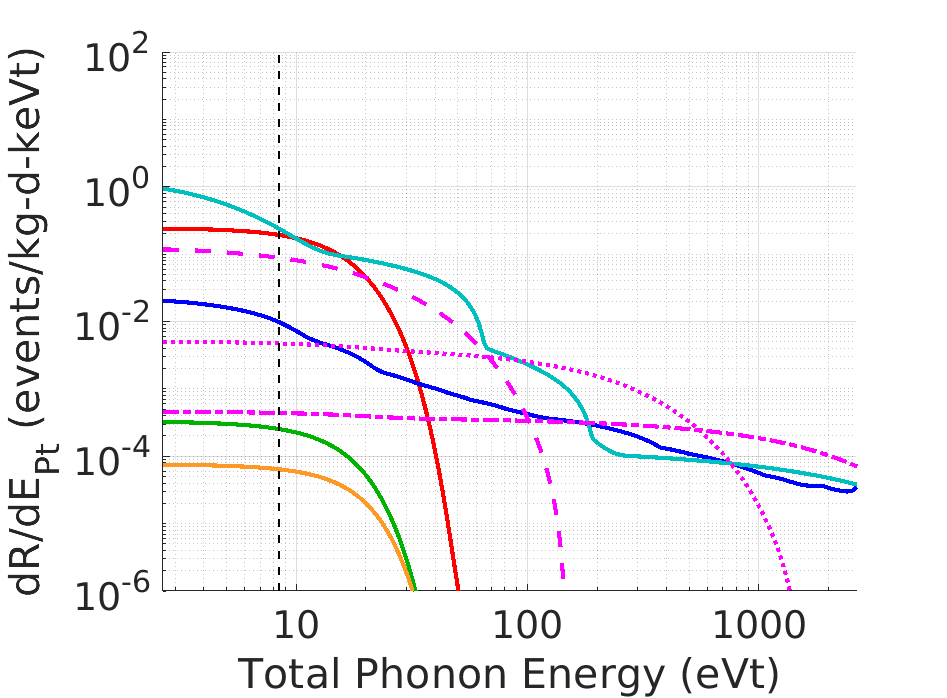}
%\red{Spectrum plot for NRDM for piZIP 10~\cmcu Si \bkgon \detB with backgrounds broken down} \\
\end{center}
\caption[Sensitivity to \GeV nucleon-coupled DM with 10~\cmcunosp~piZIP detectors]{\textbf{Sensitivity to \GeV nucleon-coupled DM with 10~\cmcunosp~piZIP detectors.}   (Left)~Median expected PLR-based 90\%~CL exclusion sensitivity, assuming ready-to-implement improvements in backgrounds (\bkgon scenario), for two Towers of 10~\cmcunosp~Ge (red-brown) and six Towers of 10~\cmcunosp~Si piZIP detectors (blue) for the \detA (thick) and \detB (thin) improvement scenarios along with the sensitivity of SuperCDMS SNOLAB (envelope of all detector types, thick black).  We also include a 20$\times$ exposure case for the Ge piZIP \detB scenario to illustrate the ultimate reach of the existing SuperCDMS SNOLAB cryostat with piZIPs.  The single-neutrino and neutrino fog sensitivity and grey shaded region are as in Figure~\ref{fig:nrdm0v_reach}.  The SNOLAB-sized Ge and Si piZIP \detB and \detC forecasted sensitivities are approximately the same as the corresponding 10~\cmcu piZIP \detA and \detB sensitivities.  We show six-Tower forecasts for Si detectors because two-Tower forecasts are exposure-limited.  We do not show 10~\cmcu piZIP \detC forecasts because the standard treatment of nuclear recoils is invalid below tens of \evrnosp~(\S\ref{sec:nucleon_coupling}).  We also show discovery potential for the \detAnosp~10~\cmcunosp~Ge case for two candidate DM masses, with legend as in Figure~\ref{fig:nrdm0v_reach}.
(Right)~Example spectrum plot showing the background spectra for the 10~\cmcu Si piZIP \bkgonnosp, \detB case and candidate nucleon-coupled DM signal spectra \underline{after} all analysis cuts described in \S\ref{sec:forecast_procedure_detail}.  Legend as in Figures~\ref{fig:snolab_reach_nrdm} and~\ref{fig:nrdmizip_reach}, including the differences for Ge and the interpretation of the shapes of the curves.  The bulk ER background rolls off at much lower energy than in the \detB iZIP case (Figure~\ref{fig:nrdmizip_reach}) because of the significantly better effective ionization energy resolution, while the surface NR background rolls off at lower energy due to the significantly better phonon energy resolution's effect on the phonon fiducial volume cut (see Table~\ref{tbl:detector_upgrade_scenarios}).  Both improvements contribute to the lower-energy roll-off of the surface ER background.}
\label{fig:nrdmpizip_reach}
\hrule
\end{figure}

The piZIP detector design provides a third attractive option for reaching the neutrino fog, as shown in Figure~\ref{fig:nrdmpizip_reach}.  Recall that the piZIP design uses ionization-yield-based NR discrimination in a manner similar to the iZIP, with an effective ionization resolution of
\begin{align}
\sigqpizip & = \frac{1}{\fdisc}\, \frac{\epseh}{e\,V_b}\, \sigpt \approx 2.5\,\sigpt  \quad\text{for $\fdisc = 0.2$}
\end{align}
%\marginpar{\red{\footnotesize Something is wrong with piZIP/iZIP comparison.  piZIP should be much better.  Waiting on checks by Osmond and Harrison.}}
Thus much of the same discussion as presented for the iZIP is valid here.  %However, what is striking is that \textit{two Towers of 10~\cmcu Ge piZIP detectors, in the high-maturity \bkgonnosp, \detA scenario, can already reach the neutrino fog down to 1~GeV!}  Because phonon resolutions are substantially better than ionization resolutions for any \detX scenario, and because of the small multiplier between \sigpttxt and \sigqpiziptxtnosp, piZIP sensitivity for a given \detX scenario is generally comparable to iZIP sensitivity at the \text{next} \detX scenario.  \textit{If the piZIP design can be demonstrated and the \fdisctxt value validated, the piZIP design is obviously preferable to the iZIP in terms of both performance and readout simplicity.}
Two Towers of 10~\cmcu Ge piZIP detectors with the \bkgonnosp, \detA upgrade scenario provides reach between single-neutrino sensitivity and the neutrino fog.  Use of Si requires six Towers to achieve comparable reach.  The \detB Ge piZIP scenario reaches the neutrino fog, again with only the \bkgon background improvement scenario, while the six-Tower \detB Si piZIP scenario is less sensitive because of the exposure limitation.  The \detB Ge piZIP scenario with 20$\times$ exposure displays the ultimate reach of the existing SuperCDMS SNOLAB cryostat with piZIPs, deep into the neutrino fog. Comparing the potential reach of various piZIP detector sizes, the behavior is much as for iZIPs: the 10~\cmcu piZIP design at a given level of detector improvement gives reach comparable to the SNOLAB-sized piZIP design at the \textit{next} level of detector improvement.  It makes sense to pursue in parallel both smaller detector sizes and improved phonon resolution for the same reasons as for the iZIP, with the hope that both succeed in the long run and can be combined.  We do not present 10~\cmcunosp~\detC piZIP projections because, below tens of \evrnosp, the standard treatment of nuclear recoils is invalid (\S\ref{sec:nucleon_coupling}).

\subsubsection{SG-3, SG-4, SG-5: eV/$\mathbf{c^2}$ Dark Photons and Axion-Like Particles (DPDM, ALPDM); \\Electron-Scattering Light (MeV/$\mathbf{c^2}$) Dark Matter (LDM)}
\label{sec:erdm}

\textit{The potential reach with SNOLAB-sized HV and small 0V detectors in the mass range 1--100~\eV for DPDM and ALPDM and 1--100~\MeV for LDM is enormous, strongly motivating these science goals.}

We group these three science goals together because the optimal energy deposition range to use in searching for all three candidates is the same, 1--100~eV deposited energy, due to the small bandgaps ($\approx 1$~eV) and the excellent expected energy resolution of upcoming generations of detectors.\footnote{For energy depositions greater than 100~eV (LDM mass above 100~\MeVnosp), noble liquids become competitive targets for DPDM, ALPDM, and heavy-mediator LDM because the typical ionization energy in these materials is tens of eV.  LXe, however, remains a poor target for light-mediator LDM because of the $1/q^2$ form factor.}  The same upgrade scenarios thus apply to all three, though with nuances in how the upgrades enable these science goals.  These are the same detectors that also enable SG-1 (nucleon-coupled sub-\GeV DM with small 0V detectors) and SG-2A (nucleon-coupled \GeV DM to the neutrino fog with SNOLAB-sized HV detectors).
% would like to explain the point about LXe better, don't have time right now

We consider both HV and 0V detectors for these science goals because they can have comparable reach while having complementary development risks.  The conservative analysis scenario adopted due to ionization leakage (\S\ref{sec:forecast_procedure}) effectively imposes a higher analysis threshold than possible based on the phonon energy resolution alone.   We will see below that, despite this limitation, the large NTL amplification of DPDM/ALPDM recoil signals ($\approx 25$--$30\times$) still provides DPDM/ALPDM mass reach not accessible to 0V detectors with comparable energy resolution (i.e., of the same mass and \detX scenario).  In fact, NTL gain is so effective that a 0V detector must have the smallest size considered here, 1~\cmcunosp, to have comparable DPDM/ALPDM mass reach as a SNOLAB-sized HV detector (for the same \detX scenario).  In that case, however, the small 0V detector option would be exposure-limited.  The compromise case, 10~\cmcunosp~0V, provides sufficient exposure but its DPDM/ALPDM mass reach is poorer.  Thus, achieving the same DPDM/ALPDM sensitivity with 0V detectors requires either a 1~\cmcu payload with 20$\times$ the two-Tower detector mass (\S\ref{sec:live_time_channel_counts}) or a 10~\cmcu payload incorporating the next generation of \detX upgrade.  Hence the risk tradeoff: 0V detectors mitigate the risk that HV ionization leakage is worse than or cannot be improved as much as expected, while HV detectors mitigate the risk that the expected advances in phonon energy resolution (10~\cmcunosp), or the capacity to read out 20$\times$ more channels (1~\cmcunosp), are not achieved as well as the risk that low-energy particle and/or environmental backgrounds (\S\ref{sec:env_backgrounds_summary}) will artificially increase the practical thresholds for the very low phonon energy threshold 0V detectors.

For LDM, the situation is somewhat different because of the kinematics of LDM scattering with the light electron.  When \mdmtxt is much larger than the electron mass, the recoil energy spectrum becomes independent of \mdmtxtnosp, with the maximum energy \emaxldmtxt set by the electron mass and the DM escape velocity, not \mdmtxtnosp.  For lighter \mdmtxtnosp, the spectrum becomes softer with a lower cutoff energy.  Therefore, if the phonon energy resolution is poor enough that the threshold is above \emaxldmtxtnosp, there is no sensitivity to LDM \text{of any mass}.  When the threshold drops below \emaxldmtxtnosp, the sensitivity increases exponentially with threshold because of the exponential tail of the recoil spectrum.  Thus, when a 0V detector's DM sensitivity turns on, it quickly surpasses that of the competing SNOLAB-sized HV detector.  The argument about complementarity of risks continues to hold though, and the less dramatic dependence of HV detector reach on the precise threshold achieved provides an additional level of risk mitigation.  This risk complementarity should be kept in mind when reviewing forecasts for HV detector sensitivity to LDM that seem obviously poorer than the 0V detector sensitivities \textit{for the same upgrade scenario} (i.e., comparing Figure~\ref{fig:ldmhv_reach} and Figure~\ref{fig:ldm0v_reach}).

\paragraph{SG-3: DPDM with HV and 0V Detectors}
\label{sec:dpdmhv0v}

\subparagraph*{SG-3A: DPDM with HV Detectors}

\begin{figure}[t!]
\begin{center}
\textbf{SG-3A: \eV Electron-Coupled DPDM with HV} \\
\includegraphics*[width=0.47\textwidth,viewport=0 0 800 568]{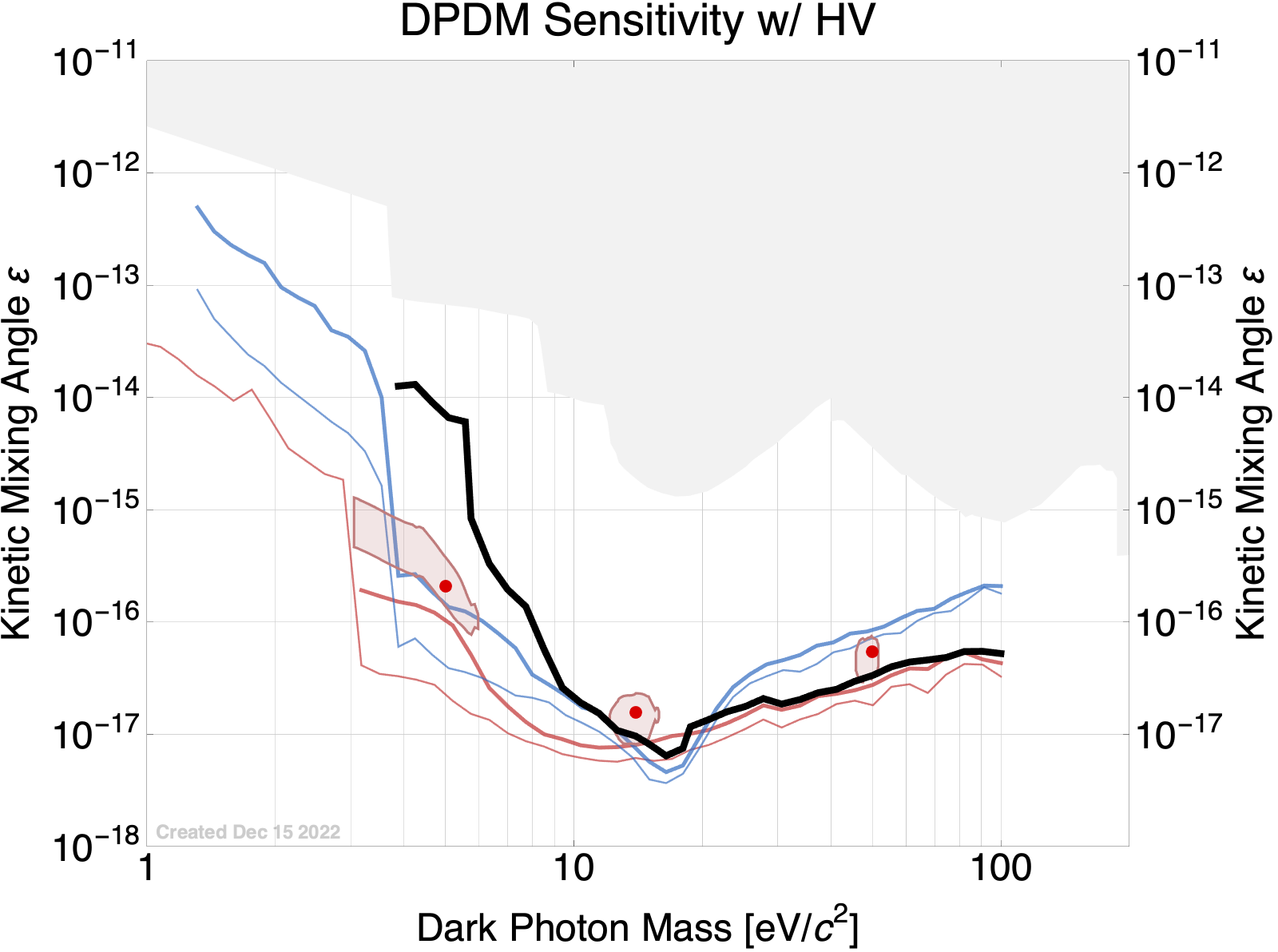} 
\hfill
% Mar 2022
%\includegraphics*[width=0.46\textwidth,viewport=0 0 430 322]{spectra/DP_HV_bkgA_detC_cons_Ge_postCuts_linear_lowE_v4.png} \\
%\includegraphics*[width=0.46\textwidth,viewport=0 0 430 322]{spectra/DP_HV_bkgA_detC_cons_Ge_postCuts_linear_midE_v4.png}
%\hfill
%\includegraphics*[width=0.46\textwidth,viewport=0 0 430 322]{spectra/DP_HV_bkgA_detC_cons_Ge_postCuts_linear_highE_v4.png} \\
% 220930
% v. modest changes: II/CT bug fix, increased neutron rate
\includegraphics*[width=0.46\textwidth,viewport=0 0 430 322]{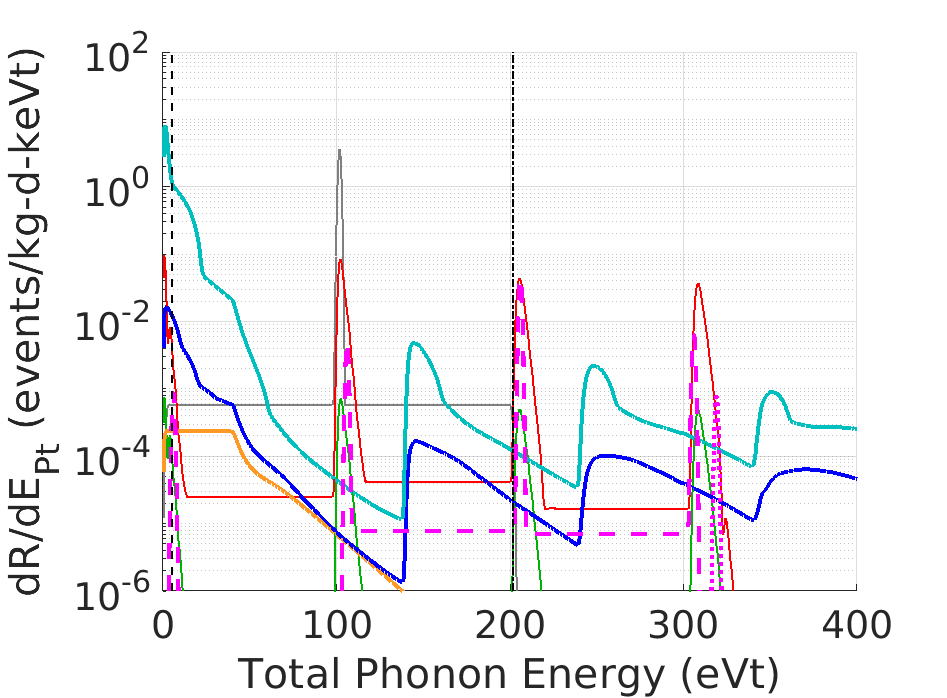} \\
\includegraphics*[width=0.46\textwidth,viewport=0 0 430 322]{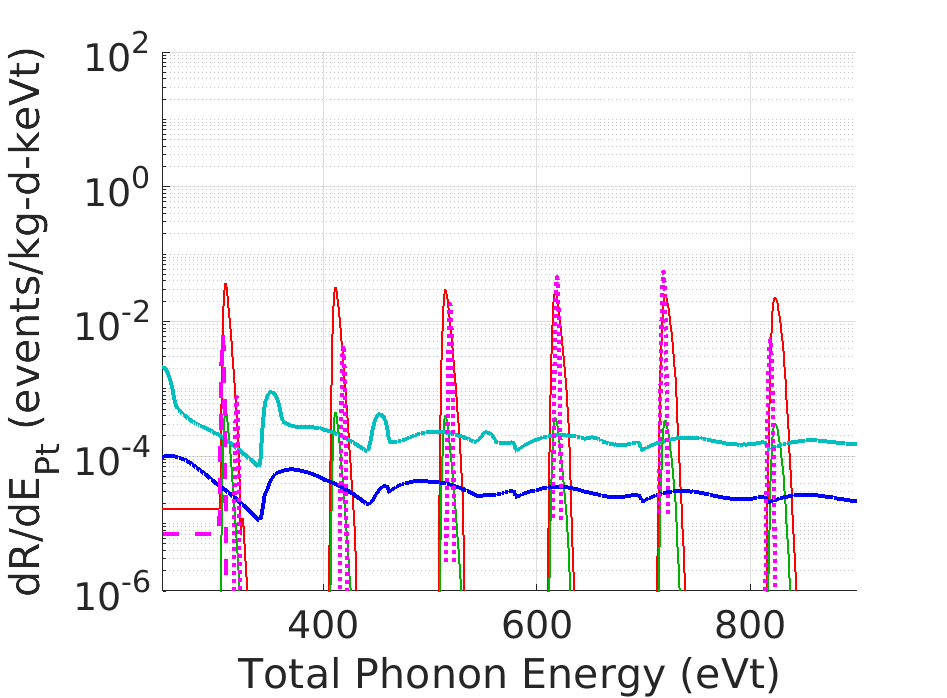}
\hfill
\includegraphics*[width=0.46\textwidth,viewport=0 0 430 322]{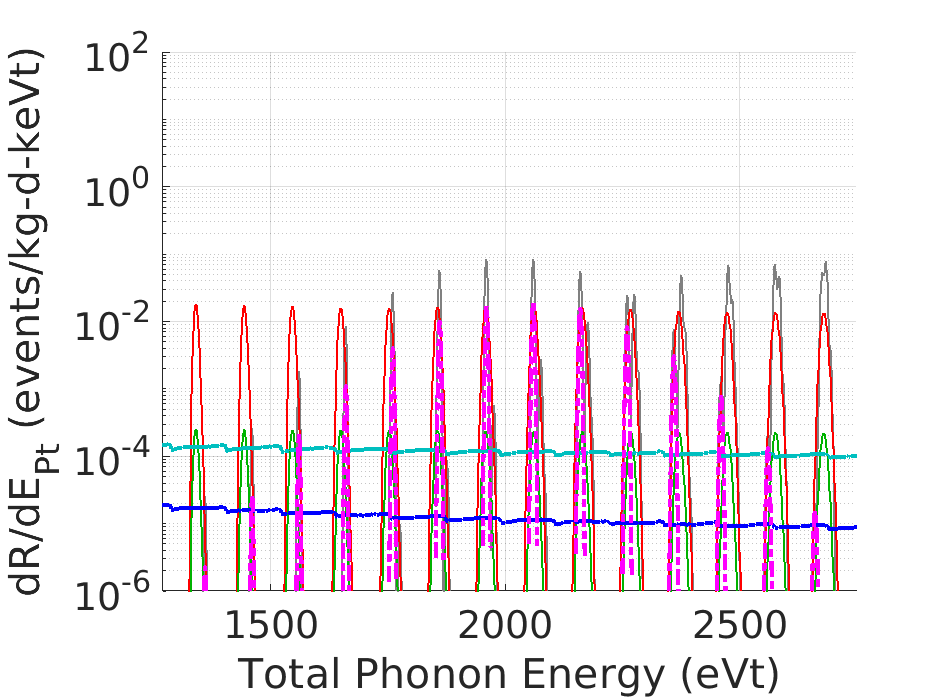} \\
%\red{Spectrum plots for DPDM for HV SNOLAB-sized Ge \bkgon \detA and \detC with backgrounds broken down} 
\end{center}
\caption[Sensitivity to dark photon dark matter (DPDM) with SNOLAB-sized HV detectors]{\textbf{Sensitivity to dark photon dark matter (DPDM) with SNOLAB-sized HV detectors.} (Top left)~Median expected PLR-based 90\%~CL exclusion sensitivity for two Towers each of SNOLAB-sized Ge (red-brown) and Si (blue) HV detectors for the two more advanced detector upgrade scenarios (\detBnosp, and \detCnosp; thick and thin) and assuming ready-to-implement improvements in backgrounds (\bkgon scenario) along with the sensitivity of SuperCDMS SNOLAB (envelope of all detector types, thick black).  Ge and Si are both useful because of their complementary sensitivities.  The \detA scenario, not shown, has sensitivity comparable to SuperCDMS SNOLAB because it assumes no improvement in ionization leakage.  
%\red{We also show the \detCnosp, \underline{\bkgth} scenarios for Ge and Si (dashed) to illustrate the modest sensitivity gain from background improvements.}  
The grey region represents already excluded parameter space.  
%The \detA and \detB scenarios for SNOLAB-sized Si HV detectors, not shown, have comparable sensitivity to Ge below \mdmtxtnosp~=~15~eV and approximately half an order of magnitude poorer sensitivity above this mass (in $\epsilon$; an order of magnitude in cross section), assuming the same ionization leakage.  
We also show discovery potential for the \detB Ge case for three candidate DM masses, with legend as in Figure~\ref{fig:nrdm0v_reach}.
(Top right and bottom)~Example spectrum plots showing the background spectra for the SNOLAB-sized Ge HV \bkgonnosp, \detC case and three candidate DPDM signal spectra.  Independent plots are made for each candidate mass ($\mathit{\mdm \approx 6}$, 20, and 60~\eVnosp) for clarity, and the DM spectra are artificially cut off at $\pm4\sigma$ from each peak.  Legend as in Figure~\ref{fig:snolab_reach_nrdm}, with the addition of cosmogenic activation spectral peaks (thin grey; also cut off at $\pm4\sigma$).  It is clear that the ability to detect a DM signal relies on it being comparable to the bulk ER (red) and cosmogenic activation (grey) backgrounds in at least one peak since the spectra are so similar.  (On close inspection, one can see the DM peaks are narrower and can be slightly displaced.)}
\label{fig:dpdmhv_reach}
\hrule
\end{figure}

Figure~\ref{fig:dpdmhv_reach} shows the projected sensitivity for DPDM with SNOLAB-sized HV detectors for the more advanced \detB and \detC cases (the \detA case has sensitivity similar to SuperCDMS SNOLAB).  Each of these stages of detector upgrade provides 1/4--1/3 decade additional mass reach, with Si and Ge having complementary sensitivity: Si is more sensitive near 15~\eV due to in-medium effects while Ge is more sensitive at lower and higher masses.  Since prior limits are weaker at lower masses, each extension in mass covers increasingly more new parameter space.  The kinetic-mixing parameter sensitivity does not improve significantly for \detC at masses for which sensitivity was not threshold-limited in the \detB scenario.  

% some of these conclusions require looking at the full set of plots, not just what is shown here
%Below $\mdm \approx 10$~eV and assuming comparable leakage, Si and Ge have comparable sensitivity, so we do not present an exhaustive set of Si predictions.  It is possible, however, that Ge may have poorer ionization leakage (it has a smaller gap energy), so we show Si forecasts for the \bkgthnosp, \detC scenario.  At higher masses, Ge is about 5$\times$ more sensitive (in $\epsilon$) than Si for any scenario due to lower backgrounds (even in the \bkgth scenario where the \hthree and \sitt backgrounds are removed).

For a given target material, the impact of background improvements (\bkg upgrades) is %greatest in the 20--100~eV mass range, with \bkgth yielding a factor of 1.5--2.5 in kinetic-mixing parameter sensitivity ($\times$2--6 in cross section) for any \detX scenario.  However, any improvements are significant over such a small portion of these detectors' regime of unique sensitivity 
in the range of 1/4--1/3 of a decade in kinetic-mixing parameter, a modest enough improvement that \textbf{particle background improvements beyond \bkgon are not strongly motivated for HV detector reach for DPDM}.  

\subparagraph*{SG-3B: DPDM with 0V Detectors}

\begin{figure}[t!]
\begin{center}
\textbf{SG-3B: \eV Electron-Coupled DPDM with 0V} \\
% 220928
%\red{Tarek, let's drop Ge 10cm3 Det B, it's no better than SuperCDMS SNOLAB.} \\
%\red{Osmond, is it possible to get allowed regions for Si 0V 1 cm3 Det B? Ge limit and allowed region is all in space to be ruled out SuperCDMS SNOLAB.} \\
%\red{SG: Yes, let's drop Ge Bkg 1 Det B 10cm3 limit and allowed regions.  But keep Ge Bkg 1 Det C 10cm3 limit and allowed regions.}\\
%\textcolor{magenta}{TS 0929: Done.}\\
% \red{Tarek to drop Ge Bkg 1 Det B 10cm3 allowed region} \\
%\green{221007 up to date} \red{switch limit plot to png} \\
%
%\red{\textbf{DPDM with 0V Sensitivity plot showing:}}
%{\footnotesize
%\begin{tabular}{l} \\
%\bkgon \detB 0V 10~\cmcu Ge \\
%\bkgon \detC 0V 10~\cmcu Ge \\
%\bkgon \detC 0V 10~\cmcu Si \\
%\bkgth \detC 0V 10~\cmcu Ge \\
%\bkgth \detC 0V 10~\cmcu Si \\
%\bkgon \detA 0V 1~\cmcu Si 20$\times$ mass \\
%\bkgon \detB 0V 1~\cmcu Si 20$\times$ mass \\
%\end{tabular}} \\
%\textbf{placeholder plots:} \\
%dot-dashed = 20x mass \\
% \includegraphics[width=0.32\textwidth]{limit_plots/dpdm_0v/0V10cm3_Ge_darkphoton_plrlimit.png}
% \includegraphics[width=0.32\textwidth]{limit_plots/dpdm_0v/0V10cm3_Si_darkphoton_plrlimit.png} 
% \includegraphics[width=0.32\textwidth]{limit_plots/dpdm_0v/0V1cm3_Si_darkphoton_plrlimit.png}
%\includegraphics*[width=0.50\textwidth,viewport=0 0 576 410]{limit_plots/dpdm_0v/DP_0V_GeSi_discovery.png}
%\includegraphics*[width=0.50\textwidth,viewport=0 0 576 410]{limit_plots/dpdm_0v/DP_0V_GeSi_discovery.pdf} 
\vspace{3pt}
\includegraphics*[width=0.47\textwidth,viewport=0 0 800 567]{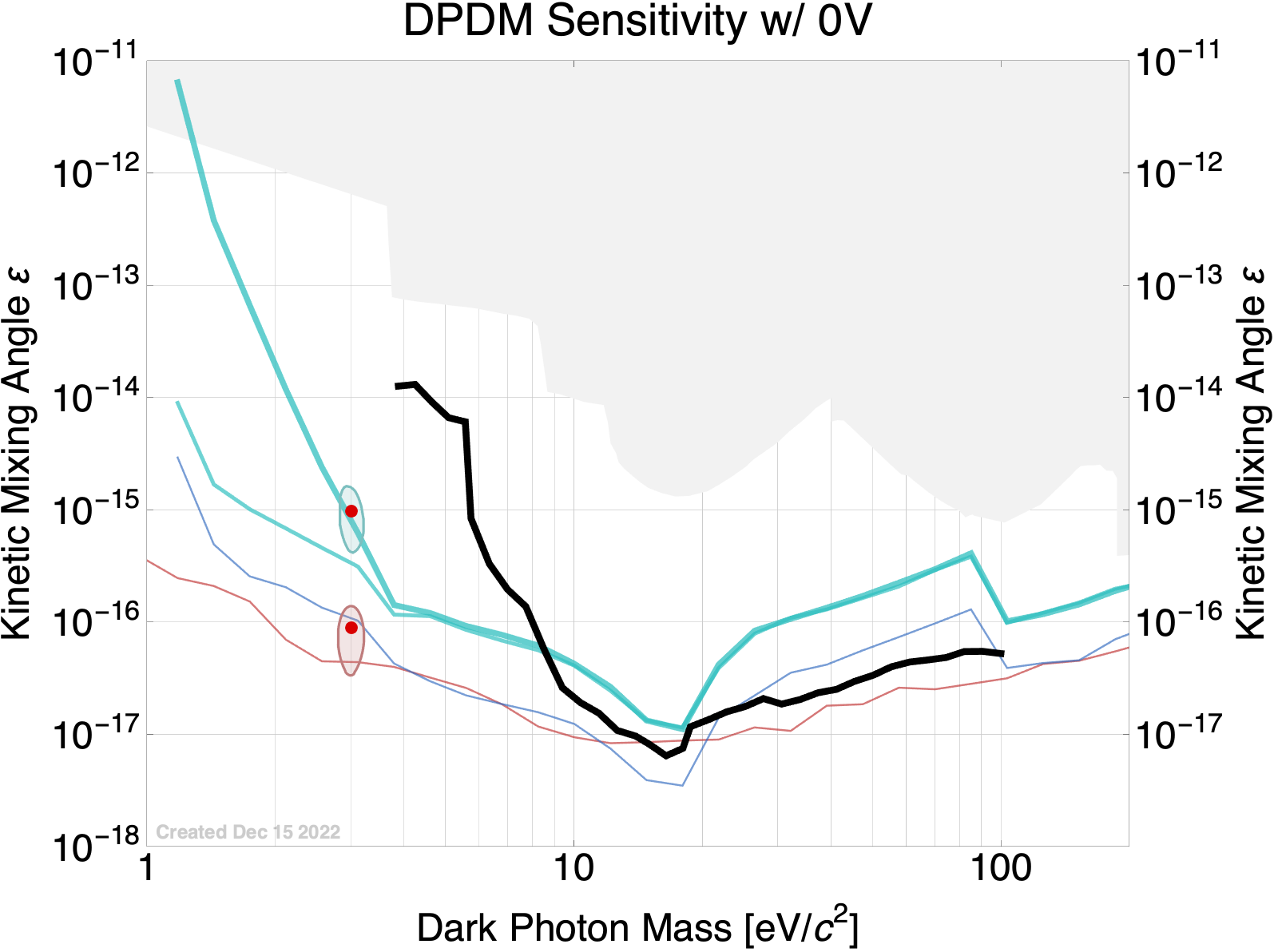} \\
%\parbox[b]{3.2in}{\red{\footnotesize
%\begin{tabular}{p{3in}} \\
%\textbf{Tarek:} \textcolor{blue}{done}\\
% Let's remove the \bkgth scenarios, they make the plot too busy \\
% Also, could you set the range to 1--100 eV to be consistent with other plots? 
% (x200 dynamic range) Reach above 100 eV is not unique (and is not even shown here). \\
%\textbf{Limit Plot (Tarek):} \\
%Could you replace \\
%\bkgon \detA 0V 1~\cmcu Si 20$\times$ mass \\
%\bkgon \detB 0V 1~\cmcu Si 20$\times$ mass \\
%with their nominal exposure versions? \textcolor{blue}{TS: Done, tweaked caption slightly by removing 20x mention} \\
% \textbf{Discovery Potential Plot:} \\
% \sout{\bkgon \detB 0V 10~\cmcu Ge 30 eV} \\
% \bkgon \detB 0V 10~\cmcu Ge 10 eV \\
% \bkgon \detC 0V 10~\cmcu Ge 3 eV \\
% \\
%\red{Tarek, I think we can just have one plot with all limits and allowed regions, it's not too busy.} 
% \textcolor{blue}{TS: replaced with png to remove contour plotting artifact but it totally messed up the view port number.}\\ 
%\vspace{20pt} \\
%\end{tabular}} \\}
%\red{Harrison: Could you modify the y scale to match to other plots (1e-6 to 1e2)?} \\
\medskip
% Mar 2022
%\includegraphics*[width=0.46\textwidth,viewport=0 0 420 322]{spectra/0V1cm3_bkgA_detC_DP_Ge_postCuts_ER_v3.png}
%\hfill
%\includegraphics*[width=0.46\textwidth,viewport=0 0 420 322]{spectra/0V1cm3_bkgA_detC_DP_Si_postCuts_ER_v2.png} \\
% 220930 No noticeable changes
\noindent
\hspace{0.45in}Ge\hspace{3.35in}Si\hfill \\
\includegraphics*[width=0.46\textwidth,viewport=0 0 420 322]{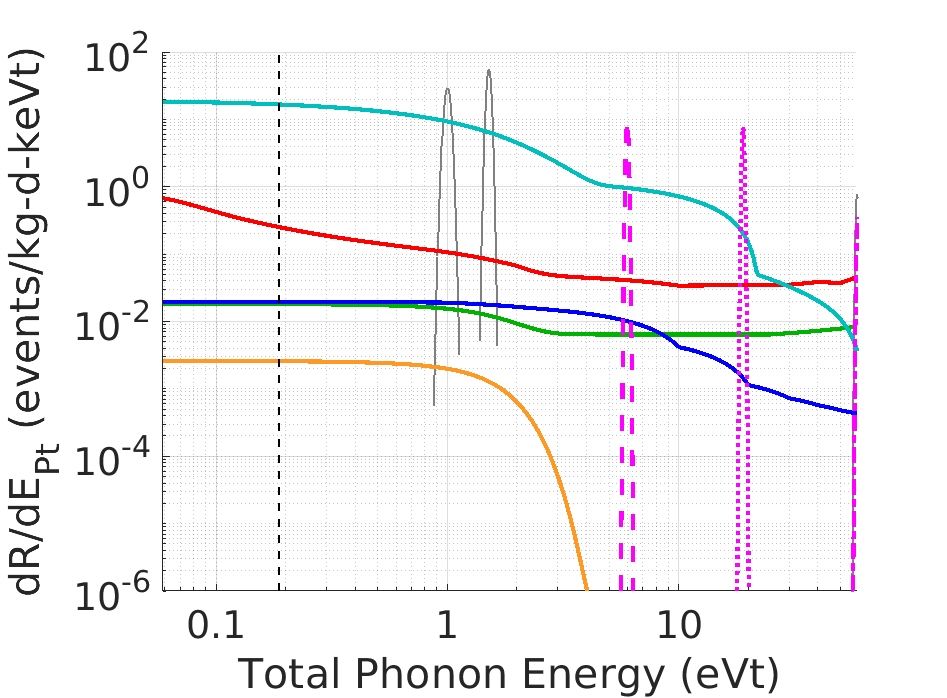}
\hfill
\includegraphics*[width=0.46\textwidth,viewport=0 0 420 322]{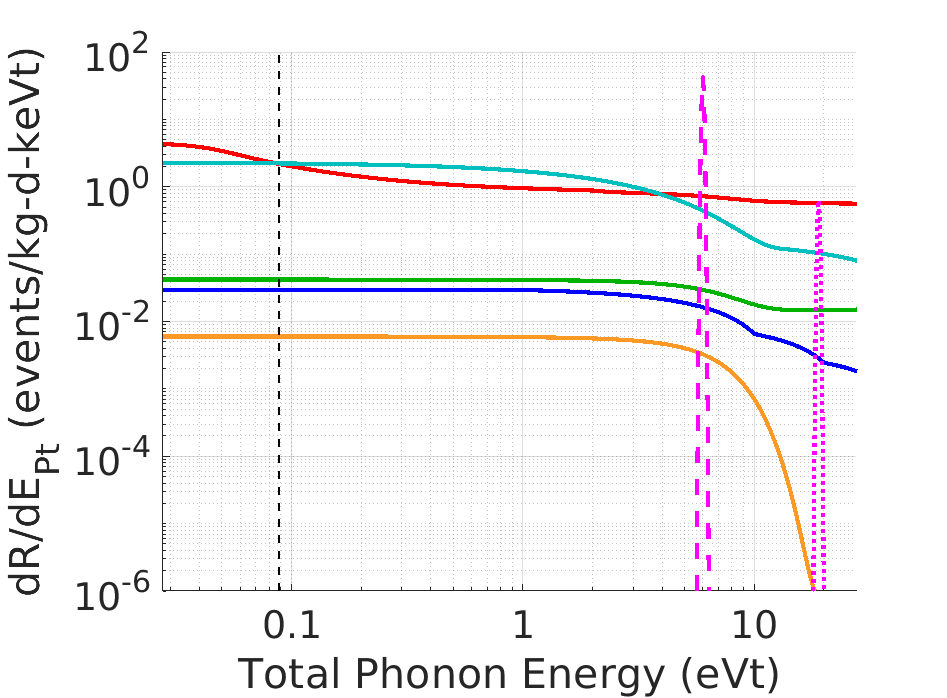} \\
%\red{Spectrum plots for DPDM and 0V 1~\cmcu Ge and Si \bkgon \detC with backgrounds broken down} 
\end{center}
% the 20x mass exposures are becoming bgnd-limited: improvement with 20x exposure is about 3x in epsilon, 
% which is < sqrt(20) (remember, epsilon ~ sqrt(rate)) so not exposure limited.
\caption[Sensitivity to dark photon dark matter (DPDM) with 10~\cmcu and 1~\cmcunosp~Ge and Si 0V detectors]{\textbf{Sensitivity to dark photon dark matter (DPDM) with 10~\cmcu and 1~\cmcunosp~Ge and Si 0V detectors.} (Top)~Median expected PLR-based 90\%~CL exclusion sensitivity, assuming ready-to-implement improvements in backgrounds (\bkgon scenario), for two Towers of: 1~\cmcunosp~Si 0V detectors for the \detA and \detB phonon-resolution scenarios (cyan, thick and thin); 10~\cmcu Ge 0V detectors for the \detCnosp~(thin) phonon-resolution improvements (red-brown, thin); and 10~\cmcunosp~Si 0V detectors for the \detC phonon-resolution scenario (blue, thin).  Also shown is the sensitivity of SuperCDMS SNOLAB (envelope of all detector types, thick black).  As for HV detectors, it is useful to consider both Ge and Si detectors because their differing dielectric functions provide complementary DPDM sensitivity.  Improving the backgrounds to the \bkgth scenario (not shown) has very modest impact, much less than an order of magnitude in $\epsilon$.  The plot illustrates how two Towers of 1~\cmcunosp~Si 0V in the \detA scenario provide an immediate factor of 2 in DPDM mass reach relative to SuperCDMS SNOLAB.
% 221006 S. Golwala
% don't bother with the 10cm3 detA cases -- nothing special about SuperCDMS SNOLAB reach, unlike ALP case
%and how the reach of HV SNOLAB-sized detectors for DPDM can be matched or exceeded by two Towers of 10~\cmcunosp~Ge and/or Si 0V detectors for the same \detX scenarios.  
We also show discovery potential for the 1~\cmcunosp~Si \detB and 10~\cmcunosp~Ge \detC cases for a candidate DM mass, with legend as in Figure~\ref{fig:nrdm0v_reach}.
(Bottom)~Example spectrum plots showing the background spectra for the \bkgonnosp, \detC case for 1~\cmcu Ge (left) and Si (right) 0V detectors and three candidate DPDM masses ($\mathit{\mdm \approx 6}$, 20, and 60~\eVnosp). Legend as in Figure~\ref{fig:snolab_reach_nrdm}, with the addition of cosmogenic activation spectral peaks (thin grey; artificially cut off at $\pm$4$\sigma$).}
\label{fig:dpdm0v_reach}
\hrule
\end{figure}

Figure~\ref{fig:dpdm0v_reach} displays the 0V detector reach for a set of scenarios we consider.  Comparing Figure~\ref{fig:dpdm0v_reach} to Figure~\ref{fig:dpdmhv_reach}, it is readily apparent that, as explained above, 1~\cmcu and 10~\cmcunosp~0V detectors can offer reach for DPDM comparable to or exceeding that of SNOLAB-sized HV detectors.  The 1~\cmcunosp~Si~0V \detA scenario provides an immediate factor of 2 in mass reach over SuperCDMS SNOLAB, and, in the \detC scenario, both Ge and Si 10~\cmcunosp~0V detectors exceed HV \detC sensitivity, reaching down fully to the gap energy  $\approx$1~eV.  Operation of both Ge and Si is advantageous because of their complementary sensitivities due to differing dielectric response functions.  The impact of \detX upgrade scenarios on mass reach is far more dramatic for 0V detectors than for HV detectors because there is no ionization leakage to detract from the improvement in phonon energy resolution. The dependence on background levels (not shown) is very weak, smaller than for the HV detectors because of the greater difference in DM and background spectral shapes: \textbf{particle background improvements beyond \bkgon are not needed for 0V detector reach for DPDM}.

\paragraph{SG-4: ALPDM with HV and 0V Detectors}
\label{sec:alpdmhv0v}

\begin{figure}[t!]
\begin{center}
\textbf{SG-4:  \eV Electron-Coupled ALPDM with HV and 0V} \\
\vspace{3pt}
\begin{tabular}{p{0.48\textwidth}p{0.48\textwidth}}
\centering SNOLAB-sized Ge HV & \centering 1~\cmcu and 10~\cmcu Ge 0V \\
\end{tabular} \\
\vspace{-12pt}
\includegraphics*[width=0.47\textwidth,viewport=0 0 800 567]{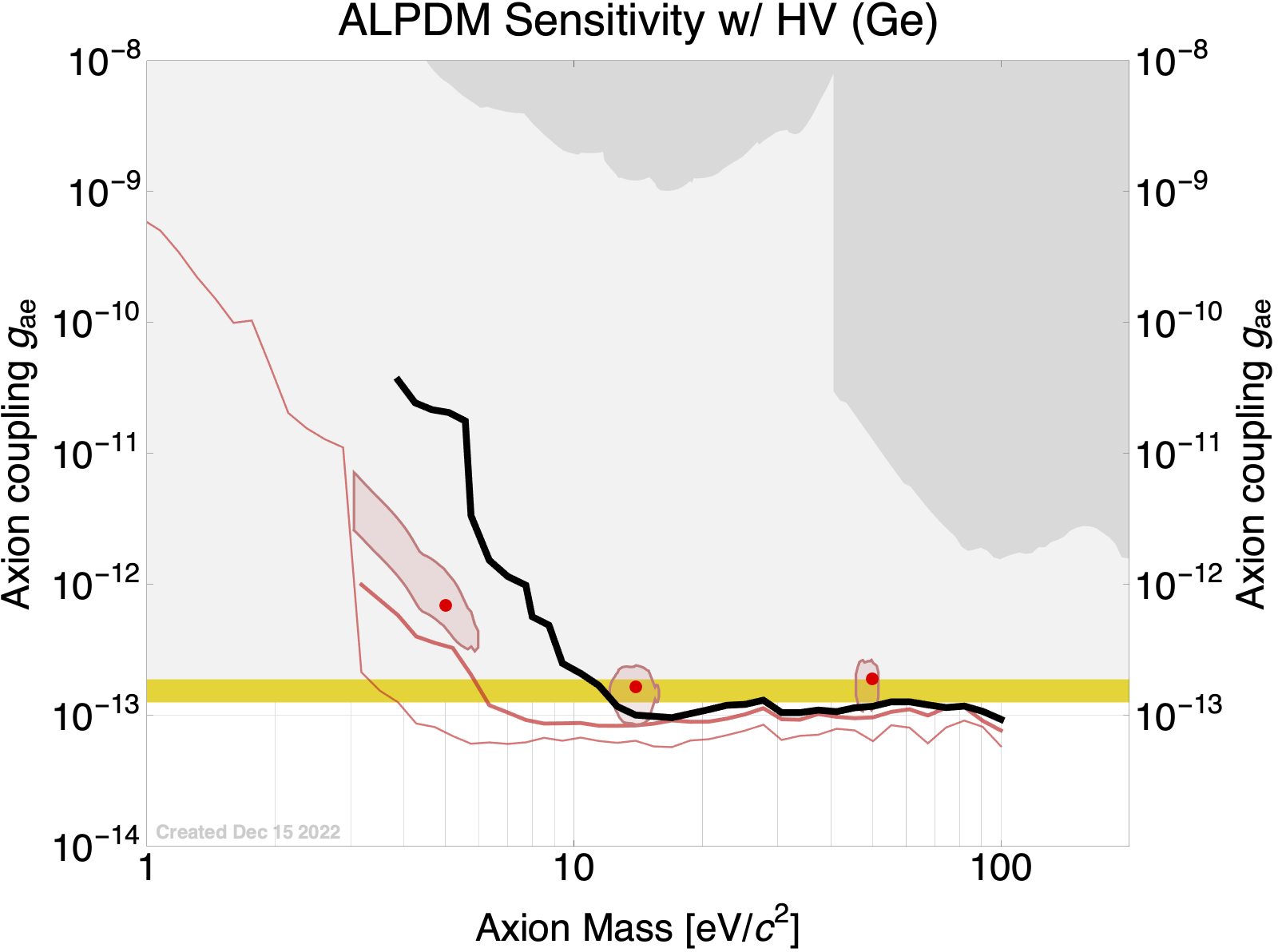}
%\red{\textbf{ALPDM with 0V Sensitivity plot showing:}}
%{\footnotesize
%\begin{tabular}{l} \\
%\bkgon \detC 0V 10~\cmcu Ge \\
%\bkgth \detA 0V 10~\cmcu Ge \\
%\bkgth \detB 0V 10~\cmcu Ge \\
%\bkgth \detC 0V 10~\cmcu Ge \\
%\bkgon \detB 0V 1~\cmcu Ge 20$\times$ mass \\
%\end{tabular}} \\
%\textbf{placeholder plots:} \\
%dot-dashed = 20x mass \\
%\includegraphics[width=0.32\textwidth]{limit_plots/alpdm_0v/0V10cm3_Ge_alp_plrlimit.png}
%\includegraphics[width=0.32\textwidth]{limit_plots/alpdm_0v/0V10cm3_Si_alp_plrlimit.png}
%\includegraphics[width=0.32\textwidth]{limit_plots/alpdm_0v/0V1cm3_Si_alp_plrlimit.png}\\
%\includegraphics[width=0.50\textwidth]{limit_plots/alpdm_0v/ALP_0V10cm3_Si.pdf} \\
\hfill
\includegraphics*[width=0.47\textwidth,viewport=0 0 800 567]{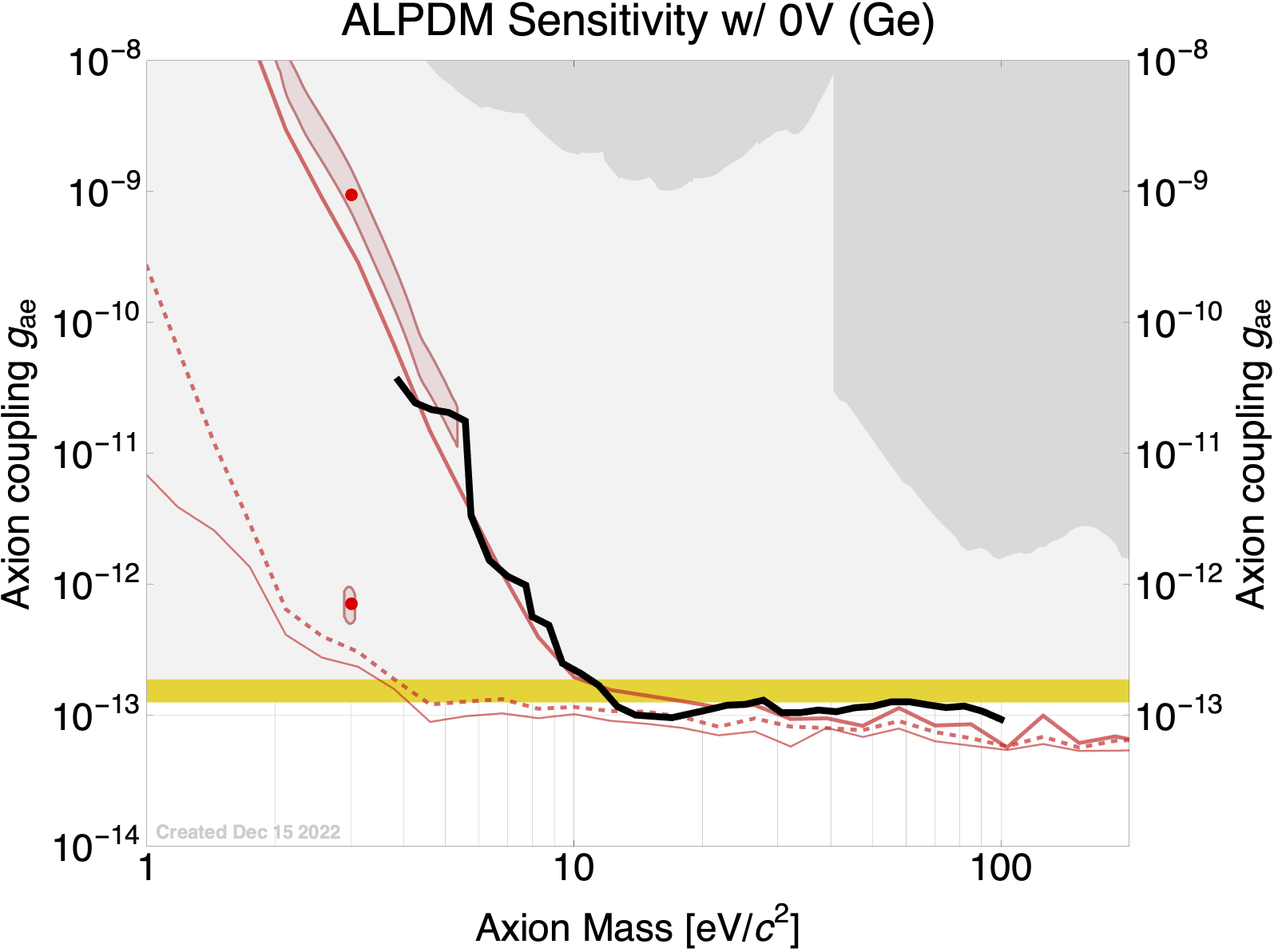} 
\iffalse
\medskip \\
\red{\textbf{Tarek}, %let's get rid of \detA since it is now superceded by SuperCDMS SNOLAB.   
I think we can just have one plot for each detector type with all limits and allowed regions, it's not too busy.}\\
\parbox[b]{3.2in}{\red{\footnotesize
\begin{tabular}{p{3in}} \\
\textbf{Discovery Potential Plot:} \\
\sout{\bkgon \detA HV SNOLAB-sized Ge 30 eV} \\
\bkgon \detB HV SNOLAB-sized Ge 10 eV \\
\bkgon \detC HV SNOLAB-sized Ge 3 eV  \\
\end{tabular}} \\}\parbox[b]{3.2in}{\red{\footnotesize
\begin{tabular}{p{3in}} \\
\textbf{Discovery Potential Plot:} \\
\sout{\bkgth \detA 0V 10~\cmcu Ge 100 eV} \\
\sout{\bkgth \detB 0V 10~\cmcu Ge 30 eV} \\
\bkgth \detB 0V 10~\cmcu Ge 10 eV \\
\bkgth \detC 0V 10~\cmcu Ge 3 eV  \\
\end{tabular}} \\} 
\fi
\end{center}
\caption[Sensitivity to axion-like particle dark matter (ALPDM) with SNOLAB-sized Ge HV and 1~\cmcu and 10~\cmcu Ge 0V detectors]{\textbf{Sensitivity to axion-like particle dark matter (ALPDM) with SNOLAB-sized Ge HV and 1~\cmcu and 10~\cmcu Ge 0V detectors.} Median expected PLR-based 90\%~CL exclusion sensitivity for \gaetxt for SNOLAB-sized Ge HV and 10~\cmcu and 1~\cmcu Ge 0V detectors along with the sensitivity of SuperCDMS SNOLAB (envelope of all detector types, thick black).  As described in the text, these plots are remappings of Figures~\ref{fig:dpdmhv_reach} and \ref{fig:dpdm0v_reach}, but we choose to focus on scenarios capable of testing the stellar cooling hint (yellow band), in particular neglecting Si detectors because they do not provide sufficient coupling strength reach.
(Left)~Sensitivity for two Towers of SNOLAB-sized Ge HV detectors for the \detB and \detC \bkgon scenario (red-brown, thick and thin).  Each scenario extends the mass reach, in particular the mass range over which the stellar cooling hint can be tested.  We also show discovery potential for the \detB scenario for various DM masses, with legend as in Figure~\ref{fig:nrdm0v_reach}.  (Right)~Sensitivity for two Towers of 10~\cmcunosp~Ge 0V detectors for the \detB and \detC~\bkgon scenarios (red-brown, thick and thin), along with the \bkgonnosp, \detBnosp~1~\cmcunosp~Ge 0V 20$\times$ mass scenario (red-brown, dashed).  The \detB scenario serves as a backup to SuperCDMS SNOLAB in testing the stellar cooling hint above $\approx$12~\eVnosp, while the \detC scenario provides significant new mass reach.  The 1~\cmcunosp~Ge 0V \detC scenario, not shown, does not provide appreciably greater mass reach than the scenarios shown due to the dielectric function of Ge.  We also show discovery potential for the 10~\cmcunosp~Ge scenarios for one candidate DM mass, with legend as in Figure~\ref{fig:nrdm0v_reach}.  The \detC contour is very narrow because of the excellent energy resolution.  Because both HV and 0V plots are remappings of the DPDM plots, most of the same comments and spectrum plots are applicable.  Astrophysical (light shaded) constraints exclude $\mathit{g_{ae} \gtrsim 2\,\times\,10^{-13}}$, while prior particle DM searches (grey shaded) have reach down to $\mathit{\approx}$$\mathit{10^{-12}}$ around 100~\eV and $\mathit{\approx}$$\mathit{10^{-9}}$ at tens of \eVnosp.  A particular interesting benchmark sensitivity to reach is $\mathit{1\,\times\,10^{-13}}$, which would test the stellar cooling hint at 1--2$\mathit{\,\times\, 10^{-13}}$ (yellow band)~\cite{stellarcooling2016}.  
%The \bkgth scenario for HV detectors can robustly do so, while there is less margin for the \bkgon scenario for HV detectors and any \bkg scenario for 0V detectors (which motivates showing their \bkgth scenarios). 
}
\hrule
\label{fig:alpdm_reach}
\end{figure}

\begin{figure}[t!]
\begin{center}
\textbf{SG-5A: \MeV Electron-Coupled LDM with HV} \\
% 220928
%\red{Tarek, let's drop Bkg3, very little gain.}\\
%\textcolor{magenta}{TS 0928 done}
%\red{Tarek, there are some funny shaded rectangles on the plot.}\\
%\red{Osmond, is it clear why the left allowed region's injected model is at the edge of the allowed region?} \\
%\green{220930 up to date} \red{switch to png} \\
%\red{Tarek to change allowed regions to HV Det B} \\
%\textcolor{magenta}{TS: 1006 Done}\\
%\red{Harrison: change spectrum horizontal axes to eV, ticks at 0, 200, ..., 1000 eV (drop 1000 eV label, but keep tick, if necessary for aesthetics) - Done} \\
%
%\red{\footnotesize \textbf{Tarek:} \textcolor{blue}{Done}
%Could you reduce the vertical scale as requested for Figure~\ref{fig:ldm0v_reach}, \\
%and also make the sharp target colors the same? \\
%Also, could you add the axis label at 100 MeV?} \\
%\red{\textbf{LDM with HV Sensitivity plot showing:}}
%{\footnotesize
%\begin{tabular}{l} \\
%\bkgon \detA HV SNOLAB-sized Ge \\
%\bkgon \detB HV SNOLAB-sized Ge \\
%\bkgon \detC HV SNOLAB-sized Ge \\
%\bkgth \detC HV SNOLAB-sized Ge \\
%\end{tabular}} \\
%\textbf{placeholder plots:} \\
% \includegraphics[width=0.32\textwidth]{limit_plots/ldm_hv/HV_Ge_cons_escattering4_plrlimit.png}
% \includegraphics[width=0.32\textwidth]{limit_plots/ldm_hv/HV_Ge_cons_escattering6_plrlimit.png} \\
\vspace{3pt}
\begin{tabular}{p{0.48\textwidth}p{0.48\textwidth}}
\centering $F(q) = 1$ & \centering $F(q) = 1/q^2$ \\
\end{tabular} \\
\vspace{-12pt}
%\includegraphics*[width=0.48\textwidth,viewport=0 0 576 410]{limit_plots/ldm_hv/LDM_F1_HV_Ge_cons.pdf}
%\hspace{0.02\textwidth}
%\includegraphics*[width=0.48\textwidth,viewport=0 0 576 410]{limit_plots/ldm_hv/LDM_Fq2_HV_Ge_cons.pdf} \\
\includegraphics*[width=0.47\textwidth,viewport=0 0 800 567]{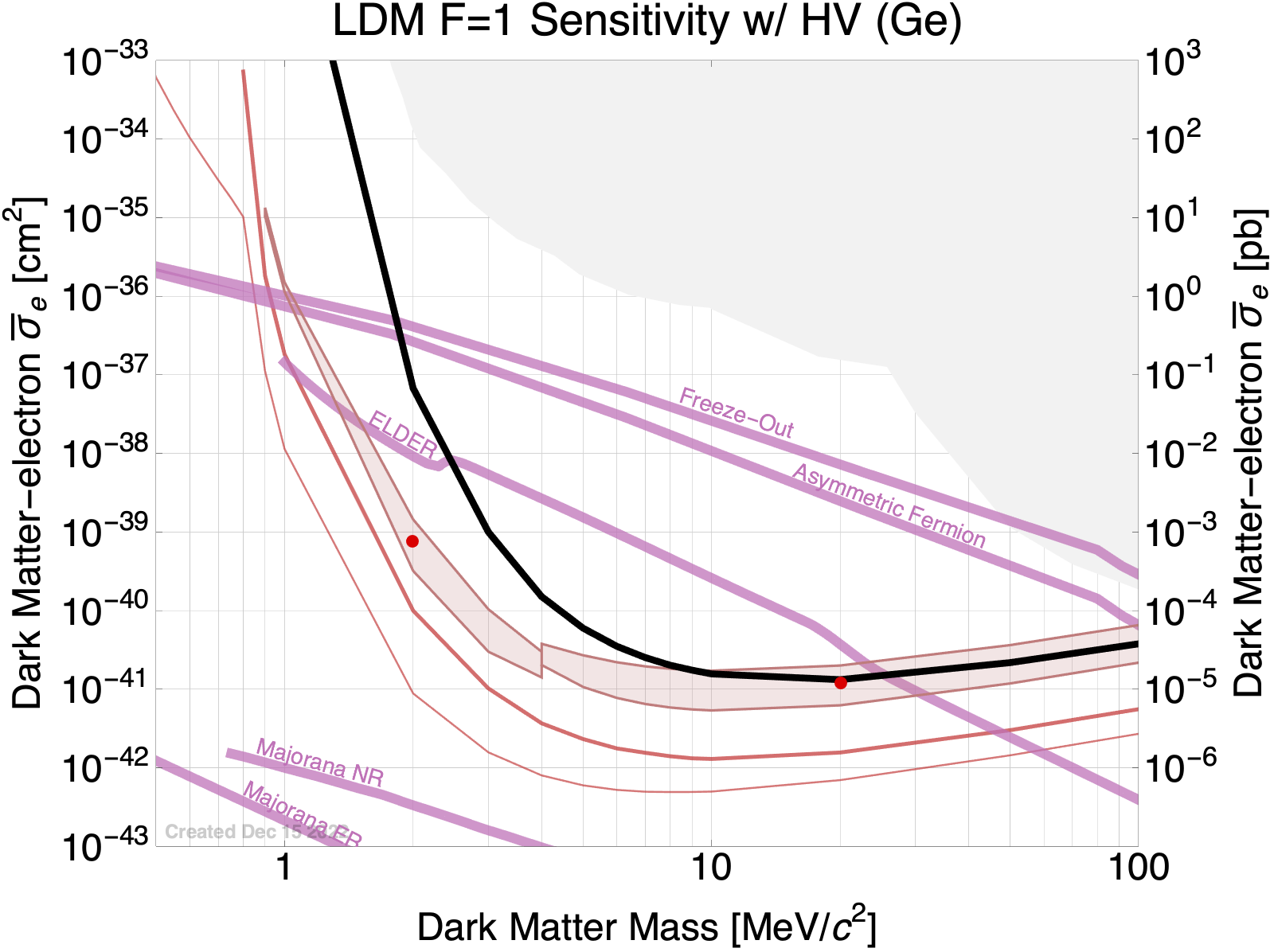}
\hspace{0.02\textwidth}
\includegraphics*[width=0.47\textwidth,viewport=0 0 800 567]{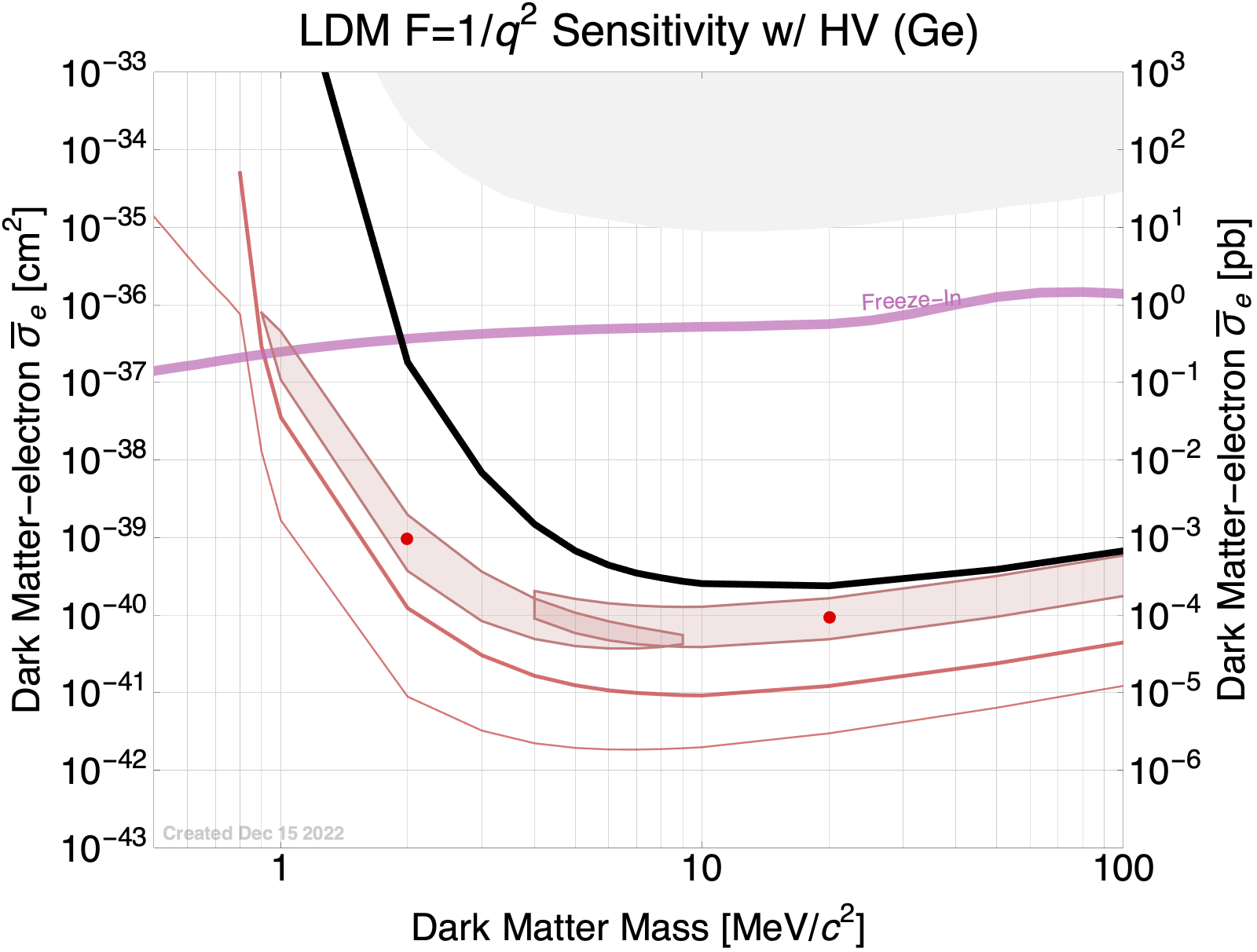} \\
%\red{\textbf{Osmond:} As we discussed on Friday, you may need to drop the 10 or 30 MeV discovery contour since we expect a mass-cross-section degeneracy when the DM mass is $>>$ the electron mass.  \\
%\red{\textbf{Tarek:} %Let's drop \detAnosp.  
%\newline It is not clear what discovery potential contours we can show on one plot since they have a big mass-cross-section degeneracy, and also not clear whether they can go on the exclusion plots or need a separate plot.  Let's see how the plots look.} \\
\iffalse
\parbox[b]{3.2in}{\red{\footnotesize
\begin{tabular}{p{3in}} \\
\textbf{Discovery Potential Plot:} \\
\sout{\bkgon \detA HV SNOLAB-sized Ge 30 \MeV Asymmetric DM line} \\
\bkgon \detB HV SNOLAB-sized Ge 10 \MeV \sout{ELDER line} \\
\bkgon \detB HV SNOLAB-sized Ge 3 \MeV \sout{ELDER line} \\
\bkgon \detC HV SNOLAB-sized Ge 1 \MeV \sout{ELDER line} \\
\vspace{20pt} \\
\end{tabular}} \\} 
\parbox[b]{3.2in}{\red{\footnotesize
\begin{tabular}{p{3in}} \\
\textbf{Discovery Potential Plot:} \\
\sout{\bkgon \detA HV SNOLAB-sized Ge 30 \MeV } \\
\sout{\bkgon \detA HV SNOLAB-sized Ge 10 \MeV Freeze-in line} \\
\bkgon \detB HV SNOLAB-sized Ge 3 \MeV \sout{Freeze-in line} \\
\bkgon \detC HV SNOLAB-sized Ge 1 \MeV \sout{Freeze-in line} \\
\vspace{20pt} \\
\end{tabular}} \\} 
\fi
\medskip
%\red{Harrison: Could you modify the y scale to match to other plots (1e-6 to 1e2)?  Are the masses in the caption correct?  Could you change the linestyle to be thin for the DM spectra so they are easier to see?} \\
%\includegraphics*[width=0.48\textwidth,viewport=0 0 420 322]{spectra/ES_HV_bkgA_detC_cons_Ge_postCuts_escattering4_linear_v3.png}
\hfill
%\includegraphics*[width=0.48\textwidth,viewport=0 0 420 322]{spectra/ES_HV_bkgA_detC_cons_Ge_postCuts_escattering6_linear_v3.png} \\
% 220930 changes:
% II/CT higher, neutron bgnd higher
\includegraphics*[width=0.48\textwidth,viewport=0 0 420 322]{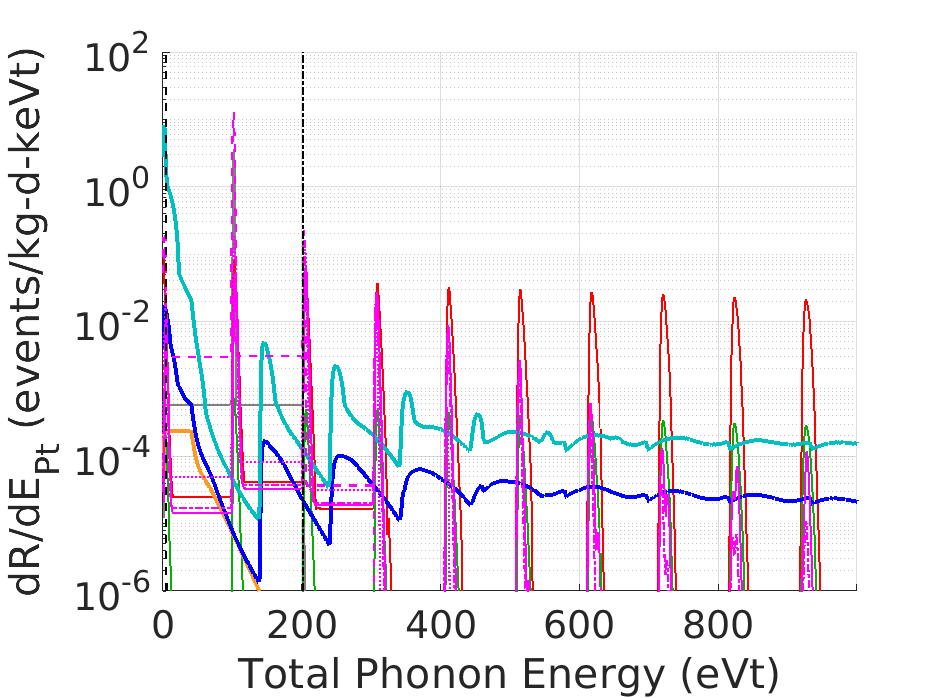}
\hfill
\includegraphics*[width=0.48\textwidth,viewport=0 0 420 322]{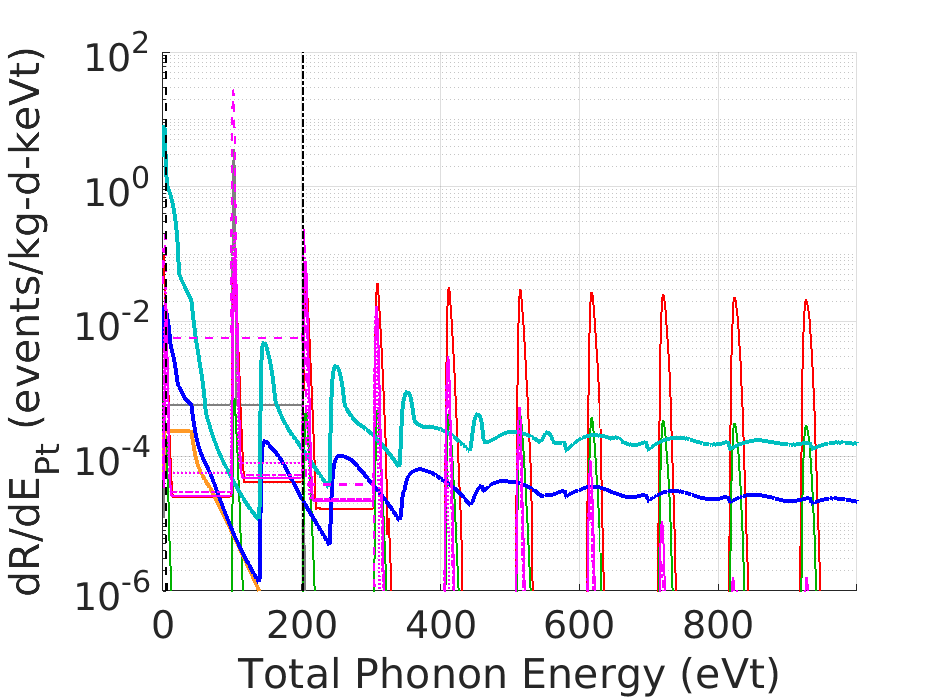} \\
%\red{Spectrum plot for Ge HV SNOLAB-sized \bkgon \detC with backgrounds broken down} 
\end{center}
\caption[Sensitivity to dark-photon-coupled light dark matter (LDM) with  SNOLAB-sized HV detectors]{\textbf{Sensitivity to dark-photon-coupled light dark matter (LDM) with  SNOLAB-sized HV detectors.} For all rows, the left column corresponds to the heavy mediator case ($F(q) = 1$) and the right column to the light mediator case ($F(q) = 1/q^2$).  The sensitivity of SuperCDMS SNOLAB (envelope of all detector types) is shown in thick black.  (Top)~Median expected PLR-based 90\%~CL exclusion sensitivity for two Towers of SNOLAB-sized Ge HV detectors for the \detB and \detC detector upgrade scenarios (red-brown, thick and thin) and assuming ready-to-implement improvements in backgrounds (\bkgon scenario).  The \detA scenario, not shown, has sensitivity comparable to SuperCDMS SNOLAB because it assumes no improvement in ionization leakage.  Sensitivity for SNOLAB-sized Si HV detectors is the same to within half an order of magnitude so is not shown.  Sharp science targets from the 2017 Cosmic Visions DM report~\cite{cvdm2017} are indicated, many of which can be tested over a wide mass range ($\mathit{M_{A'} = 3 \mdm}$ for the heavy mediator case).  We also show discovery potential for the \bkgonnosp, \detB SNOLAB-sized Ge HV scenarios, with legend as in Figure~\ref{fig:nrdm0v_reach}. The elongation of the contours is due to the mass-cross-section degeneracy.  
(Bottom)~Example spectrum plots showing the background spectra for the SNOLAB-sized Ge HV \bkgonnosp, \detC case.  Legend as in Figures~\ref{fig:snolab_reach_nrdm} and~\ref{fig:dpdmhv_reach}.  The candidate DM masses are $\mathit{\mdm \approx 1}$ (dashed), 3 (dotted), 10 (long/short dash), and 30~(solid)~\MeVnosp.  They are difficult to distinguish given the narrowness of the spectral peaks; on close inspection, it is apparent the higher-mass spectra extend to higher energy.}
\label{fig:ldmhv_reach}
\hrule
\end{figure}

\begin{figure}[t!]
\begin{center}
\textbf{SG-5B: \MeV Electron-Coupled LDM with 0V} \\
\vspace{3pt}
\begin{tabular}{p{0.48\textwidth}p{0.48\textwidth}}
\centering $F(q) = 1$ & \centering $F(q) = 1/q^2$ \\
\end{tabular} \\
\vspace{-12pt}
%\includegraphics*[width=0.48\textwidth,viewport=0 0 576 410]{limit_plots/ldm_0v/LDM_F1_0V_Si_discovery.pdf}
%\hspace{0.02\textwidth}
%\includegraphics*[width=0.48\textwidth,viewport=0 0 576 410]{limit_plots/ldm_0v/LDM_Fq2_0V_Si_discovery.pdf} \\
%\includegraphics*[width=0.48\textwidth,viewport=0 0 576 410]{limit_plots/ldm_0v/LDM_F1_0V_Si_discovery_v2.pdf} 
%\hfill
%\includegraphics*[width=0.48\textwidth,viewport=0 0 576 410]{limit_plots/ldm_0v/LDM_Fq2_0V_Si_discovery_v2.pdf}\\
\includegraphics*[width=0.47\textwidth,viewport=0 0 800 567]{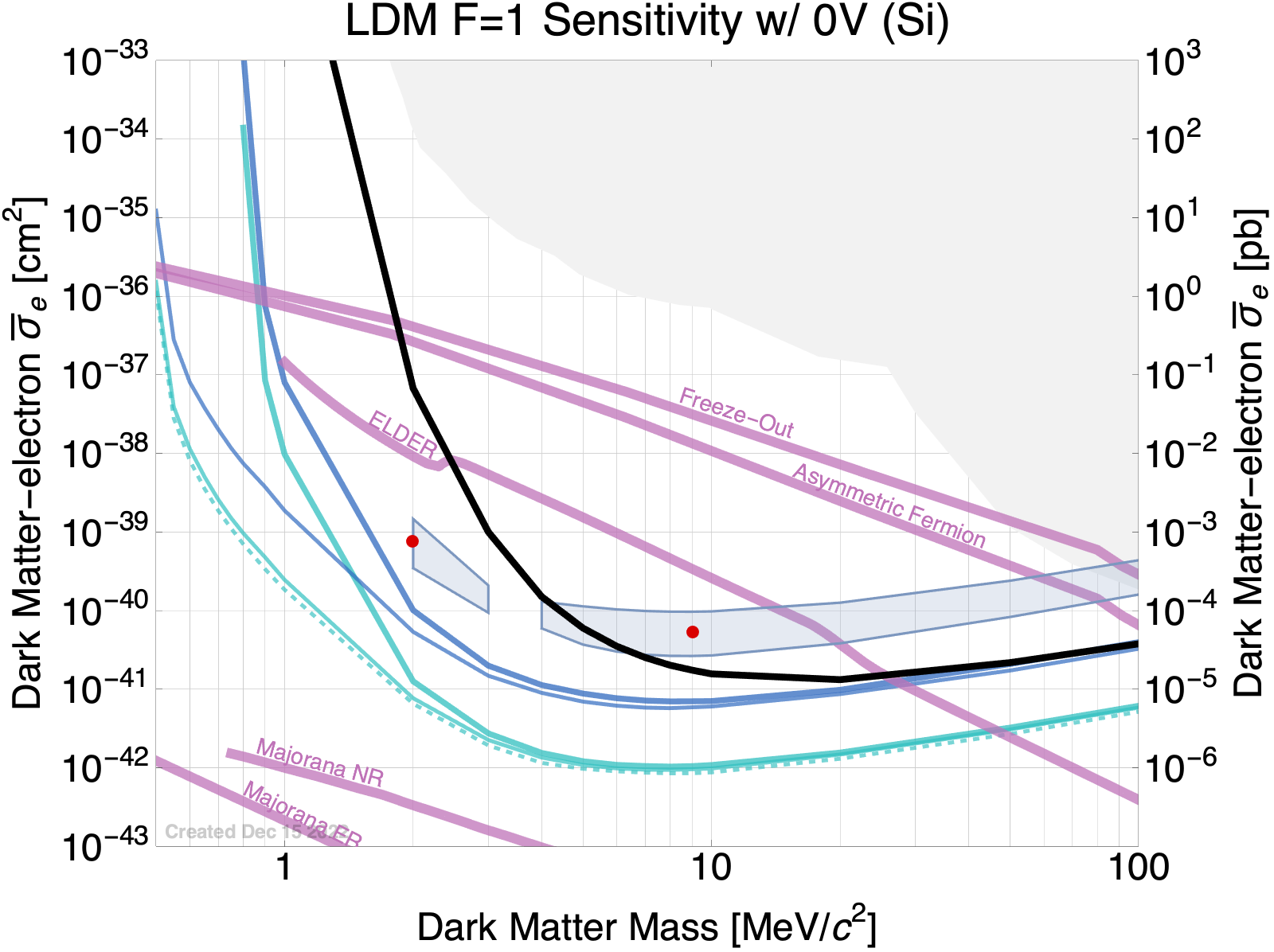} 
\hspace{0.02\textwidth}
\includegraphics*[width=0.47\textwidth,viewport=0 0 800 567]{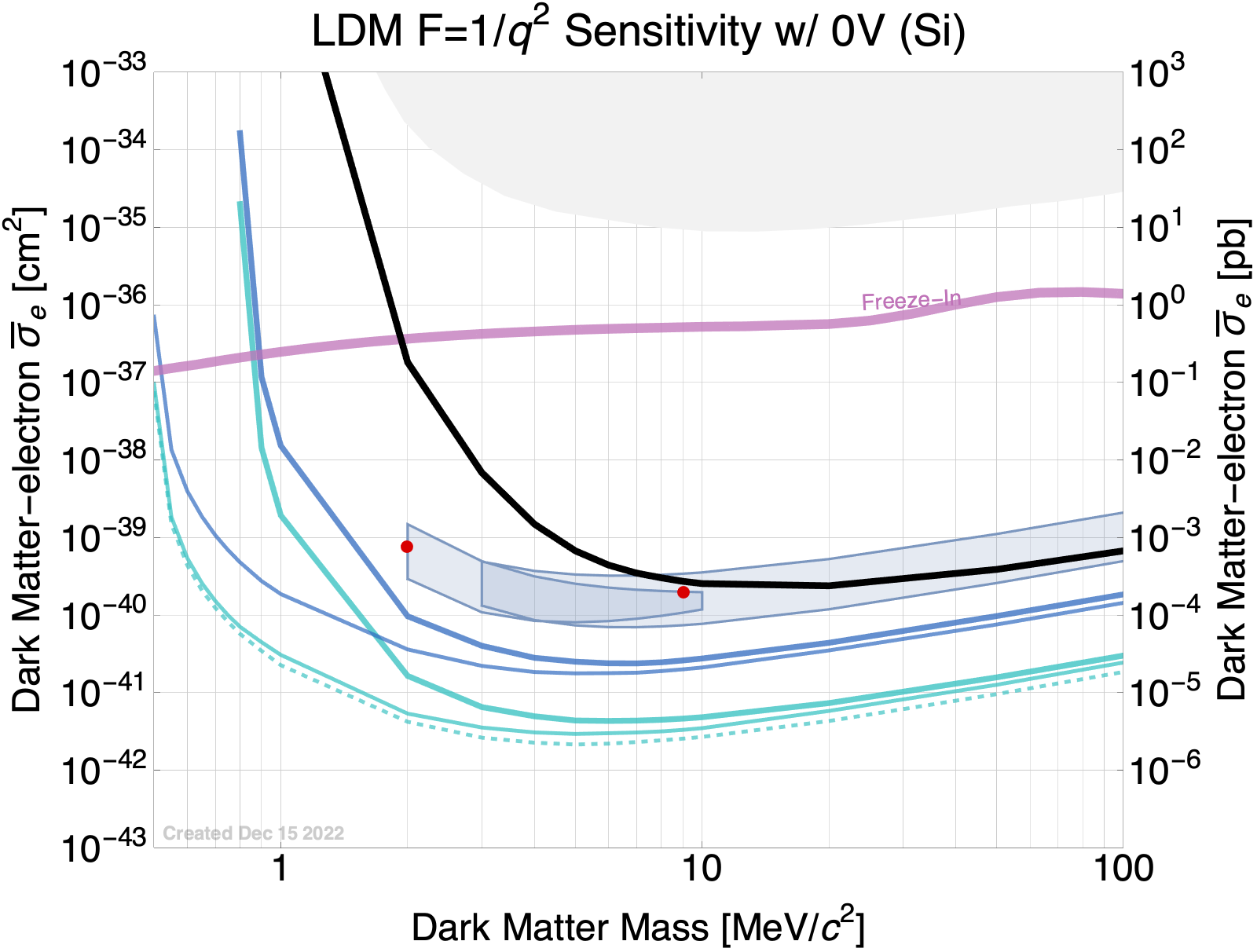}\\
\iffalse
\red{\textbf{Tarek:} See notes on prior LDM plot.} \\
\parbox[b]{3.2in}{\red{\footnotesize
\begin{tabular}{p{3in}} \\
\textbf{Discovery Potential Plot:} \\
\bkgon \detA 0V 1~\cmcu Si 30 \MeV Asymmetric DM line \\
\bkgon \detA 0V 1~\cmcu Si 10 \MeV ELDER line \\
\bkgon \detB 0V 1~\cmcu Si  3 \MeV ELDER line \\
\bkgon \detB 0V 1~\cmcu Si  1 \MeV ELDER line \\
\vspace{60pt} \\
\end{tabular}} \\} 
\parbox[b]{3.2in}{\red{\footnotesize
\begin{tabular}{p{3in}} \\
\textbf{Discovery Potential Plot:} \\
\bkgon \detA 0V 1~\cmcu Si 30 \MeV Freeze-in line \\
\bkgon \detA 0V 1~\cmcu Si 10 \MeV Freeze-in line \\
\bkgon \detB 0V 1~\cmcu Si  3 \MeV Freeze-in line \\
\bkgon \detB 0V 1~\cmcu Si  1 \MeV Freeze-in line \\
\vspace{60pt} \\
\end{tabular}} \\} 
\fi
\medskip
%\red{\footnotesize \textbf{Harrison} Could you add 1 MeV LDM mass to the plots? \\}
% Mar 2022
%\includegraphics*[width=0.46\textwidth,viewport=0 0 420 322]{spectra/0V1cm3_bkgA_detB_ES_escattering4_Si_postCuts_ER_v3.png}
%\hfill\includegraphics*[width=0.46\textwidth,viewport=0 0 420 322]{spectra/0V1cm3_bkgA_detB_ES_escattering6_Si_postCuts_ER_v3.png} \\
% 220930 changes:
% higher neutron bgnd
\includegraphics*[width=0.46\textwidth,viewport=0 0 420 322]{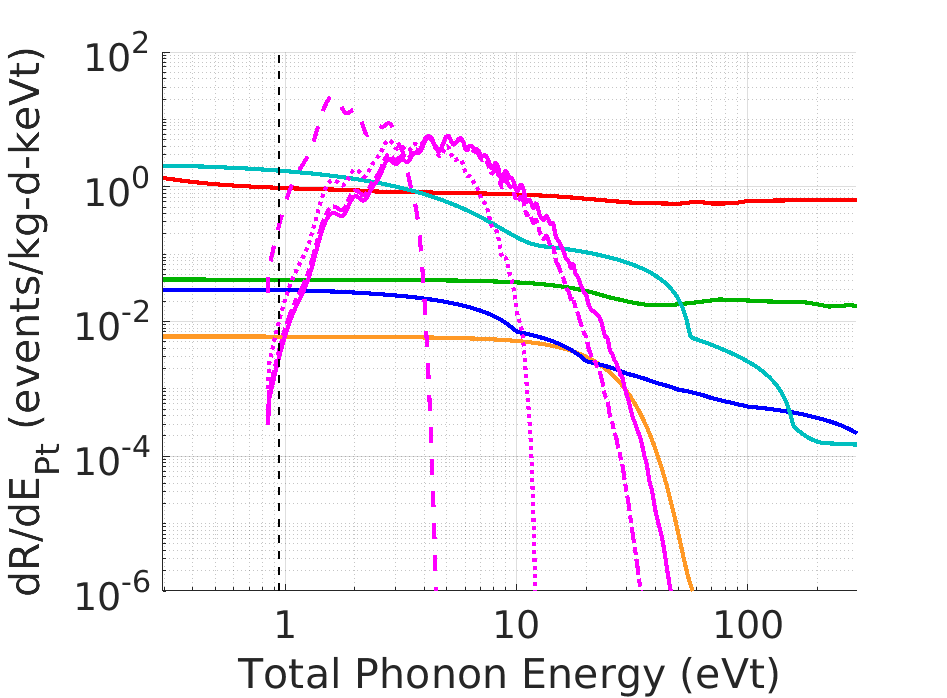}
\hfill
\includegraphics*[width=0.46\textwidth,viewport=0 0 420 322]{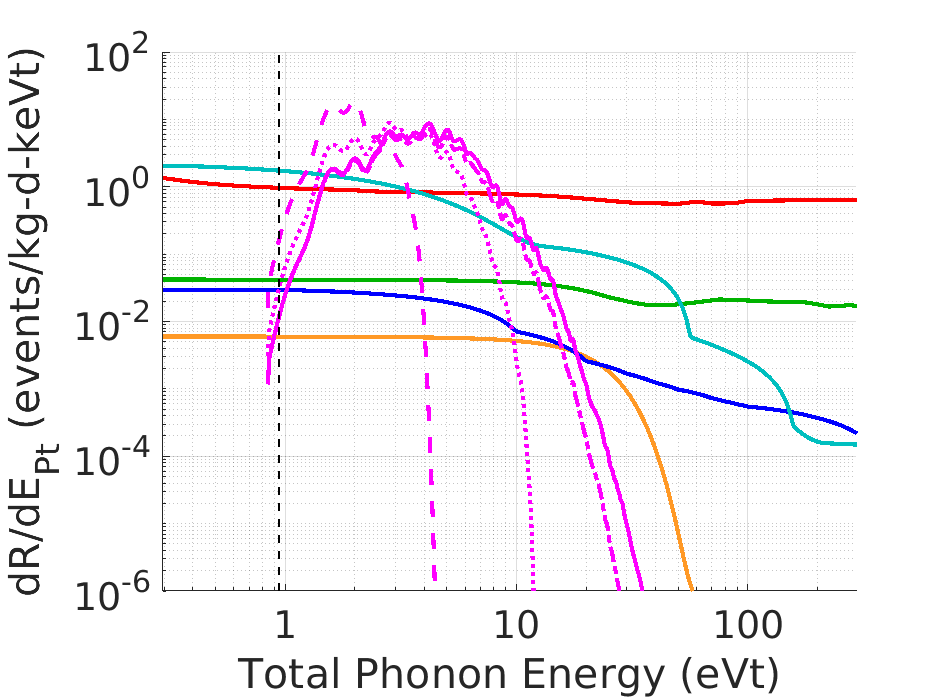} \\
%\red{Spectrum plot for Si 0V 1 \cmcu \bkgon \detB with backgrounds broken down, LDM spectra for three masses and both form factors} 
\end{center}
\caption[Sensitivity to dark-photon-coupled light dark matter (LDM) with 1~\cmcu and 10~\cmcu Si 0V detectors]{\textbf{Sensitivity to dark-photon-coupled light dark matter (LDM) with 1~\cmcu and 10~\cmcu Si 0V detectors} For all rows, the left column corresponds to $F(q) = 1$ and the right column to $F(q) = 1/q^2$.  The sensitivity of SuperCDMS SNOLAB (envelope of all detector types) is shown in thick black. (Top)~Median expected PLR-based 90\%~CL exclusion sensitivity, assuming ready-to-implement improvements in backgrounds (\bkgon scenario), for: two Towers of  1~\cmcu Si 0V detectors for the two nearer-term phonon-resolution improvements (\detA and \detBnosp; blue, thick and thin); the same two scenarios but with 20$\times$ larger payload mass (cyan, thick and thin); and, two Towers of 10~\cmcu Si 0V detectors for the \detC phonon-resolution scenario (cyan, thin, dashed).  This selection of curves optimizes sensitivity as a function of \detX scenario.  The \detA and \detB 1~\cmcunosp~Ge scenarios have much poorer sensitivity due to higher thresholds and so are omitted.  The \detA and \detB 1~\cmcunosp~Si scenarios are exposure-limited, hence we show the 20$\times$ mass cases also to demonstrate the reach possible at these states of detector evolution.  Lastly, 10~\cmcunosp~Si is shown for the \detC scenario because, at that level of detector advancement, the phonon resolution for the larger detector is sufficient to see the entire LDM recoil energy spectrum and the large single-detector mass is the most efficient way to obtain enough exposure to become background-limited.  Sharp science targets, the same as in Figure~\ref{fig:ldmhv_reach}, are again indicated.
We also show discovery potential for the two-Tower, \bkgonnosp, \detAnosp, 1~\cmcunosp~Si~0V scenarios, with legend as in Figure~\ref{fig:nrdm0v_reach}.  The elongation of the contours is due to the mass-cross-section degeneracy noted already.  The allowed region for the 2~\MeV case is artificially truncated at low mass because of the coarseness of the mass grid available from \qedarknosp; in actuality, the contour should close between 1 and 2~\MeVnosp.
(Bottom)~Example spectrum plots showing the background spectra for the \bkgonnosp, \detBnosp, 1~\cmcunosp~Si 0V case and candidate DM signal spectra.  Legend as in Figure~\ref{fig:snolab_reach_nrdm}.  The candidate DM masses are $\mathit{\mdm \approx 1}$, 3, 10, and 30~\MeVnosp.  The ripples in the DM spectra are due to  scatter in the recoil spectra provided by \qedarknosp, convolved with the resolution function, and  are not expected to significantly impact these forecasts.}
%\hrule
\label{fig:ldm0v_reach}
\end{figure}

Sensitivity for axion-like-particle DM utilizes the DPDM calculation, scaling it by a mass-dependent factor to undo the dielectric-loss-function dependence on energy, which applies for DPDM because of kinetic mixing but not for ALPDM.  Figure~\ref{fig:alpdm_reach} displays this reach for both HV and 0V detectors.  Astrophysical constraints substantially exceed the sensitivity of prior particle DM searches, excluding $g_{ae} \gtrsim 2\,\times\,10^{-13}$ up to about 10~\keV mass.  There is a hint from stellar cooling estimates that is consistent with the existence of an ALP in the range 1--$2\,\times\,10^{-13}$~\cite{stellarcooling2016}.  Reaching a sensitivity of $1\,\times\,10^{-13}$ is thus quite interesting, which is possible in some cases.  In all cases, the sensitivity substantially exceeds that of prior experiments, which had sensitivity only down to $g_{ae} \approx 10^{-12}$ near 100~\eV and provided much weaker constraints at lower masses.

\subparagraph*{SG-4A: ALPDM with HV Detectors} SNOLAB-sized HV detectors can have excellent reach for ALPDM above 3~\eVnosp, improving on current particle DM search constraints by 1--4 orders of magnitude in $g_{ae}$ (2--8 orders of magnitude in cross section).  The \detB and \detC scenarios for SNOLAB-sized Ge HV detectors can reach the $g_{ae} = 10^{-13}$ benchmark, exceeding astrophysical constraints and testing the stellar cooling hint.  
% 220927 Drop Bkg3 discussion
%The \bkgonnosp, \detC (and \bkgtwnosp, \detCnosp) scenario does not, however, have much margin, reaching down to $g_{ae} = 0.6\,\times\,10^{-13}$ at the lowest masses but degrading to $g_{ae} = 0.9\,\times\,10^{-13}$, while the \bkgthnosp, \detC scenario reaches to 0.3--0.5$\,\times\,10^{-13}$ for all masses in the 3--100~eV range.  While the \bkgth scenario offers more robust sensitivity, it is probably not warranted to consider it at this point given the high cost and the potential of the \bkgonnosp, \detC scenario to reach the benchmark.  

SNOLAB-sized Si HV detectors can exclude comparable amounts of parameter space untested by particle DM searches, though, with their comparable or slightly poorer mass reach and their similar or higher background levels, they are less sensitive than Ge detectors.  This modest degradation is enough to prevent them from  exceeding the important $g_{ae} = 10^{-13}$ sensitivity goal except over a very limited range of masses.  

\subparagraph*{SG-4B: ALPDM with 0V Detectors} 
%\marginpar{\red{\footnotesize Would like to explain the turnup in sensitivity at 3 eV better.}} 
0V detectors are also able to exclude significant ALPDM parameter space untested by particle DM searches.  Exceeding the $g_{ae} = 10^{-13}$ benchmark is, however, somewhat more challenging and comes with less margin.  The benchmark can be reached with 10~\cmcu Ge detectors in the \bkgonnosp, \detC scenario or with a 20$\times$ mass payload of 1~\cmcu Ge 0V detectors in the \bkgonnosp, \detB scenario.  The sensitivity of Si 0V detectors is always poorer than for Ge because there is no enhancement of dielectric response function to compensate for the higher bulk ER background and smaller exposure with Si.

\paragraph{SG-5: LDM with HV and 0V Detectors}
\label{sec:ldmhv0v}

We use \texttt{QEDark}~\cite{qedark} to calculate recoil spectra for dark-photon-coupled light dark matter scattering with electrons.  The gridding in mass is coarse, which is visible in discovery potential allowed regions.  Future work will use more sophisticated recoil spectrum calculations now available (EXCEED-DM~\cite{exceed-dm_2021}, DarkELF~\cite{knapen_darkelf}) and finer gridding.

\subparagraph*{SG-5A: LDM with HV Detectors}

Figure~\ref{fig:ldmhv_reach} shows the projected sensitivity for LDM with HV detectors.  For LDM, phonon energy resolution and ionization leakage improvements have impact on both mass and cross section reach.  Each upgrade yields a factor of 2 improvement in mass reach.  The gains in cross-section reach are dramatic because, as noted above, the spectrum is a falling exponential whose mean energy has very weak dependence on \mdmtxtnosp.  The gains are more dramatic for $F(q) = 1/q^2$ than for $F(q) = 1$ for the same reason.  For the heavy (light) mediator case, each upgrade scenario yields a cross-section sensitivity gain of 4--5 orders of magnitude for masses already probed, and of course substantially more for masses previously impacted by energy threshold.  The SNOLAB-sized, Ge HV \detB scenario tests the heavy mediator ELDER, SIMP, Elastic Scalar, and Asymmetric Fermion models over most of the 1--100~\MeV mass range.  

For LDM, the impact of background improvements is even smaller than for DPDM, with the improvement from \bkgon to \bkgth yielding about half an order of magnitude in cross section for the same reasons as noted above: the spectrum is a falling exponential and thus sensitivity is less dependent on the flat bulk ER background level.   Thus, \textbf{particle background improvements beyond \bkgon are not needed for HV detector reach for LDM}.  

\subparagraph*{SG-5B: LDM with 0V Detectors}

Immediately, in the \detA scenario, 1~\cmcu Si 0V detectors can probe significant new parameter space beyond current limits and SuperCDMS SNOLAB, testing ELDER, SIMP, Elastic Scalar, and Asymmetric Fermion (Freeze-in) benchmark models down to 1~\MeVnosp.  The \detB upgrade scenario yields roughly a factor of 2 reach to lower mass.  The \detC upgrade scenario is exposure-limited and offers no improvement over the \detB scenario.

As noted earlier, the LDM reach of 0V detectors is a much less continuous function of detector parameters than for HV detectors, and sensitivity is driven much more strongly by threshold than by any other parameter, and even more so for $F(q) = 1/q^2$ than $F(q) = 1$.  The larger, 10~\cmsq detectors have no sensitivity until the \detB scenario is realized, but at that point they have mass reach comparable to the \detA 1~\cmcu Si 0V scenario and cross-section reach comparable to its 20$\times$ mass option.  This scenario tests the heavy mediator ELDER, SIMP, Elastic Scalar, and Asymmetric Fermion models over most of the 1--100~\MeV mass range.  The 10~\cmcunosp~Si 0V \detC upgrade option offers the same mass reach as the \detB 1~\cmcu Si 0V option but again cross-section reach comparable to that found for the \detB 1~\cmcu Si 0V 20$\times$ mass option.

Ge detectors offer poorer reach than Si detectors: while the number of electrons per detector is larger, the phonon energy resolution is usually a factor of 2 worse for a given detector size and upgrade option.

As for HV detectors,  the impact of background improvements is quite small, no more than half an order of magnitude in cross section.   Thus, \textbf{particle background improvements beyond \bkgon are not needed for 0V detector reach for LDM}.

\clearpage

\section{Novel Directions}
\label{sec:novel_forecasts}

%The upgrade scenarios discussed above enable the SuperCDMS technology in the SNOLAB facility to probe many square decades of unexplored parameter space, covering 6 decades in mass (1--100~eV for DPDM and ALPDM, 1--100~\MeV for LDM, and 0.05--5~GeV for nucleon-scattering DM), reaching the neutrino fog in the 0.5--5~GeV mass range, and testing a variety of benchmark models and sharp targets.  
%We nevertheless 
We provide here more detail on status and plans for the novel directions discussed in \S\ref{sec:novel_directions}.  We group them by the development effort and discuss how they apply to the three different directions discussed:
\begin{description}
\item[Future Direction 1 (FD-1)] New channels and new targets for new regimes in energy deposition
\item[Future Direction 2 (FD-2)] Addressing ionization leakage with new materials
\item[Future Direction 3 (FD-3)] Diurnal modulation
\end{description}
While we do not have the experienced-based models of backgrounds and detector performance to make realistic sensitivity forecasts for these directions, their potential motivates deploying prototype detectors in the SuperCDMS SNOLAB facility to develop such models.  

%\red{Please shout if any activities have been missed here!}

\subsection{Diamond and Silicon Carbide: FD-1, FD-2, FD-3}

%\red{Nader: anything activity to add here for you?  Your theory papers were already discussed in \S\ref{sec:novel_directions}.}

SuperCDMS members at SLAC and Stanford (N.~Kurinsky, P.~Brink, R.~Partridge, B.~Cabrera) have been pursuing the development of diamond and SiC targets.  As noted earlier, these materials may offer numerous advantages: the high sound speeds offer a coupling to acoustic phonons at higher energy thresholds than other materials, the large gap may enable high-voltage operation with much smaller ionization leakage than Si and Ge, and intrinsic anisotropies in the crystal structure may offer diurnal modulation signatures in ionization yield and/or phonon production. In particular, SiC's multiple stable unit cells and highly polar structure make it an attractive target for optical-phonon-coupled DPDM~\cite{griffin_SiC_2021}.

This work is currently being pursued via Brink and Kurinsky in conjunction with C.~Kenney using DOE detector development funds at SLAC through the KA-25 program, which will result in first devices and demonstrations of phonon collection on these materials. Initial devices will use a SuperCDMS-style phonon sensor, with an eye to developments in alternative charge and phonon sensors.  Success with these prototypes would motivate a request for funding for prototype deployment in SuperCDMS SNOLAB.

\subsection{Lower-Gap Materials: FD-1, FD-3}

N.~Kurinsky (SLAC) is pursuing the development of meV gap materials based on magnetic ordering as part of the SPLENDOR project. Substrates with magnetically ordered lattices can have their bandgap tuned both intrinsically, by changing the admixture of elements in the unit cell, and extrinsically, by using a large magnetic field. Initial R\&D is pursuing both charge and phonon readout targeted primarily at DPDM and LDM searches. With additional neutron-scattering measurements, sensitivity can also be expanded to nuclear recoil searches. Materials currently under investigation include $\mathrm{Eu_5In_2Sb_6}$ with a gap of $\sim$40~meV, described more in~\cite{magnetic_order}, as well as newer materials such as $\mathrm{La_3Cd_2As_6}$, expected to have O(100~meV) gap.

This work is happening in collaboration with LANL, who undertake crystal growth and do materials characterization, and A.~Phipps at CSU East Bay, who has extensive experience in low-noise ionization readout development. Material response characterization is being undertaken at UIUC by P.~Abbamonte, with subsequent loss functions interpreted by a theory collaboration between LANL and Y.~Kahn at UIUC.

\subsection{Alternative Phonon Sensors: FD-1}

A number of groups in the SuperCDMS Collaboration have been exploring alternative phonon sensors that may provide advantages in sensitivity or multiplexability over TESs.  

One approach is Kinetic Inductance Detectors (KIDs)~\cite{jonas_arcmp}.  In these devices, breaking of Cooper pairs causes a change in the inductance of a thin superconducting film that is fashioned into a lumped-element or transmission-line LC resonator.  The change in inductance can be observed by coupling the resonator to a radiofrequency (RF) feedline and monitoring the complex transmission of the network at the KID's resonant frequency, \frtxtnosp.  Changes in \frtxt are proportional to the quasiparticle density, providing a direct measure of phonon energy collected in the film via Cooper-pair breaking.  The RF readout is highly amenable to multiplexing because the resonators have high intrinsic quality factor, $\Qr \gtrsim 10^4$, permitting many resonators to be connected to a single feedline.  Readout involves generating a frequency comb of tones at room temperature tuned to the array's \frtxt values, sending it through the array, amplifying with a cryogenic low-noise amplifier, and demultiplexing the comb at room temperature.  Each resonator can consist of a single metal film, with no quasiparticle trapping required and thus no issue of interface transmission.  There may also be a long-term benefit from the readout being, in principle, non-dissipative.  In practice, however, reaching this regime may require very low readout powers, necessitating more sophisticated readout schemes.  Groups at Caltech (S.~Golwala), FNAL (L.~Hsu), SLAC (N.~Kurinsky), and TRIUMF (W.~Rau) are pursuing this work, the last in collaboration with Institute N\'eel in Grenoble.

Another approach is Cooper-pair-box (CPB) qubits.   A Cooper-pair box is a sub-($\mu$m)$^3$ volume of superconductor coupled by a Josephson tunnel junction or junctions\footnote{A tunnel junction is a junction between two metals with a thin (few nm) insulating barrier to prevent standard current flow.  ``Josephson'' is only intended to indicate that Cooper pairs, as well as quasiparticles, can tunnel through the barrier.} ($C_J \sim$~fF) to a superconducting reservoir held at ground voltage.  In isolation, the CPB Hamiltonian's eigenstates are also charge eigenstates $n$~\cite{bouchiat1998}.  Introduction of anharmonicity via the Josephson junction(s) causes the eigenvalue spacing to become unequal, rendering any pair of states --- especially, $n = 0$ and 2 --- a two-level system uniquely addressable for quantum computational operations.  Many implementations monitor and/or control the CPB state by coupling the electric field of a superconducting resonator to the CPB electric dipole moment, which then enables the system to manifest the QIS Jaynes-Cummings Hamiltonian~\cite{blais2004, wallraff2004}.  This Hamiltonian permits ``quantum non-demolition'' (QND) monitoring of the CPB quantum state: the coupled resonator-CPB system's resonant frequency depends on the CPB charge state $n$ and level spacing $\hbar\,\omega_a$, yet, with appropriate design and operational choices, measuring the resonant frequency does not collapse the CPB state.  The mode of operation would be to use the reservoir as a phonon absorber, with tunneling of quasiparticles to the CPB creating the signal.  Such a scheme might incorporate quasiparticle trapping to enhance the tunneling probability.   This CPB-qubit shares the multiplexability of KIDs or any superconducting resonator.  This approach is being pursued at Caltech (Golwala, K.~Ramanathan) and SLAC (Kurinsky).

\subsection{Cooperative Development of Sapphire and Quartz with TESSERACT: \\ FD-1, FD-3}

The TESSERACT Collaboration~\cite{tesseract} is developing three different approaches for DM detection, all sharing phonon sensor heritage with SuperCDMS:
\begin{description}
\item[LHe target:] This approach uses a LHe target in which particle interactions can create photon-emitting excimers as well as phonons and rotons.  The photons would be detected by a SuperCDMS-like Si 0V detector instrumented with phonon sensors, residing in the liquid. The phonons and rotons would be detected by causing quantum evaporation of He atoms at the liquid surface, which would then impinge on a similar detector, now residing in the space above the liquid.  This approach is oriented toward detection of low-energy nuclear recoils, benefiting from the low mass of the He nucleus and from direct creation of phonons and rotons by DM scatters. 

\item[GaAs target:] This approach seeks to employ the scintillator GaAs, a polar crystal, to obtain sensitivity to DPDM and DP-coupled LDM.  Above roughly 1~eV energy deposition (about 1~\MeV LDM mass), the material benefits from the high scintillation yield of GaAs~\cite{derenzo_gaas}, providing two detection channels: phonons via sensors deposited on the GaAs target, and photons via a separate Si 0V light detector.  Two simultaneous detection channels may provide a way to reach the necessary eV threshold while vetoing environmental backgrounds expected at those energies.  Below $\sim$1~eV, the scintillation channel becomes unavailable but optical phonon creation in the polar crystal provides reach down to 9~keV LDM mass~\cite{griffin2020} (7~meV DPDM mass~\cite{knapen_gaas}).

\item[\sapphirenosp/\quartz targets:] This approach is similar to GaAs without scintillation: \sapphire and \quartz provide reach down to 5~keV LDM mass~\cite{griffin2020} (45 and 35~meV DPDM mass, respectively~\cite{knapen_darkelf}) via optical phonons. 

\end{description}
The LHe target experiment is termed HeRALD while the polar crystal target experiment goes under the rubric SPICE.  Development of phonon sensors, exploration of coupling to optical and acoustic phonons, and diurnal modulation approaches are areas of potential cooperation.  Additionally, SPICE could consider deployment in the SuperCDMS SNOLAB facility, where it could be easily accommodated, in principle.  Groups at UC Berkeley (M.~Pyle) and Texas A\&M University (R.~Mahapatra) are members of both TESSERACT and SuperCDMS SNOLAB, enhancing the potential for collaboration.

\subsection{Cooperative Mitigation of Low-Energy Particle and Environmental Backgrounds with TESSERACT: FD-1}

As noted earlier in \S\ref{sec:env_backgrounds_summary}, it is already evident that various low-energy particle and environmental backgrounds may dominate over conventional particle backgrounds as detectors reach eV sensitivity.  A good amount of this information has emerged from the SuperCDMS SNOLAB-TESSERACT nexus, where experience with operating prototype 0V detectors using SuperCDMS SNOLAB hardware and test facilities has prompted TESSERACT detector lead M.~Pyle to eliminate sources of secondary low-energy photons, such as cirlex mounting clamps, and to develop techniques to decouple both cryogenic systems and detectors from environmental vibrations (including their own cryocoolers).  Efforts were made in the design of the NEXUS\footnote{Northwestern EXperimental Underground Site, situated in the NuMI tunnel at FNAL} and CUTE\footnote{Cryogenic Underground TEst facility} test facilities, both belonging to SuperCDMS member institutions and with roles in SuperCDMS SNOLAB, to provide some level of vibration decoupling.  It is evident from data in hand that these efforts were helpful but that vibrations remain important.  While the coupling between SuperCDMS SNOLAB and TESSERACT is strongest due to the closely allied detector technology, numerous other experiments (e.g., CRESST, EDELWEISS, MINER, NUCLEUS, RICOCHET~\cite{excess2021whitepaper}) together form a community beginning to be profoundly concerned with these new backgrounds.

\clearpage

\section{Implementation}

A key conclusion of the above forecasting work is that significant background reductions are not needed to obtain exciting new science reach.  This immediately implies that the bulk of the existing SuperCDMS SNOLAB facility is ready, as-is, for installation of new generations of detectors. Moreover, the modularity of SuperCDMS SNOLAB's detector deployment system --- the detector Towers --- makes installation of prototype detectors for pre-testing, as well as installation of new elements of the detector payload, relatively straightforward and undisruptive.  This mode of operation was in fact employed to upgrade from the CDMS~II experiment to SuperCDMS Soudan by replacing the original payload with the first deployment of iZIP detectors.
%The availability of two underground test facilities outfitted for these Towers, NEXUS and CUTE, offers the ability to test prototype detectors at key points.
The availability of a deep underground test facility outfitted for these Towers, CUTE\footnote{CUTE is an underground cryogenic test facility at SNOLAB.  It is compatible with SuperCDMS SNOLAB Towers and offers a background level of $<$7 events/kg/keV/day.  Above-ground testing of detectors is limited by pileup and high backgrounds, while CUTE offers a low-background environment in which most detector performance characteristic can be fully evaluated.}, offers the ability to test prototype detectors at key points.

%\subsubsection{Expected Funding Opportunities}
\subsection{Expected Funding Opportunities}

% $10M may be too large a number -- SuperCDMS SNOLAB Tower subsystem
% will be $11M, but that included a lot of first-time engineering
% we may not need to redo.  Also, we probably need collaboration ok
% for stating a number, and stating a number that is not well thought
% out may put us at a disadvantage in the future.
%The construction cost of a full detector payload upgrade project will likely be at the \$10M level, large enough that a new funding solicitation for dark matter experiments in this cost range will be required. 
The construction cost of a full detector payload upgrade project will likely be large enough that a new funding solicitation for dark matter experiments will be required. 
If the last decade is taken as guidance, such a solicitation might occur twice in the coming decade.  The first would ensue from the Snowmass/P5 process that provides the context for this white paper, in the vein of the DOE/NSF \textit{Second Generation Dark Matter Experiments} program that funded ADMX, LZ, and SuperCDMS SNOLAB soon after P5 in 2014.
% Prisca suggest avoiding accidentally implying we are suggesting a P5 downselection process.  We just want P5 to say there should be a new DM experiment solicitation.
%\footnote{The analogy is not perfect; the G2 FOA actually came out in 2012, prior to P5, and offered 1 year of \rnd and pre-conceptual design funding, but P5's high-level scientific guidance was used in the down-selection that occurred following P5, and P5's endorsement for a significant expansion of G2 funding was also followed.}.  
P5 is expected to deliver its report in mid-2023, in time to impact FY2025 budgets~\cite{doe_hepap_202203}.  Thus, that year would be the earliest opportunity for a ``small dark matter experiment'' solicitation, though it is possible that the agencies would, as in the last decade, provide limited \rnd funding in FY2024~\cite{doe_hepap_202111}.  (The significantly different cost scale of a potential G3 liquid noble experiment, well above \$100M, would necessitate a separate and independent funding process.)  A second solicitation might follow a few years later after a mid-decade review of the dark matter phenomenology landscape and experimental portfolio, much like the \textit{Cosmic Visions 2017} workshop~\cite{cvdm2017} led to the \textit{Basic Research Needs Study on Dark Matter New Initiatives} (DMNI) in 2018~\cite{dmbrn2018} and from there to the DOE DMNI solicitation in 2019.  

%In terms of integrating the generic upgrade plan discussed above
In terms of integrating an upgrade plan with the above funding expectations, a central question is: What level of maturity should an upgrade plan achieve to be ready for a solicitation?  Again, past history provides some guidance.  The Dark Matter Generation 2 process included 1 year of \rnd funding in FY2013. 
% again, avoid the down-selection implication
% for a broader portfolio of experiments prior to down-selection.  
The DMNI solicitation explicitly included a track with a a 2-year technical \rnd phase followed by 2 years of project development prior to entering formal project phase.  The conclusion to draw is that the initial proposal need not be risk-free and based on a full underground detector demonstration; those demonstrations can come later during the pre-CD-3 steps of the project, and perhaps are not necessary at all if the project plan is sufficiently sound.  

\clearpage

%\subsubsection{Upgrade Plan}
\subsection{Potential Upgrade Plan and Timeline}

Given the above funding expectations, we stipulate that the nearest-term upgrade should be competitive for a FY2025 funding solicitation and that we should also be prepared for a mid-decade solicitation in FY2030.  
%Incorporating the cost-benefit discussion regarding underground testing, we anticipate the following sequence:
An upgrade timeline that meshes with that sequencing is shown below. %in Figure~\ref{fig:timeline}

%\begin{figure}|[b!]
\begin{center}
\vspace{-6pt}
\includegraphics*[width=\textwidth,viewport=68 155 538 722,page=4]{tables/LTPSnowmassTable221001.pdf}
\end{center}
\vspace{-6pt}

\clearpage

\section*{Acknowledgements}

The SuperCDMS collaboration gratefully acknowledges SNOLAB and its staff for their assistance in developing the design for SuperCDMS SNOLAB.  Funding and support toward the work presented here were received from the National Science Foundation, the U.S. Department of Energy (DOE), Fermilab URA Visiting Scholar Grant No.~15-S-33, NSERC Canada, the Canada First Excellence Research  Fund, the Arthur B. McDonald Institute (Canada), the Department of Atomic Energy (DAE) and the Department of Science and Technology (DST) of the Government of India, the DFG (Germany; Project No.~420484612), and under Germany’s Excellence Strategy - EXC 2121 ``Quantum Universe" – 390833306.   SNOLAB operations are supported by the Canada Foundation for Innovation and the Province of Ontario, with underground access provided by Vale Canada Limited at the Creighton mine site.  Femilab is operated by Fermi Research Alliance, LLC,  SLAC is operated by Stanford University, and PNNL is operated by the Battelle Memorial Institute for the U.S. Department of Energy under contracts DE-AC02-37407CH11359, DE-AC02-76SF00515, and DE-AC05-76RL01830, respectively.  Most of the computations presented here were conducted in the Resnick High Performance Computing Center, a facility supported by Resnick Sustainability Institute at the California Institute of Technology.

\renewcommand{\em}{\it}
\bibliographystyle{prsty_title}
% PLEASE ADD ALL BIBLIOGRAPHY INFO TO ltptf_bib SO THAT WE CAN MAINTAIN CONSISTENCY BETWEEN INTERNAL AND EXTERNAL REPORTS.  OVERLEAF LINK SYNCHRONIZES THEM.
\bibliography{ltptf}

\clearpage

\appendix

\section{Forecasting Procedure}
\label{sec:forecast_procedure_detail}

\subsection{Low-Energy Particle Backgrounds}
\label{sec:low_energy_particle_backgrounds}

Coherent photonuclear scattering is, as first pointed out by \cite{robinson2017}, a process not modeled in standard GEANT4 simulations but one that should be included as we consider energy depositions in the eV range: \cite{robinson2017} shows that the background from this effect is comparable in rate to Compton scattering in the 10--100~eV range for Ge, exceeds a flat extrapolation of Compton scattering in the 1--10~eV range by 1--2 orders of magnitude for Ge and Si, and continues to rise dramatically below 1~eV.  Work has been ongoing in the literature to explore other mechanisms.  For example, \cite{berghaus_essig_coherent_photon_scattering_phonons_2022} show that photon-ion scattering, in particular coherent Rayleigh scattering of photons with electrons, causes direct phonon creation in the 1--100~meV range.  This previously unconsidered process yields a new and significant background at these low energies, though it can be mitigated by an active veto.

Another  class of backgrounds not currently modeled in GEANT4 is secondary photon generation from  transition radiation and Cerenkov radiation as well as luminescence photons and phonons produced by recombination, as explored in \cite{du_essig_secondary_photon_bgnd_2022}.  This work showed that such secondaries could explain eV-energy background levels in SuperCDMS gram-scale prototype detectors~\cite{Amaral:2020ryn}.  While it is not clear whether this model was correct in detail, a later reworking of the detector packaging, which removed nearby fiberglass (FR4) material, resulted in significantly lower background rates.  The implication is that eV-scale secondaries, generated by high-energy photons emitted from radiocontamination in the FR4, were indeed important in this case.   It should be noted, however, that the SuperCDMS SNOLAB detector hardware uses much lower radiocontamination materials, as well as much less dielectric material, than~\cite{Amaral:2020ryn}.  Thus, while such backgrounds should be considered, the dramatically high backgrounds of \cite{Amaral:2020ryn} are not applicable to the SuperCDMS SNOLAB experimental apparatus.

For our work here, we include an approximate model of coherent photonuclear scattering.  We do not consider coherent photon-electron Rayleigh scattering because the lowest threshold we consider is $\sim 0.5$~eV (Table~\ref{tbl:detector_upgrade_scenarios}; \S\ref{sec:nrdm_0v}).  We do not yet consider secondary photon backgrounds, though that is an area of active simulation development work that should soon result in their inclusion.

\subsection{Continuous and Discrete Treatments of Ionization Production}

We use two different treatments for handling ionization production and NTL phonon generation depending on the type of detector.  For iZIP and piZIP detectors, the effective ionization resolution is not good enough to resolve single electron-hole pairs, so we treat ionization production as a continuous variable and assume statistical fluctuations are smaller than measurement noise.  For 0V detectors, ionization production is irrelevant.  For HV detectors, on the other hand, we approach or achieve sufficient energy resolution to resolve single electron-hole pairs thanks to NTL amplification, and thus we must treat ionization production as discretized.  

\subsection{Threshold and Ionization Leakage}
\label{sec:threshold}
\label{sec:ionization_leakage}

For 0V, iZIP, and piZIP detectors, we assume a conservative, experienced-based trigger threshold of $7\sigpt$.  For HV detectors, ionization leakage (\S\ref{sec:ionization_leakage_upgrades}) causes substantial event rates in the first two to three electron-hole-pair peaks.  In current data, the spectrum in the region between the peaks is not understood in terms of known backgrounds and leakage convolved with impact ionization (II) and charge trapping (CT) effects (\S\ref{sec:ionization_leakage_upgrades}).  We attribute this to an insufficient understanding of II and CT and/or presence of unmodeled low-energy particle and environmental backgrounds.  We thus conservatively set an analysis threshold just above the first peak at which the leakage rate exceeds by a factor of 10 the continuum background (evaluated at 1~k\evrnosp).

\subsection{Ionization Yield}
\label{sec:ionization_yield}

For electron recoils in Ge and Si, we assume that the mean energy to create an electron-hole pair is $\epseh = 3.0$~eV in Ge (3.8~eV in Si).  For nuclear recoils in Ge, we use the model of Sarkis~et al.~\cite{sarkis_yield2020}, which approaches Lindhard above 1~k\evnr and fits most existing data but incorporates a physically 
motivated cutoff at tens of \evnrnosp.  For Si, we use the recent IMPACT direct measurement of nuclear recoil ionization yield.\footnote{These results are in preparation for publication.  IMPACT is described in, e.g.,~\cite{supercdms_impact}.}  No cutoff in yield at low recoil energy is imposed.

\subsection{Ionization Production and Collection}
\label{sec:ionization_production}
\label{sec:ionization_collection}

For the continuous model, a given recoil energy and type results in a specific value of ionization production, a continuous variable.  We define mean values of ionization collection for various regions in the detector and assign each a fraction of the volume: bulk (low-$r$ and $z$), $z$ surface, and $r$ surface.  These regions have slightly different volumes for bulk electron and nuclear recoils.  Surface electron and nuclear recoils are assigned only to the surface $z$ or surface $r$ regions.  Given  the ionization yield and collection for each event type and region, we are able to define a mapping from input recoil energy spectra (\ertxtnosp, from a dark matter model or particle Monte Carlo simulations) to spectra in phonon energy (\epttxtnosp).

For the discrete model, which is used only for HV detectors, we first account for Fano (statistical) fluctuations in ionization production by using the ionization yield function to provide the mean of the appropriate statistical distribution (Gaussian or binomial), and then we distribute the rate for a given \ertxt bin among a range of values of \nehtxt based on these distributions.  For bulk and $z$ surface events, we account for charge trapping and impact ionization during propagation and NTL phonon generation using full probability distribution functions for $\neh < 10$ and using mean behavior for $\neh \ge 10$.  For $r$ surface events, we account for the trapping of charges at the radial sidewall, using a uniform PDF between 0 and 1/2 for the ionization collection and thus NTL phonon generation.  We do not account for charge trapping and impact ionization because the PDF is already quite smooth.  All these statistical, charge trapping, and impact ionization effects break the one-to-one mapping between \ertxt and \epttxt and distribute events away from the mean \epttxt that would be expected for a given recoil type and \ertxtnosp.  

\subsection{Application of Position Fiducialization and Ionization Yield Cuts and Evaluation of their Acceptance}

For both the continuous and discrete ionization cases and for most detector types, we use position information to define a fiducial volume that excludes as much of the surface backgrounds as is feasible.\footnote{We do not do so for 0V detectors because the dominant backgrounds are bulk ER, neutron, or neutrino events, which are spatially distributed in the detector in the same manner as dark matter events.  These detectors rely entirely on spectral shape discrimination between dark matter and backgrounds.}  For iZIP detectors, we do this on the basis of the asymmetry of the ionization and phonon collection on the two sides of the detector and the partitioning between radial inner and outer electrodes/sensors.  For piZIP detectors, we do the same but only using the phonon signal.  For HV detectors, we use the radial partitionining of the phonon signal, which, being dominated by NTL phonons, reflects the radial partitioning of ionization collection.

For the continuous ionization case, we assume mean values, derived from prior data, for these fiducialization parameters for the three different regions and for the different types of recoils.  We calculate the measurement noise on the fiducialization parameters as a function of these mean values.  The noise derives from the total phonon and ionization energy resolution numbers provided in Table~\ref{tbl:detector_upgrade_scenarios}.  We then set cuts designed to provide 2$\sigma$ acceptance (efficiency) for bulk nuclear recoils and calculate the acceptance (misidentification) of these cuts for all backgrounds in the three regions.   We do the same for ionization yield based on the assumed ionization yield functions for nuclear and electron recoils, the noise on the parameter (derived from the ionization and phonon energy resolutions (iZIP) or the phonon energy resolution alone (piZIP)).  We multiply the raw spectra (as a function of \epttxtnosp) by these energy-dependent acceptances to arrive at the final processed spectra that will be fed into the simulations used to determine expected sensitivity.

For the discrete ionization case, we endeavor to calculate the full distributions of $z$ and $r$ fiducialization parameters based on geometrical models for the propagation of phonons.  (There is no ionization yield cut, of course.)  Phonons are only assumed to carry position information up to their first bounce off a detector surface, and the fraction of phonons that are absorbed on their first bounce is based on prior data.  The models use the solid angle subtended by the sensor from the phonon emission point and the $1/r^2$ fall-off in phonon flux.  The recoil phonons are assumed to be emitted from the interaction point, while the NTL phonons are assumed to be emitted isotropically from the entire drift track.  Thus, the $z$ fiducialization parameter depends only on the recoil phonon energy (\ertxtnosp) while the $r$ fiducialization parameters depends on both recoil and NTL phonon energies (differently).  The $z$ parameter, which is linear in $z$, is assumed to have a uniform PDF to reflect the uniform PDF in $z$ event position for bulk events.  Its maximum value is the product of the ``first-bounce'' fraction and the recoil energy.  The PDF for the $r$ parameter is determined by calculating the $r$ parameter's value at every point in the detector (using the same maximum first-bounce fraction and now the geometrical models for recoil and NTL phonon propagation as a function of $(r,z)$), and then assuming a uniform distribution of dark matter events over the detector and marginalizing over $(r,z)$ to obtain the PDF.  The PDFs are smeared by the energy-dependent resolutions in the two parameters and the cuts set to obtain 90\% acceptance for the $z$ cut and 95\% for the $r$ cut.  

In the discrete ionization case, the \textit{actual} acceptance for signal and backgrounds for these cuts is evaluated in a much more sophisticated manner.  A number of effects have to be accounted for:
\begin{itemize}
\item For a given value of recoil energy (\ertxtnosp), Fano fluctuations lead to a distribution of potential \nehtxt values.  Even with full ionization collection, the result is a distribution of NTL phonon energy (\entltxtnosp) and thus total phonon energy (\epttxtnosp) values for a given \ertxtnosp.
\item Charge trapping and impact ionization then cause even a single value of \nehtxt to give rise to a distribution of \entltxt and thus \epttxtnosp.
\item The acceptance of the $z$ and $r$ fiducialization cuts for $r$ surface events, for even a single \ertxt bin and \nehtxt value, depends additionally on \entltxt because \entltxt varies with the $z$ position of these events as noted above.  This causes a modulation of the acceptance for such backgrounds as a function of \epttxt at fixed \ertxt and \nehtxtnosp.  The situation is less complicated for bulk events  because we defined the cut to have an energy-independent fixed acceptance for bulk events.  We can extend the same assumption to $z$ surface events because the $\er \ll \entl$ and it is the former that determines $z$ asymmetry.
\end{itemize}
The distribution of rate from a given \ertxt bin to a range of \epttxt bins, including the modulation of the acceptance of $r$ surface events as a function of energy, is calculated by an appropriate set of convolutions, stepping in \ertxt and then looping over all potential values of \nehtxt to accumulate the output spectra as a function of phonon energy.

\subsection{Generation of Experimental Realizations and Determination of Expected Sensitivity}

With the above spectra for dark matter and backgrounds, as a function of total phonon energy (\epttxtnosp) and accounting for the acceptance of the fiducial volume cut and, where necessary, the ionization yield cut, we may generate simulated event spectra for a specific exposure.  To determine exclusion sensitivity, we generate these spectra with no dark matter signal added.  To assess discovery potential, we add a known dark matter signal with sufficiently large cross section to obtain allowed regions that exclude zero cross section at approximately 3$\sigma$ (99.7\%) CL.

To determine allowed regions for a given experimental realization, we use a standard profile likelihood-ratio method.  That is, we consider the likelihood ratio
\begin{align}
R(s) = \frac{\Lcal\left(d\, \Big|\, s, \bhathat\right)}
            {\Lcal\left(d\, \Big|\, \shat, \bhat\right)}
\end{align}
where $s$ is the signal model being tested, the pair \shattxtnosp, \bhattxt is the set of signal and background model parameters that maximizes the likelihood (best fits the data), and \bhathattxt is the set of background model parameters that maximizes the likelihood when the signal model parameters are fixed to the values being tested, $s$.  The signal model has two parameters, mass and cross section.  The background parameters are the normalizations of three components of the background: neutrinos, neutrons, and the sum of all other backgrounds.  The likelihood includes a 10\% rms Gaussian prior on these normalizations.  The quantity $D(s) = -2 \ln R(s)$ is assumed to follow a $\chi^2$ distribution for two degrees of freedom, the mass and overall rate parameters of the signal model, $s$.  A model $s$ is in the X\%~CL allowed region for a given experimental realization $d$ if its $D$ statistic is among the X\% most likely (smallest) values of the $\chi^2$ distribution.  For each mass, we search in $s$ to find the boundaries of the X\%~CL allowed region.  There may be only one value, yielding an upper limit, or there may be two values, excluding zero cross section and providing lower and upper bounds on the cross section.

We average over realizations (typically, 100--200) as follows:
\begin{itemize}
\item For exclusion limits, we simply take the median upper limit at each mass.
\item For discovery cases, the situation is more challenging because most realizations yield a closed contour that moves around from realization to realization.  (The contour closes at any mass for which the upper and lower limits coincide.  For masses outside the contour, the upper and lower limits are undefined.)  Each such contour is representative of the constraint that one would obtain in a given experimental realization.  Therefore, we plot the closed contour for the experimental realization whose best fit model $s$ is closest to the input model, effectively centering its contour on the input model.  We find that this contour contains the realizations at a level consistent with the CL chosen.  The contour thus displays a representative allowed region while also indicating how much it might move around.
\end{itemize}
These two choices are approximately consistent in the sense that, for an exclusion limit, the majority (but not necessarily all) best-fit models will have zero cross section.  All of these are equally close to the zero cross section input model, and therefore the median upper limit is almost certainly within this group.

\end{document}

%% file: ltptf_macros.tex
\newcommand{\red}[1]{{\textcolor{red}{#1}}}

\newcommand{\qedarknosp}{\texttt{QEDark}}
\newcommand{\qedark}{\qedark\ }

\newcommand{\rndnosp}{R\&D}
\newcommand{\rnd}{\rndnosp\ }
\newcommand{\bkgnosp}{\texttt{Bkg}}
\newcommand{\bkg}{\bkgnosp\ }

\newcommand{\bkgonnosp}{\texttt{Bkg~1}}
\newcommand{\bkgon}{\bkgonnosp\ }
\newcommand{\bkgtwnosp}{\texttt{Bkg~2}}
\newcommand{\bkgtw}{\bkgtwnosp\ }
\newcommand{\bkgthnosp}{\texttt{Bkg~3}}
\newcommand{\bkgth}{\bkgthnosp\ }
\newcommand{\detXnosp}{\texttt{Det}}
\newcommand{\detX}{\detXnosp\ }
\newcommand{\detAnosp}{\texttt{Det~A}}
\newcommand{\detA}{\detAnosp\ }
\newcommand{\detBnosp}{\texttt{Det~B}}
\newcommand{\detB}{\detBnosp\ }
\newcommand{\detCnosp}{\texttt{Det~C}}
\newcommand{\detC}{\detCnosp\ }

\newcommand{\sapphirenosp}{Al$_2$O$_3$}
\newcommand{\quartznosp}{SiO$_2$}
\newcommand{\cawofournosp}{CaWO$_4$}

\newcommand{\sapphire}{\sapphirenosp\ }
\newcommand{\quartz}{\quartznosp\ }
\newcommand{\cawofour}{\cawofournosp\ }

\usepackage{xspace}
\newcommand{\cevnsnosp}{CE$\nu$NS}
\newcommand{\cevns}{\cevnsnosp\xspace}
\newcommand{\mdm}{M_\chi}
\newcommand{\mdmtxtnosp}{$\mdm$}
\newcommand{\mdmtxt}{\mdmtxtnosp\ }
\newcommand{\gae}{g_{ae}}
\newcommand{\gaetxtnosp}{$\gae$}
\newcommand{\gaetxt}{\gaetxtnosp\ }

\newcommand{\hthreenosp}{$^{3}$H}
\newcommand{\besevennosp}{$^{7}$Be}
\newcommand{\beightnosp}{$^{8}$B}
\newcommand{\cftnosp}{$^{14}$C}

\newcommand{\sittnosp}{$^{32}$Si}
\newcommand{\artnnosp}{$^{39}$Ar}

\newcommand{\pbtsnosp}{$^{206}$Pb}
\newcommand{\pbttnosp}{$^{210}$Pb}
\newcommand{\uttenosp}{$^{238}$U}
\newcommand{\thtttnosp}{$^{232}$Th}

\newcommand{\hthree}{\hthreenosp\ }
\newcommand{\beseven}{\besevennosp\ }
\newcommand{\beight}{\beightnosp\ }
\newcommand{\cft}{\cftnosp\ }

\newcommand{\sitt}{\sittnosp\ }
\newcommand{\artn}{\artnnosp\ }

\newcommand{\pbtt}{\pbttnosp\ }

\newcommand{\utte}{\uttenosp\ }

\newcommand{\cmsqnosp}{cm$^2$}
\newcommand{\cmsq}{\cmsqnosp\ }
\newcommand{\cmcunosp}{cm$^3$}
\newcommand{\cmcu}{\cmcunosp\ }
\newcommand{\mcunosp}{m$^3$}

\newcommand{\bqcmsqnosp}{Bq/\cmsqnosp}
\newcommand{\bqcmsq}{\bqcmsqnosp\ }
\newcommand{\bqmcunosp}{Bq/\mcunosp}

\newcommand{\evrnosp}{eV$_r$}
\newcommand{\evr}{\evrnosp\ }
\newcommand{\evnrnosp}{eV$_{nr}$}
\newcommand{\evnr}{\evnrnosp\ }
\newcommand{\eveenosp}{eV$_{ee}$}

\newcommand{\GeVnosp}{GeV/c$^2$}
\newcommand{\MeVnosp}{MeV/c$^2$}
\newcommand{\keVnosp}{keV/c$^2$}
\newcommand{\eVnosp}{eV/c$^2$}
\newcommand{\meVnosp}{meV/c$^2$}

\newcommand{\GeV}{\GeVnosp\ }
\newcommand{\MeV}{\MeVnosp\ }
\newcommand{\keV}{\keVnosp\ }
\newcommand{\eV}{\eVnosp\ }
\newcommand{\meV}{\meVnosp\ }

\newcommand{\emaxldm}{E_\text{max}^\text{LDM}}
\newcommand{\emaxldmtxtnosp}{$\emaxldm$}
\newcommand{\emaxldmtxt}{\emaxldmtxtnosp\ }
\newcommand{\ephonon}{E_\text{phonon}}
\newcommand{\ephonontxtnosp}{$\ephonon$}

\newcommand{\er}{E_\text{r}}
\newcommand{\ertxtnosp}{$\er$}
\newcommand{\ertxt}{\ertxtnosp\ }

\newcommand{\ept}{E_\text{pt}}
\newcommand{\epttxtnosp}{$\ept$}
\newcommand{\epttxt}{\epttxtnosp\ }

\newcommand{\nleak}{N_\text{leak}}
\newcommand{\nleaktxtnosp}{$\nleak$}

\newcommand{\neh}{N_\text{eh}}
\newcommand{\nehtxtnosp}{$\neh$}
\newcommand{\nehtxt}{\nehtxtnosp\ }

\newcommand{\vb}{V_\text{b}}
\newcommand{\vbtxtnosp}{$\vb$}
\newcommand{\vbtxt}{\vbtxtnosp\ }
\newcommand{\evb}{e\vb}
\newcommand{\evbtxtnosp}{$\evb$}
\newcommand{\evbtxt}{\evbtxtnosp\ }
\newcommand{\sigpt}{\sigma_\text{pt}}
\newcommand{\sigpttxtnosp}{$\sigpt$}
\newcommand{\sigpttxt}{\sigpttxtnosp\ }

\newcommand{\fsig}{f_\sigma}
\newcommand{\fsigtxtnosp}{$\fsig$}
\newcommand{\fsigtxt}{\fsigtxtnosp\ }
\newcommand{\entl}{E_\text{NTL}}
\newcommand{\entltxtnosp}{$\entl$}
\newcommand{\entltxt}{\entltxtnosp\ }

\newcommand{\epseh}{\epsilon_\text{eh}}
\newcommand{\epsehtxtnosp}{$\epseh$}
\newcommand{\epsehtxt}{\epsehtxtnosp\ }

\newcommand{\sigt}{\sigma_\text{t}}
\newcommand{\sigttxtnosp}{$\sigt$}

\newcommand{\fdisc}{f_\text{disc}}
\newcommand{\fdisctxtnosp}{$\fdisc$}
\newcommand{\fdisctxt}{\fdisctxtnosp\ }

\newcommand{\sigqpizip}{\sigma_\text{Q}^\text{piZIP}}
\newcommand{\sigqpiziptxtnosp}{$\sigqpizip$}
\newcommand{\sigqpiziptxt}{\sigqpiziptxtnosp\ }

\newcommand{\Tc}{T_\text{c}}
\newcommand{\Tctxtnosp}{$\Tc$}
\newcommand{\Tctxt}{\Tctxtnosp\ }

\newcommand{\Lcal}{\mathcal{L}}

\newcommand{\shat}{\widehat{s}}
\newcommand{\shattxtnosp}{$\shat$}

\newcommand{\bhat}{\widehat{b}}
\newcommand{\bhattxtnosp}{$\bhat$}
\newcommand{\bhattxt}{\bhattxtnosp\ }
\newcommand{\bhathat}{\widehat{\bhat}}
\newcommand{\bhathattxtnosp}{$\bhathat$}
\newcommand{\bhathattxt}{\bhathattxtnosp\ }

\newcommand{\fr}{f_r}
\newcommand{\frtxtnosp}{$\fr$}
\newcommand{\frtxt}{\frtxtnosp\ }
\newcommand{\Qr}{Q_r}

%% file: authors.tex
\author[1,2]{M.F.~Albakry} \affil[1]{Department of Physics \& Astronomy, University of British Columbia, Vancouver, BC V6T 1Z1, Canada} \affil[2]{TRIUMF, Vancouver, BC V6T 2A3, Canada}
\author[3]{I.~Alkhatib} \affil[3]{Department of Physics, University of Toronto, Toronto, ON M5S 1A7, Canada}
\author[4]{D.W.P.~Amaral} \affil[4]{Department of Physics, Durham University, Durham DH1 3LE, UK}
\author[5]{T.~Aralis} \affil[5]{Division of Physics, Mathematics, \& Astronomy, California Institute of Technology, Pasadena, CA 91125, USA}
\author[6]{T.~Aramaki} \affil[6]{Department of Physics, Northeastern University, 360 Huntington Avenue, Boston, MA 02115, USA}
\author[7]{I.J.~Arnquist} \affil[7]{Pacific Northwest National Laboratory, Richland, WA 99352, USA}
\author[8]{I.~Ataee~Langroudy} \affil[8]{Department of Physics and Astronomy, and the Mitchell Institute for Fundamental Physics and Astronomy, \newline Texas A\&M University, College Station, TX 77843, USA}
\author[8]{E.~Azadbakht} %\affil{Department of Physics and Astronomy, and the Mitchell Institute for Fundamental Physics and Astronomy, Texas A\&M University, College Station, TX 77843, USA}
\author[9]{S.~Banik} \affil[9]{School of Physical Sciences, National Institute of Science Education and Research, HBNI, Jatni - 752050, India}
\author[10]{C.~Bathurst} \affil[10]{Department of Physics, University of Florida, Gainesville, FL 32611, USA}
% ...